\documentclass[]{aa}
\usepackage{amssymb,graphicx,natbib,txfonts}

\authorrunning{Pace et al.}

\titlerunning{Statistical properties of SZ and X-ray cluster detections}

\begin{document}

\title{Statistical properties of SZ and X-ray cluster detections}

\author{Francesco Pace\inst{1},
  Matteo Maturi\inst{1},
  Matthias Bartelmann\inst{1},
  Nico Cappelluti\inst{2},
  Klaus Dolag\inst{3},
  Massimo Meneghetti\inst{4},
  Lauro Moscardini\inst{5,6}}

\institute{$^1$ ITA, Zentrum f\"ur Astronomie, Universit\"at Heidelberg,
  Albert \"Uberle Str. 2, D-69120 Heidelberg, Germany \\
  $^2$ Max Planck Institut f\"ur Extraterrestrische Physik, D-85478
  Garching, Germany \\
  $^3$ Max Planck Institut f\"ur Astrophysik, D-85478
  Garching, Germany \\
  $^4$  INAF-Osservatorio Astronomico di Bologna, Via Ranzani 1,
  I-40127 Bologna, Italy \\
  $^5$  Dipartimento di Astronomia, Universit\`a di Bologna, Via Ranzani 1,
  I-40127 Bologna, Italy \\
  $^6$ INFN-National Institute for Nuclear Physics, Sezione di Bologna, Viale
  Berti Pichat 6/2, I-40127 Bologna, Italy}

\date{\emph{Astronomy \& Astrophysics, submitted}}

\abstract{}{We calibrate the number density, completeness, reliability
  and the lower mass limit of galaxy-cluster detections through their
  thermal SZ signal, and compare them to X-ray cluster detections.}{We
  simulate maps of the thermal SZ effect and the X-ray emission from
  light cones constructed in a large, hydrodynamical, cosmological
  simulation volume, including realistic noise contributions. The maps
  are convolved with linear, optimised, single- and multi-band filters
  to identify local peaks and their signal-to-noise ratios. The
  resulting peak catalogues are then compared to the halo population
  in the simulation volume to identify true and spurious
  detections.}{Multi-band filtering improves the statistics of SZ
  cluster detections considerably compared to single-band
  filtering. Observations with the characteristics of ACT detect
  clusters with masses $M\ge6-9\times10^{13}\,M_\odot/h$, quite
  independent of redshift, reach $50\%$ completeness at
  $\sim10^{14}\,M_\odot/h$ and 100\% completeness at
  $\sim2\times10^{14}\,M_\odot/h$. Samples are contaminated by only a
  few per cent spurious detections. This is broadly comparable to
  X-ray cluster detections with XMM-\textit{Newton} with 100~ks
  exposure time in the soft band, except that the mass limit for X-ray
  detections increases much more steeply with redshift than for SZ
  detections. A comparison of true and filtered signals in the SZ and
  X-ray maps confirms that the filters introduce at most a negligible
  bias.}{}


\maketitle

\section{Introduction}
Galaxy clusters efficiently trace the large-scale structure of the Universe
\citep{DJEIetal2005.1,GHU2006.1}, and their mass function strongly depends on
the cosmological parameters
\citep{JWERABA2003.1,ESEetal2007.1,WFAZHA2007.1}. They can be easily studied
at different frequencies, ranging from microwaves to X-rays, and via weak and
strong gravitational lensing
\citep{DCLetal2004.1,JFHEDNSP2005.1,JYTAZHFA2005.1,JMCOetal2006.1}.
The Sunyaev-Zel'dovich (hereafter SZ) effect and the X-ray emission
are complementary, and combining them permits individually impossible
constraints. Moreover, the SZ effect allows cluster detections to
larger distances than X-ray emission. In turn, gravitational lensing
probes the total mass rather than the gas, and is most sensitive at
intermediate distances between the sources and the observer.

Expecting wide-field data in the near future, many authors developed
different techniques to detect clusters through their thermal SZ signal
\citep[][]{JMDIetal2002.1, MLOetal2006.1, DHEetal2002.1,
  DHEetal2002.2} and investigated the statistical properties of SZ
detections in simulated maps \citep[see,
  e.g.,][]{MLOetal2006.1,JBMEetal2006.1,CVAMWH2006.1,BMSCMBA2006.1}. These
studies were based on analytic or semi-analytic cluster models applied to
cosmological simulations, but gas physics was not included, which is known to
affect the appearance of clusters \citep[see,
  e.g.][]{SKAetal2004.1,EPUetal2005.1,ABOetal2007.1}. Physical and
statistical properties of X-ray galaxy clusters have been extensively
studied \citep{SBOetal2004.1,MROetal2006.2,PMAetal2004.1}, showing
that X-ray cluster properties are very sensitive to gas physics
\citep[see, e.g.][]{JECAetal2002.1}.

In this paper, we investigate the statistical properties of cluster detections
in simulated maps via single- and multi-band matched filters, applied to X-ray
and thermal SZ maps. We study completeness and contamination of cluster
catalogues and the lowest detectable cluster mass. We also cross-correlate
X-ray and SZ cluster detections. The maps are simulated by projecting the
outputs of a cosmological hydrodynamical simulation \citep{SBOetal2004.1} of
the concordance $\Lambda$CDM model along deep light cones. The simulation
contains gravitation and gas dynamics and includes also several physical
processes that affect the baryonic component, such as a star formation,
supernova feedback, radiative cooling and a simple reionisation
scenario.

Our final goal is to optimise cluster detections in terms of reliability and
completeness with linear filtering techniques applied to multi-wavelength
data. Such cluster catalogues provide information on the underlying cosmology,
the statistical properties of the cluster population and its redshift
evolution, and thus on the non-linear structure growth in the universe.

We summarise the SZ effect, the X-ray emission and our models for them in
Sect.~\ref{sect:signal}. The cosmological simulation is briefly described in
Sect.~\ref{sect:simulations} together with our procedure for creating SZ and
X-ray maps including noise. The single and multi-band filters are described
in Sect.~\ref{sect:Filter}. In Sect.~\ref{sect:detections}, we analyse the
properties of the synthetic catalogues obtained by applying the single- and
multi-band filters, while the correlation between SZ and X-ray detections is
discussed in Sect.~\ref{sect:correlation}. Our main results are summarised in
Sect.~\ref{sect:conclusions}.

\section{The observables}
\label{sect:signal}

\subsection{The SZ effect}

The SZ effect is due to the inverse Compton scattering of CMB photons off the
intracluster electrons. The thermal SZ (tSZ) effect is caused by the thermal
motion of the electrons, the kinetic SZ (kSZ) effect by the bulk motion of the
clusters. The kSZ effect is typically an order of magnitude weaker than the
tSZ effect. The tSZ effect is quantified by the dimension-less Compton
$y$-parameter
\begin{equation}\label{eq:tSZ}
  y=\frac{k_B\sigma_T}{m_ec^2}\int dl\,n_eT_e\;,
\end{equation}
integrating over the product of the electron number density $n_e$ and their
temperature $T_e$, where $k_B$ is the Boltzmann constant, $\sigma_T$ the
Thompson cross-section, $m_e$ the electron rest mass and $c$ the speed of
light. The tSZ effect changes the CMB brightness temperature by
\begin{equation}
  \frac{\Delta T}{T}=yg_{\nu}(x)\;,
\end{equation}
where $g_{\nu}(x)$ is the tSZ spectrum
\begin{equation}\label{eq:gx}
  g_{\nu}(x)=\frac{x}{\tanh{(x/2)}}-4
\end{equation}
as a function of the dimension-less CMB photon energy $x\equiv
h\nu/(k_BT_{CMB})$. In the Rayleigh-Jeans limit ($x\ll 1$) $g_{\nu}(x)\simeq
-2$, and the tSZ effect vanishes at $\nu\simeq 217$~GHz.

The kSZ effect causes brightness-temperature fluctuations
\begin{equation}\label{eq:kSZ}
  \frac{\Delta T}{T}=-b=-\frac{\sigma_T}{c}\int dl\,n_e v_r\;,
\end{equation}
where $v_r$ is the radial component of the cluster velocity, defined positive
for receding clusters. The kSZ effect does not depend on frequency and is thus
best estimated from observations near $\nu\simeq 217$~GHz, where the tSZ
effect vanishes.

\subsection{The X-ray emission}
\label{sect:X-ray}

X-ray emission of galaxy clusters due to thermal free-free emission
(bremsstrahlung) has the bolometric emissivity
\begin{equation}
  \epsilon_{\rm ff}\approx 1.2\times 10^{-27} T^{1/2}n_P^2g(T)
\label{eq:eps_bol}
\end{equation}
in cgs units, where the gas temperature $T$ is given in keV, $n_P$ is the
proton density in cgs units, and $g(T)$ is the Gaunt factor. The thermal
free-free emission spectrum is $f(\nu)=e^{-h\nu/k_BT}g(\nu,T)$, where
$\gamma=Z/Z_\odot$ is a parameter depending on the metallicity
($Z\approx0.3\,Z_\odot$) \citep[see
  e.g.][]{PROetal2004.1,YHAetal2004.1}.

The X-ray luminosity in the energy interval [$E_1$,$E_2$] is given by $L_{\rm
  band}=L_X^{\rm bol}\,F_{\rm band}^{[E_1,E_2]}$, where $L_X^{\rm
  bol}=\epsilon_{\rm ff}V$, and $V$ is the emitting volume. The band-function
is defined as
\begin{equation}\label{eq:band}
  F_{\rm  band}^{[E_1,E_2]}(T)=\int_{x_{1,0}(1+z)}^{x_{2,0}(1+z)}
  C\,f(x)\,dx \;,
\end{equation}
where $x$ is defined as above and $C$ is a normalisation constant defined by
$C\,\int_0^\infty f(x)dx=1$. The subscript 0 in the integration boundaries
means that the band limits are taken in the observer's rest-frame, thus the
$(1+z)$ term in Eq.~(\ref{eq:band}) represents the K-correction. We shall
consider three different bands: soft [$(E_1, E_2)=(0.5,2)$~keV], hard [$(E_1,
  E_2)=(2,4)$~keV], hardest [$(E_1, E_2)=(4,10)$~keV].

\section{The simulations}
\label{sect:simulations}

\subsection{The cosmological simulation}

To create realistic mock catalogues of SZ and X-ray halos, we use the outputs
of a hydrodynamical cosmological simulation \citep{SBOetal2004.1} carried out
using the {\small GADGET-2} code \citep{VSP2005.1}. Here we briefly report the
main characteristics of the simulation. The SZ effect and X-ray properties of
objects identified in this simulation are described in other studies
\citep[e.g.][]{ADIetal2005.1, MROetal2007.1, SETetal2004.1, GMUetal2004.1,
  ERAetal2005.1, MROetal2006.1}, where further detail can be found.

The simulation assumes a concordance $\Lambda$CDM model with matter density
parameter $\Omega_m=0.3$, cosmological constant $\Omega_{\Lambda}=0.7$, and
Hubble parameter $h\equiv (H_0/100~\rm{km/s/Mpc})=0.7$. The power spectrum is
normalised to $\sigma_8=0.8$. These cosmological parameters are consistent
with recent observational estimates derived from CMB and weak lensing data
\citep[e.g.][]{DNSPetal2007.1,HHOetal2006.1}.

The computational box with a comoving side length of 192 Mpc/h is filled with
$480^3$ particles each of dark matter and gas. The mass resolution is thus
$4.6\times10^9 M_{\odot}/h$ and $6.9\times10^8 M_{\odot}/h$ for dark matter
and gas, respectively.  The Plummer-equivalent gravitational softening is set
to $\epsilon_{Pl}=7.5~\rm{kpc/h}$ comoving for $0\le z\le 2$, and switches to
physical units at higher redshifts. The gas component is treated including
several physical processes: a hybrid multi-phase model for star formation in
the interstellar medium \citep{VSPLHE2003.1}, radiative cooling within an
optically thin gas consisting of $76\%$ of hydrogen and $24\%$ of helium by
mass, supernova feedback to model galactic outflows, and heating by a
time-dependent, photoionising uniform UV background given by quasars
reionising the Universe at $z\approx 6$ \citep{FHAPMA1996.1}. The output of
the simulation consists of one hundred snapshots, logarithmically equidistant
in redshift between $z=9$ and $z=0$. We use these snapshots to create our mock
light-cones.

\subsection{Construction of the halo catalogues}

To identify the halos in each snapshot, we run a friends-of-friends algorithm
with a linking length of $0.15$ times the mean particle separation to identify
particles belonging to the same group. Next, we identify the halo centre as
the position of the particle at the local minimum of the gravitational
potential. The final halo catalogue contains the positions, the virial masses,
radii and redshifts of all halos in each snapshot. More details can be
found in \cite{FPAetal2007.1}.

\begin{figure}[!ht]
  \centerline{\includegraphics[angle=-90,width=0.9\hsize]{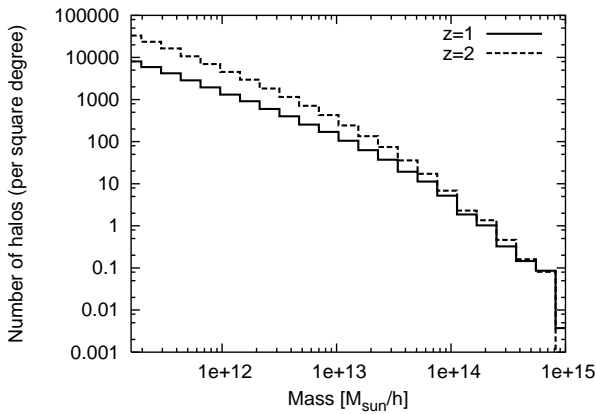}}
  \caption{Average number of halos per mass bin per square degree. The solid
    and dashed lines show the numbers for light-cones extending to $z=1$ and
    $z=2$, respectively. The number of high-mass halos is approximately the
    same in both case, while the number of low-mass halos is higher for the
    deeper cone. The values are averaged over eleven different realisations.}
  \label{fig:1}
\end{figure}

Figure~\ref{fig:1} shows the mass function of the halos contained in light
cones for limiting redshifts of $z=1$ and $z=2$, averaged over eleven
different realisations and normalised to one square degree. Light-cones
reaching $z=2$ contain more low-mass halos than light-cones up to $z=1$
because of their larger volume, while the number of massive halos is almost
unchanged since most of them are at low redshifts.

\subsection{Construction of the simulated SZ and X-rays light-cones}

To construct a realistic three-dimensional distribution of matter, we
construct several light cones by stacking the snapshots of our cosmological
simulation at different redshifts, observing two aspects: First, we shift and
rotate each snapshot to make it completely independent from the
others. Second, we cut the boxes appropriately to avoid including matter from
more than one snapshot, since contiguous snapshots partially overlap in
redshift.

To create the maps, we proceed in two steps. We first project gas particles on
a two-dimensional grid, and then stack the resulting planes summing the
contribution from each individual plane to obtain the final map. Projecting
the gas particles on a regular grid, we take the smoothing length of each
particle into account and use the SPH \citep[Smoothed Particle Hydrodynamics,
  see e.g.][]{JJMO1992.1} interpolating kernel from the original simulation
\citep{JJMOJCLA1985.1}. Particles are distributed over grid cells with a
fractional weight calculated from the overlap of the spline kernel with the
grid cell.

Since each particle has its individual smoothing length, the number of pixels
involved in each projection varies, which produces appropriately smooth
maps. This algorithm was applied by \cite{EPUetal2005.1} to study the impact
of gas physics on strong lensing by clusters and by \cite{ACdaSIetal2000.1} to
create simulated SZ maps.

We repeated this procedure for each snapshot of the simulation, and then
summed the contributions of all planes to obtain the final map. We created
eleven different maps including matter up to redshift $z=1$ and $z=2$ with a
resolution of $2048\times 2048$ pixels. The opening angle of the light-cone is
determined by the last plane of the pile, because its dimension is fixed in
comoving units: it corresponds to $4.9$ and $3.1$ degrees, for light-cones up
to $z=1$ and $z=2$, respectively.

We built our light-cones in the same way as for gravitational-lensing maps,
i.e.~projecting the whole box on a regular grid and then selecting the portion
of the plane enclosed by the light cone (see Fig.~1 in
\cite{FPAetal2007.1}). In other words, our planes all have the same
number of pixels with constant comoving size, chosen according to the
simulation resolution. Other authors \citep[see
  e.g.][]{ACdaSIetal2000.1,MROetal2006.1} prefer to construct the
light-cones keeping the angular resolution of the pixels constant,
such that each plane contributes the same amount of pixels. We avoid
this because then the pixel resolution would exceed the physical
resolution of the cosmological simulation, especially at low redshift.

\subsubsection{Simulating the SZ maps}

In order to create a map for the SZ effect we replace the line-of-sight
integrals in Eqs.~(\ref{eq:tSZ}) and (\ref{eq:kSZ}) by sums over gas
particles. The contributions to the tSZ and kSZ effects of the $i$-th particle
are thus
\begin{eqnarray}
  \label{tSZ_i}
  y_i &=& \frac{1}{L_{\rm pix}^2}\frac{k_B\sigma_T}{m_ec^2}n_{e,i}T_i\\
  \label{kSZ_i}
  b_i &=& \frac{1}{L_{\rm pix}^2} \frac{\sigma_T}{c}n_{e,i}v_{r,i} \;,
\end{eqnarray}
respectively, where $L_{\rm pix}$ is the physical size of a pixel at the
distance of the corresponding plane. We approximate the radial velocity
$v_{r,i}$ with the velocity component along the $z$-direction and relate the
temperature of the gas particles $T$ (in Kelvin) to their internal energy per
unit of mass ($U$ in km$^2$/s$^2$) through
\begin{equation}
  T=10^6\times\frac{2}{3k_B}m_p\mu~U
\end{equation}
as for a monatomic ideal gas, where $m_p$ is the proton mass and $\mu$ is the
mean molecular weight
\begin{equation}\label{mu}
  \mu=\frac{1+4y_{He}}{1+y_{He}+n_e} \;.
\end{equation}
We adopt $y_{He}\approx 0.08$.

\subsubsection{Simulating the X-ray maps}

We create the light-cone maps for the X-ray emission in the three energy bands
mentioned above: soft ($0.5-2$~keV), hard ($2-4$~keV) and hardest
($4-10$~keV). We model the contribution of each particle to the X-ray signal
in the soft and hard bands adopting the MeKaL model \citep{MeKaL1995.1} as
implemented in XSPEC \citep{KAAR1996.1}. It returns the emission spectrum of
the hot diffuse plasma and is particularly suited for the soft band where the
influence of the metal line emission is important. Using this model, the X-ray
luminosity is
\begin{equation}
  L_{X,i}=(m_p\mu)^{-2}m_i\rho_i x_e\Lambda(T_i,Z_i,E_1',E_2') \;,
\end{equation}
where $x_e\equiv n_e/n_H$ is the ratio between the number density of free
electrons and hydrogen nuclei. The cooling function $\Lambda$ depends on the
particle temperature, on the metallicity and on the energy band
$[E_1,E_2]$.

For the hardest band, we can use the model described in \cite{SBOetal1999.1}
because the influence of metal lines is negligible there, and a simple
power-law parametrisation suffices. It is then possible to parametrise X-ray
luminosity of the $i$-th particle in the hardest band by
\begin{equation}
  L_{X,i}^{hardest}\approx 1.7\times 10^{42} \frac{m_i\rho_i\sqrt{T_i}}{\mu^2}
  ~\textrm{erg/s} \;,
\end{equation}
where $m_i$, $\rho_i$ and $T_i$ are its mass, its density, and its temperature
in keV.

For each gas particle, the X-ray flux is
\begin{equation}\label{eq:Xint}
  I_{X,i}=\frac{L_{X,i}}{4\pi d_L(z)^2}\,\mathrm{erg/s/cm^2} \;,
\end{equation}
where $d_L(z)$ is the luminosity distance of the particle from the
observer. As for the SZ maps, we project the X-ray flux on a regular grid
using the SPH kernel.

\subsection{Simulating observations}

We simulate observations with specific instruments, namely the Atacama
Cosmology Telescope (ACT) for the SZ effect because of its high angular
resolution and sensitivity, and the space telescopes XMM-{\em Newton} and
Chandra for the X-ray emission, assuming exposure times of 30 and 100 ks.

\subsubsection{Noise inclusion into the SZ map}

\begin{figure}[!ht]
  \centerline{\includegraphics[angle=-90,width=0.9\hsize]{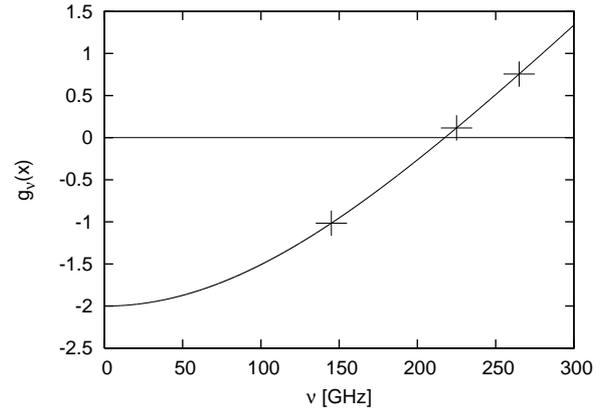}}
  \caption{The spectrum of the tSZ effect, with the ACT frequency bands
    overplotted (crosses).}
  \label{fig:2}
\end{figure}

Figure~\ref{fig:2} shows the tSZ spectrum $g_\nu(x)$. The crosses indicate the
three frequency bands of ACT. The $\nu=225$~GHz channel is close to the
frequency where $g_\nu(x)$ vanishes and the dominant signal will be due to the
kSZ effect. Nonetheless, we shall see that this channel is very important for
our multi-band frequency filter because it contains precious information on
the noise due to the primary CMB anisotropies.

The noise in tSZ observations has three main contributions: the CMB radiation,
the instrumental noise, and the kSZ effect. The noise due to the primary CMB
anisotropies and to the instrument can be modelled as two independent Gaussian
random fields. We used the CMB power spectrum as computed with CMBEASY
\citep{MDO2005.1}, and the instrumental noise power spectrum
\begin{equation}\label{eq:Inoise}
  C_l^{noise}=w^{-1}\exp\left[\frac{l(l+1)FWHM^2}{8\ln 2}\right]\;,
\end{equation}
where $FWHM$ is the full width at half maximum of the instrumental beam in
arcmins, $l$ represents the multipole order, and $w^{-1}\equiv(\Delta
T/T~FWHM)^2$. The quantity $\Delta T/T$ gives the sensitivity of the
instrument on the scale of the beam \citep{LKN1995.1}. Finally, the maps are
convolved with the instrumental beam. Table~\ref{tab:ACT} summaries some
parameters used to mimic ACT observations.

\begin{table}[!ht]
  \begin{center}
    \begin{tabular}{ccc}
      \hline
      \hline
      Central Band Frequency  & FWHM & $\Delta T$/beam \\
      (GHz) &  (arcmin) & ($\mu$K)        \\
      \hline
      145   & 1.7  & 2               \\
      225   & 1.1  & 3.3             \\
      265   & 0.93 & 4.7             \\
      \hline
    \end{tabular}
  \end{center}
  \caption{Parameters used to produce mock observations with ACT. In
    the first column we show the central bands frequencies
    at which ACT operates; second and third columns list
    the corresponding FWHM and sensitivity, respectively.}
  \label{tab:ACT}
\end{table}

\begin{figure*}[!ht]
  \centerline{\includegraphics[width=\hsize]{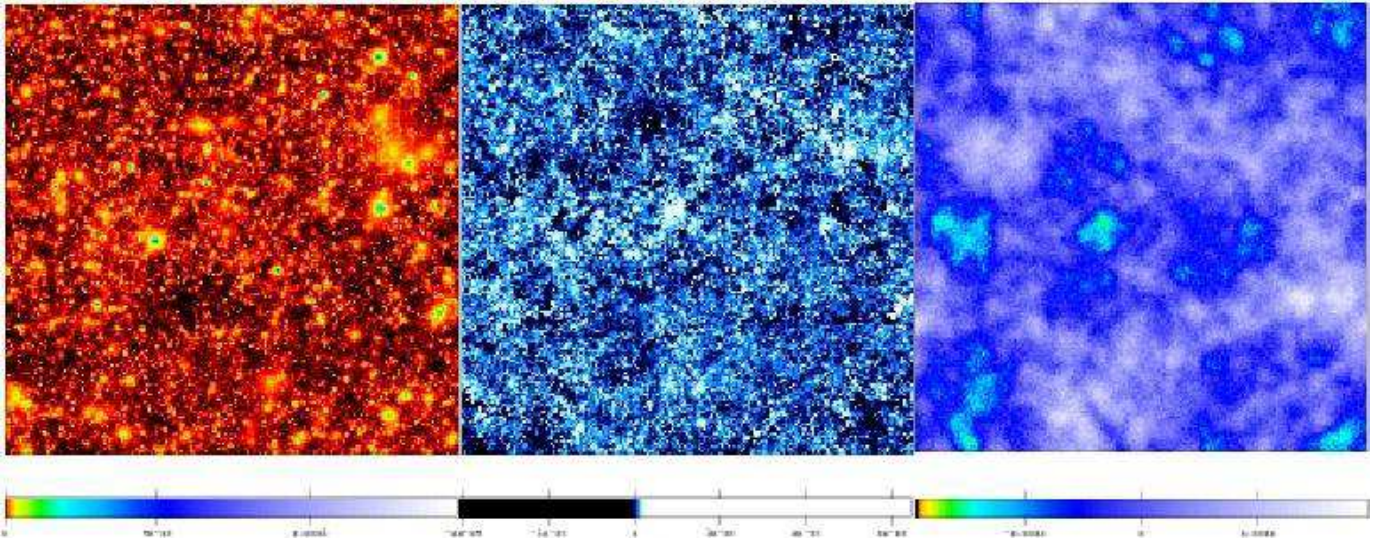}}
  \caption{Examples of SZ maps including cosmic structures to $z=1$. The left
  panel shows the simulated tSZ map, the central panel the kSZ map, and the
  right panel shows a simulated observation of the same field in the 145~GHz
  channel of ACT, including the kSZ effect, the CMB signal and the
  instrumental noise. The size of the maps is $\approx 4.9$~deg$^2$.}
  \label{fig:3}
\end{figure*}

In Fig.~\ref{fig:3} we show an example of the SZ simulations. The right panel
displays the simulated observation with the $\nu=145$~GHz channel of ACT,
including the kSZ effect, the instrumental noise and the primary CMB
fluctuations. The left and central panel show instead a simulated map
for the tSZ and kSZ effect, respectively.

\subsubsection{Noise inclusion into the X-ray maps}

In order to transform our X-ray flux maps into count rates, we first multiply
them by the energy conversion factor of the instrument. For each energy band,
this factor is computed assuming a spectral model, consisting of thermal
free-free emission with $k_BT=4~\rm{keV}$, convolved with the response
matrices of the EPIC-PN detector on-board XMM-{\em Newton} or with the
ACIS-I array on-board {\em Chandra}.  Assuming an exposure time
(either 30 or 100 ks), we transform our count-rate maps into
photon-count maps. We then added the background of an X-ray
observation, given by the sum of the detector noise, the unresolved
X-ray background, and the particle background. Finally, we convolved
the resulting maps with the point-spread function (PSF) of the
instrument (5 and 0.5 arcsec for XMM-{\em Newton} and Chandra,
respectively). In order to reproduce the observational noise, we added
Poissonian noise according to the local photon flux. The noise levels
used in the simulated observations with XMM-{\em Newton} and Chandra
are summarised in Tab.~\ref{tab:Xtel}.

\begin{table}[!ht]
  \begin{center}
    \begin{tabular}{cccc}
      \hline
      \hline
      Instrument & Band & Noise                & Effective area \\
                 &      & (counts/sec/deg$^2$) & (at 1.4 keV)     \\
      \hline
      XMM-{\em Newton}  & $0.5-2$~keV &  24.624 &       \\
      (EPIC-PN)            & $2-4$~keV   &  11.    & $\backsim 1200$
      cm$^2$ \\
                        & $2-10$~keV  &  19.64  &       \\
      \hline
      Chandra    & $0.5-2$~keV &        2.96        &       \\
      (ACIS-I)   & $2-4$~keV   &        2.47        & $\backsim
      600$ cm$^2$ \\
                 & $4-10$~keV  &        7.4         &       \\
      \hline
    \end{tabular}
  \end{center}
  \caption{XMM-{\em Newton} and Chandra noise levels and effective
  detector areas for the EPIC and ACIS-I instruments, respectively.}
  \label{tab:Xtel}
\end{table}

\begin{figure}[!ht]
  \centerline{\includegraphics[width=\hsize]{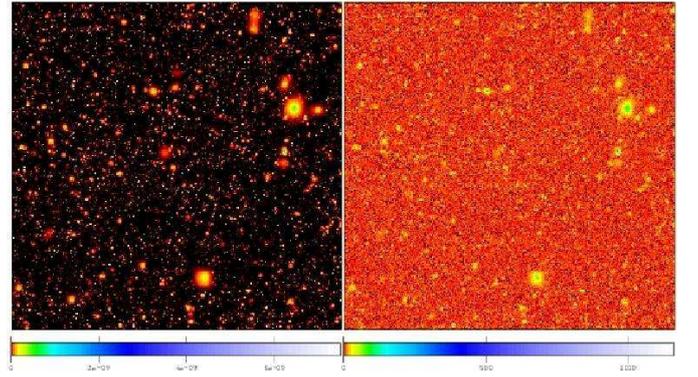}}
  \caption{Examples of X-ray maps. The left panel shows the X-ray flux
    in the soft band for the same field displayed in Fig.~\ref{fig:3};
    the right panel represents a simulated observation with XMM-{\em
    Newton} assuming an exposure time of 30 ks. The size of the image
    is the same as in Fig.~\ref{fig:3}.}
  \label{fig:4}
\end{figure}

In Fig.~\ref{fig:4} we give an example of X-ray simulation in the soft
band. The left panel shows a synthetic X-ray map, while the
corresponding simulated observation with XMM-{\em Newton}, including
the background, instrumental and Poissonian noises is shown in the
right panel. An exposure time of 30 ks is adopted here. The maps are
based on the same mass distribution underlying Fig.~\ref{fig:3}.

\section{Filtering method}
\label{sect:Filter}

We now describe the matched filter used to detect clusters in the simulated
maps. For the X-ray observations, the noise is not correlated across bands or
with the noise in SZ observations, thus the multi-band filter described above
will be applied only to the SZ observations.

\subsection{Single-band filter}
\label{sect:sbf}

The single-band filter $\Psi$ we use is analogous to the optimal filter
proposed by \cite{MMAetal2005.1}. We suppose that the observational data
$D(\vec{\theta})$ are given by
\begin{equation}
  D(\vec{\theta})=S(\vec{\theta})+N(\vec{\theta})\;,
\end{equation}
where $S(\vec{\theta})=A\tau(\vec{\theta})$ is the signal from
sources whose spatial shape is modeled by $\tau$, and $N(\vec{\theta})$
represents the noise. The estimated amplitude of the linearly filtered
signal is

\begin{equation}
  A_{\rm est}(\vec\theta)=\int~d^2 \theta'
  D(\vec\theta')\Psi(|\vec\theta'-\vec\theta|)\;.
  \label{eq:estimate}
\end{equation}

Imposing the conditions that the optimal filter be unbiased
($\langle A_{est}-A\rangle=0$) and that its variance
($\sigma^2=\langle(A_{\rm est}-A)^2\rangle$) be minimal, its shape
follows from a variational problem with the two previously described
constraints \citep[see][]{MGHAMTE1996.1},
\begin{equation}
  \hat{\Psi}(\vec k)=
  \frac{1}{(2\pi)^2}\left[
    \int\frac{|\hat{\tau}(\vec k)|^2}{P_N(k)}~d^2k\right]^{-1}
  \frac{\hat{\tau}(\vec k)}{P_N(k)} \;,
  \label{eq:SBfilter}
\end{equation}
where the hats denote Fourier transforms. Equation~(\ref{eq:SBfilter}) shows
that the shape of the filter $\Psi$ is determined by the shape of the signal,
$\tau$, and by the power spectrum of the noise, $P_N$.

\subsection{A simple model for galaxy clusters}

We model the gas distribution for the filter template by the truncated
King profile \citep{IKI1962.1}
\begin{equation}\label{eq:Kprofile}
  \rho(x)=\frac{1}{1+x^2}\sqrt{\frac{|(r_t/r_c)^2-x^2|}{1+x^2}}
  \quad\mathrm{for}\quad x\le r_t\;,
\end{equation}
where $r_c$ is the core radius, $r_t$ is the truncation radius defined as ten
times the virial radius $r_v$, and $x=r/r_c$. Although based on the simple
assumption that gas follows a static and isothermal distribution, this profile
is justified by simulations and X-ray observations \citep[see
  e.g.][]{SBOetal2004.1,AFIetal2001.1}.

The tSZ effect (Eq.~\ref{eq:tSZ}) is proportional to the density $\rho$ and
normalised to the Compton-$Y$ parameter
\begin{equation}
  Y=\int\,y\,d\Omega
\end{equation}
integrated over the solid angle $\Omega$.

The X-ray luminosity is proportional to the square of the electron
density and thus to the square of the profile (\ref{eq:Kprofile}). It is
normalised such that the bolometric luminosity follows the
empirical relation
\begin{equation}
  L_{X,bol}=2.99\times
  10^{44}h^{-2}\left(\frac{T}{6~keV}\right)^{2.6}\;
  erg/s
\end{equation}
\citep{TKIYSU1997.1}, where $T=T_{\rm vir}(r_s)$ is the X-ray temperature. The
luminosity in a given band is computed using Eq.~(\ref{eq:band}). We
do not assume any explicit redshift dependence.

\subsection{Filter dependence on the template}

\begin{figure}[!ht]
  \centering{
    \includegraphics[angle=-90,width=0.49\hsize]{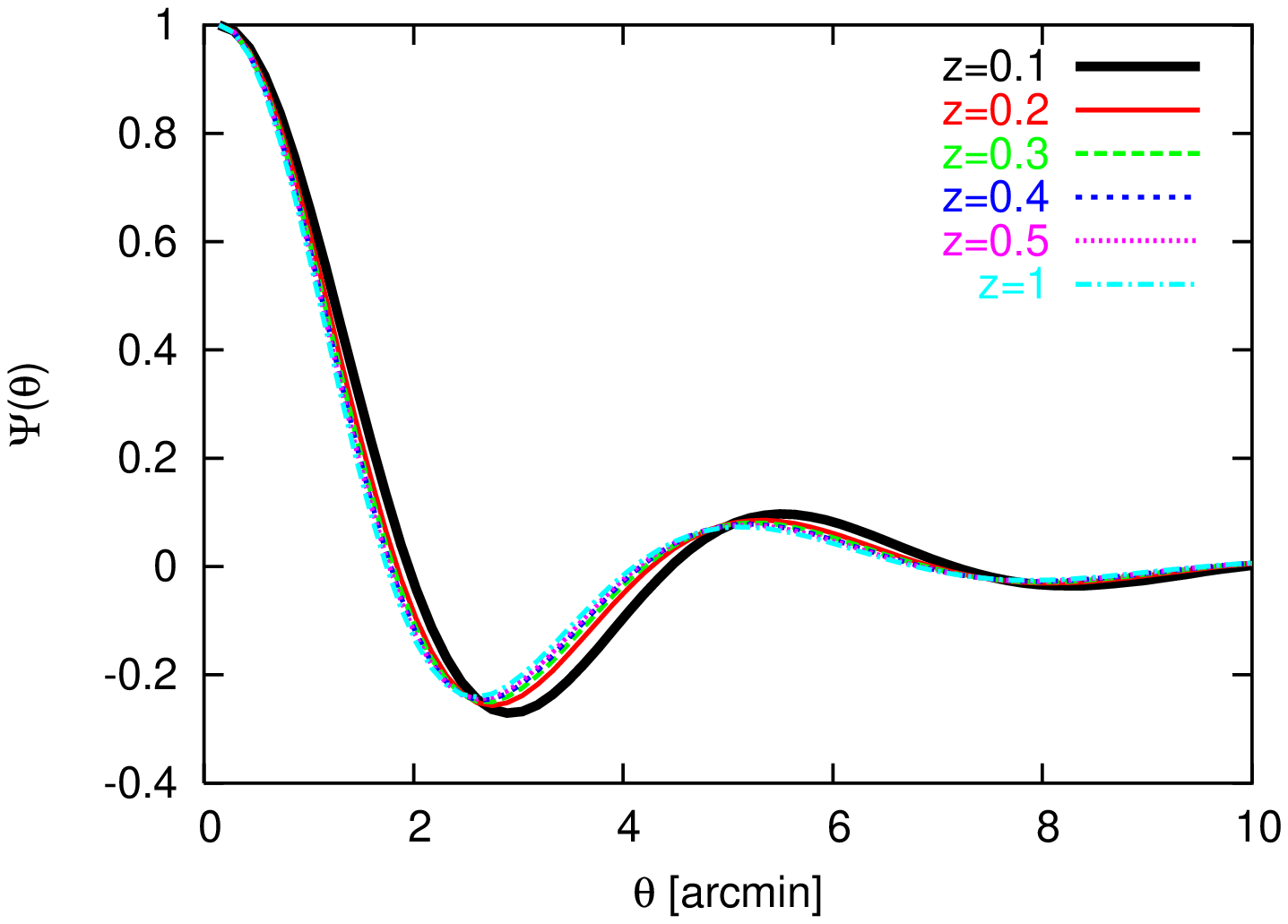}
    \includegraphics[angle=-90,width=0.49\hsize]{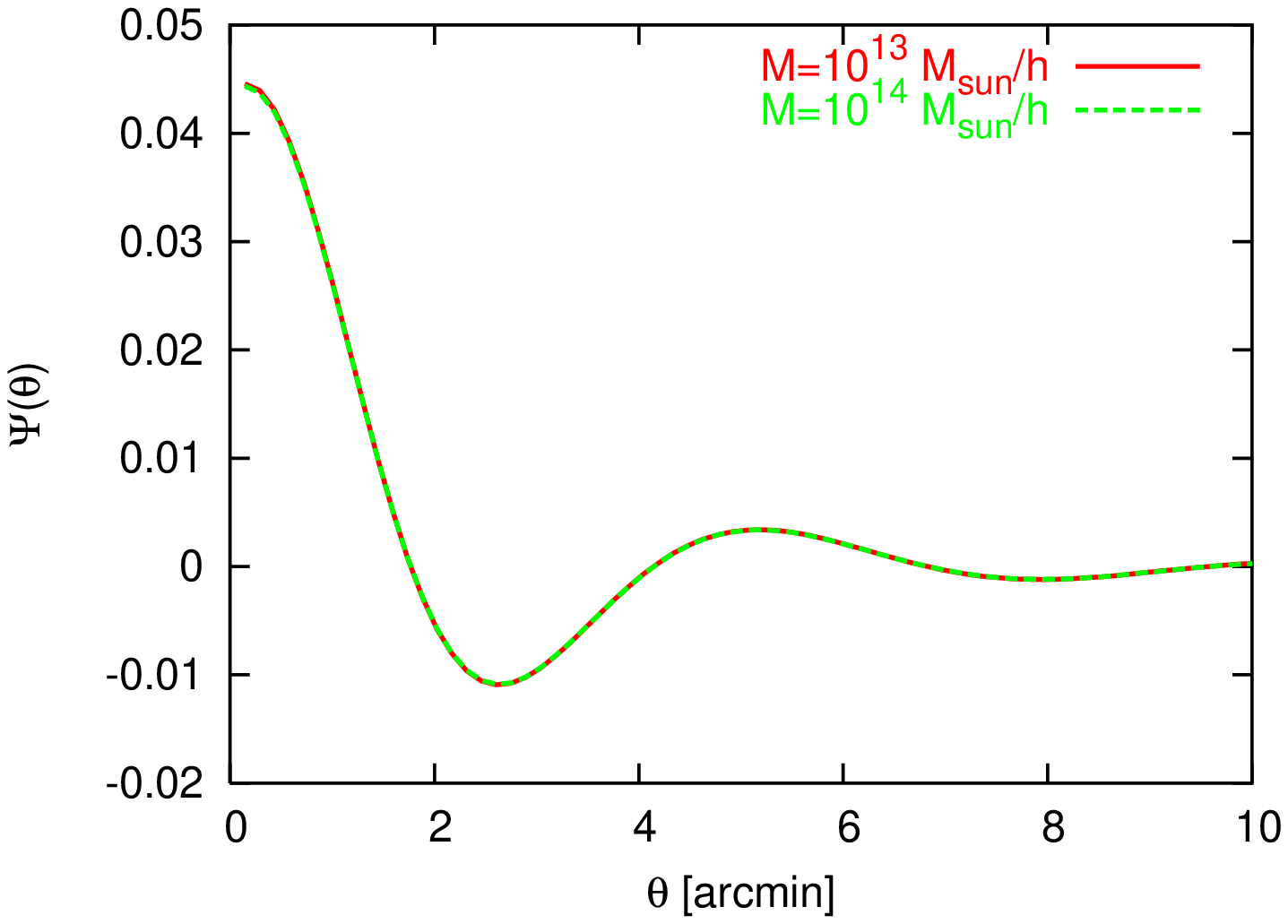}\hfill
    \includegraphics[angle=-90,width=0.49\hsize]{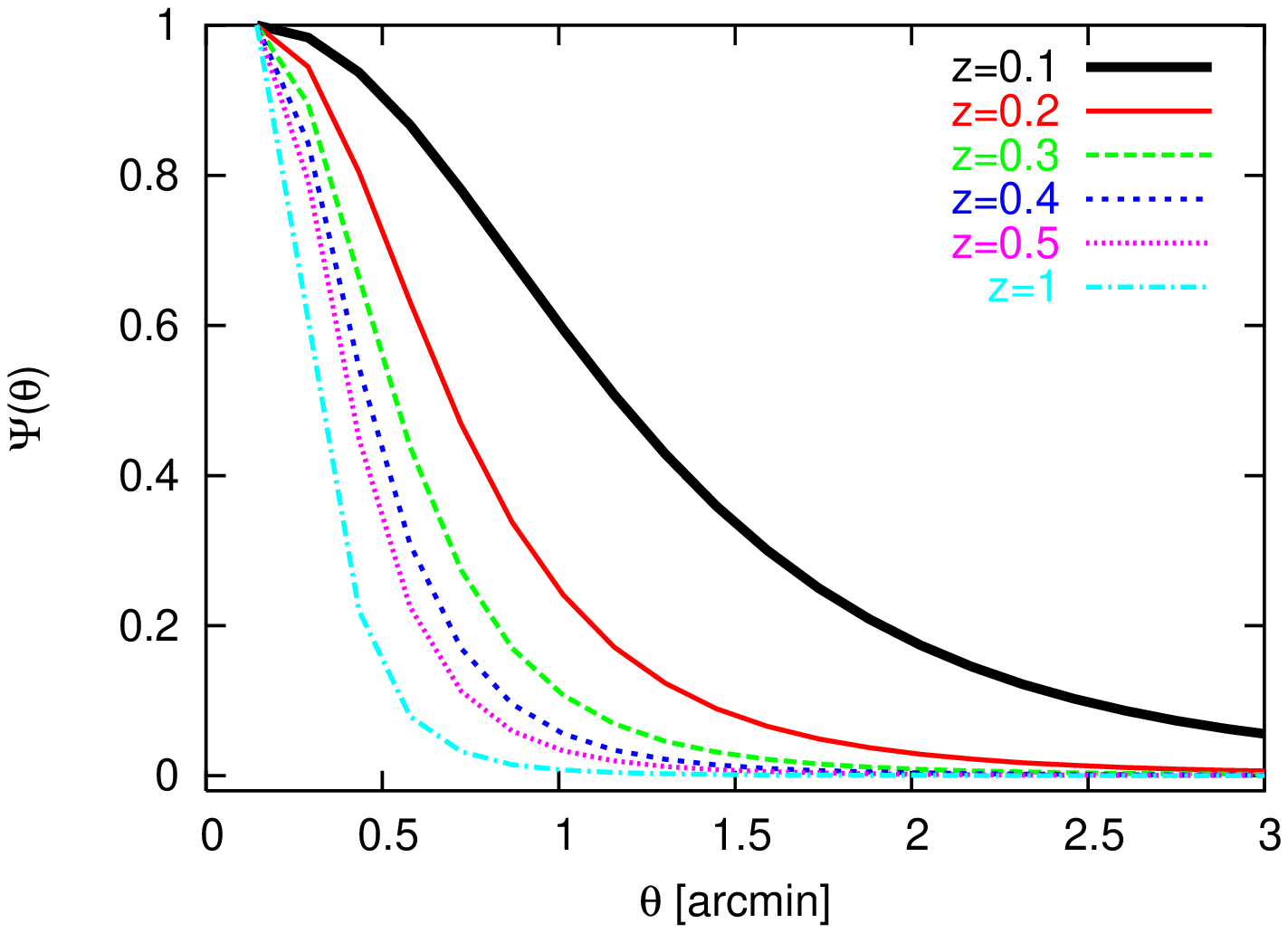}
    \includegraphics[angle=-90,width=0.49\hsize]{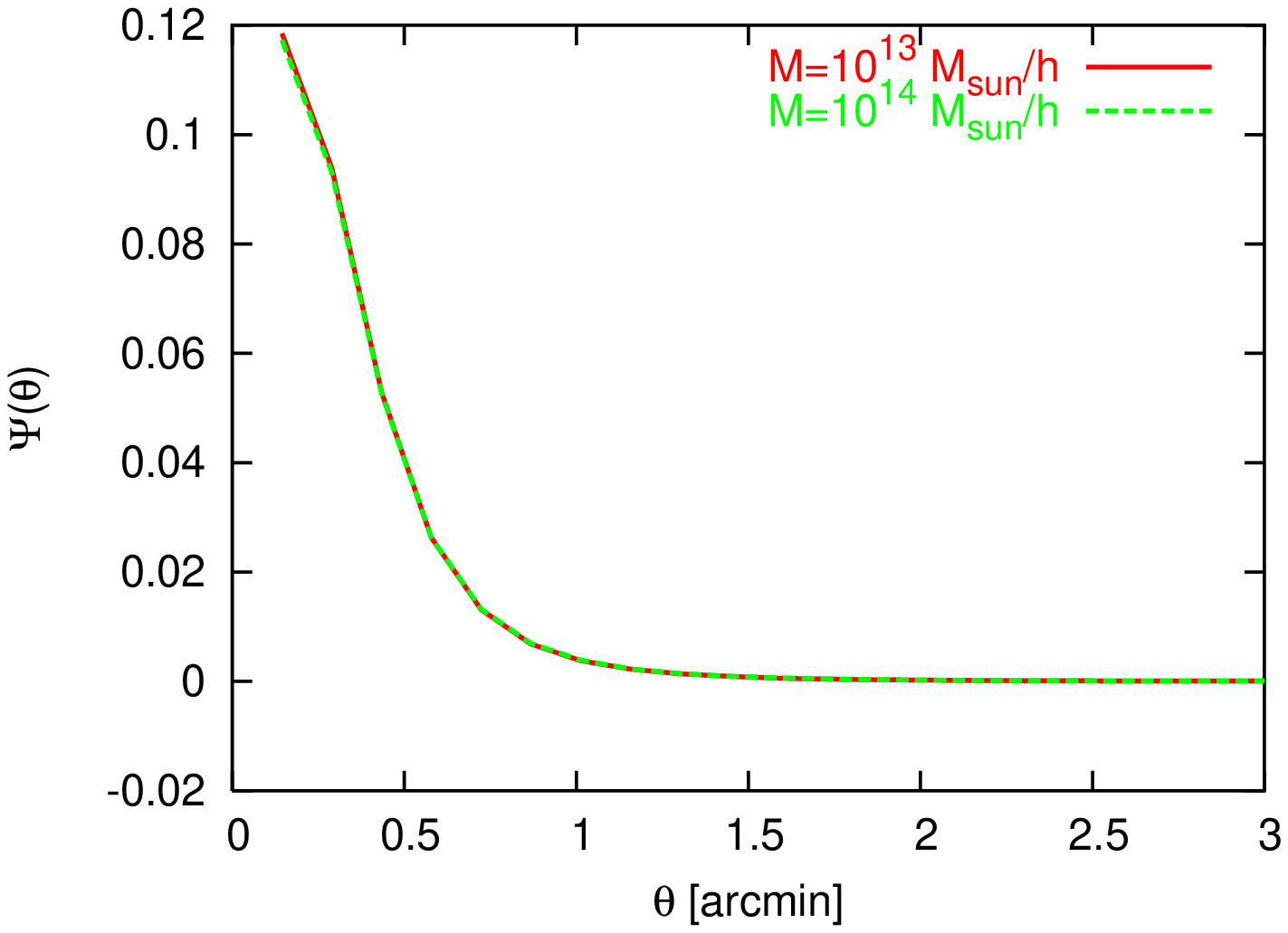}\hfill}
  \caption{The matched filter. The left panels show the filter
    profiles normalized to unity (for a better understanding) for
    different redshifts for a fixed halo mass of
    $M=10^{14}~M_\odot/h$. The right panels show the filter profile
    for different halo masses at a fixed redshift of $z=0.5$. SZ and
    X-ray filter profiles are presented in the upper and lower panels,
    respectively.}
  \label{fig:5}
\end{figure}

Figure~\ref{fig:5} shows the profile of the filter used for the SZ
observations at 145~GHz (upper panels) and for the X-ray observations in the
soft band (lower panels). In the left panels, we assume a template with mass
$M=10^{14}\,M_\odot/h$ and illustrate how the filter depends on the cluster
redshift. In the right panels, we fix the redshift of the template to $z=0.5$
and show how the profile depends on cluster mass, based on the two cases
$M=10^{13}\,M_\odot/h$ (solid line) and $M=10^{14}\,M_\odot/h$ (dashed
line). The upper right panel demonstrates that the shape of the SZ filter is
insensitive to both the mass and the redshift of the template, while only the
normalisation changes.

\begin{figure}[!t]
  \centerline{\includegraphics[angle=-90,width=0.8\hsize]{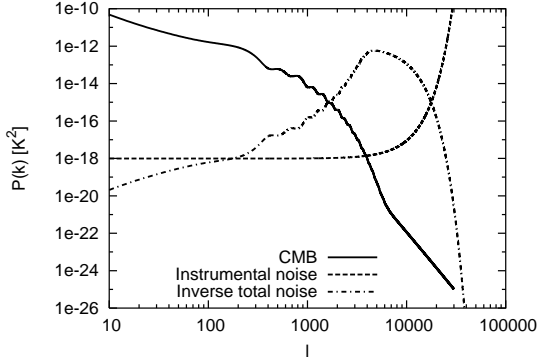}}
  \caption{The noise power spectra. The solid line shows the CMB power
  spectrum, the dashed line the instrumental noise, and the dotted-dashed line
  the inverse of the total noise, given by the sum of CMB and instrumental
  noise. The filter is proportional to the inverse of the total power
  spectrum.}
  \label{fig:6}
\end{figure}

This implies that its sensitivity is independent of the particular choice of
the template. This characteristic results from the shape of the noise power
spectrum, combined from instrumental noise and the CMB: since their power
spectra are independent of both redshift and mass, the minimum of the noise
(and therefore the maximum of the filter) is always at the same
wavenumbers. Due to the particular shape of the noise power spectrum (shown in
Fig.~\ref{fig:6}), the filter needs to oscillate because of the lack of
information in a single band of the global characteristics of the background
noise. Unlike the SZ filter, the X-ray filter (see the lower panels of
Fig.~\ref{fig:5}) is simply proportional to the template because the noise is
white. This implies that structures in the signal do not change. Thus, varying
the template, the X-ray filter changes accordingly.

\subsection{The multi-band matched optimal filter}
\label{sect:mbf}

The multi-band matched filter allows combining information from different
bands. For its complete derivation, we refer to \cite{BMSCetal2006.1} and
\cite{JBMEetal2006.1} and references therein. Generalising the single-band
case, the data obtained at a given frequency $\nu$ are the sum of the signal
$S_\nu(\vec{\theta})$ and the noise $N_\nu(\vec{\theta})$,
\begin{equation}
  D_\nu(\vec{\theta})=S_\nu(\vec{\theta})+N_\nu(\vec{\theta})\;.
\end{equation}
The signal is modelled as
\begin{equation}
  S_\nu(\vec{\theta})=Af_\nu\tau_\nu(\vec{\theta})\;,
\end{equation}
where $A$ is the (band-independent) amplitude, $f_\nu$ is the frequency
dependence of the amplitude and $\tau_\nu(\vec{\theta})$ is the spatial
profile (normalised to unity). The index $\nu$ runs from $1$ to the number of
available bands, $M$. We assume that the background noise has zero mean in
each band and that its statistical properties are fully characterised by the
correlation functions
\begin{equation}\label{eq:correlation}
  C_{\nu_1\nu_2} = \langle
  \hat{N}_{\nu_1}(\vec{k})\hat{N}_{\nu_2}(\vec{k}')^* \rangle=
  (2\pi)^2\delta(\vec{k}-\vec{k}')P_{l,\nu_1\nu_2}(k)\;,
\end{equation}
where $P_{l,\nu_1\nu_2}(k)$ is the cross-power spectrum.

To measure the signal amplitude $A$, we define a linear estimator for each
band,
\begin{equation}
  A_{\rm est,\nu}(\vec\theta)=\int d^2 \theta'
  D_\nu(\vec\theta')\Psi_\nu(|\vec\theta'-\vec\theta|) \;,
\end{equation}
and a total estimate by
\begin{equation}
  A_{\rm est}=\sum_{\nu=1}^MA_{\rm est,\nu}(\vec\theta) \;.
  \label{eq:Aest}
\end{equation}
In analogy with the single-band filter,
$\boldmath{\Psi}(\vec{\theta})=\left[\Psi_\nu(\vec{\theta})\right]$ is the
optimal filter, minimising the estimated variance,
$\sigma^2=\langle(A_{\rm est}-\langle A\rangle)^2\rangle$, and
avoiding bias, $b\equiv\langle A_{\rm est}-A\rangle=0$.\\
In deriving the filter, it is convenient to define the filter vector
$\bar{\boldmath \Psi}(\vec{\theta})=\left[\Psi_\nu(\vec{\theta})\right]$ and
the signal vector
$\bar{\mathbf{F}}(\vec{\theta})=\left[F_\nu(\vec{\theta})\right]$. We further
define the cross-power matrix
$\mathbf{C}^{(l)}=\left[C_{\nu_1\nu_2}^{(l)}\right]$.

\cite{BMSCetal2006.1} show that the filter satisfying these conditions is
\begin{equation}
  \boldmath{\Psi}=\alpha\mathbf{C}^{-1}\mathbf{F} \;,
  \label{eq:mtc_filter}
\end{equation}
where $\mathbf{C}^{-1}$ is the inverse of the matrix $\mathbf{C}$, and the
normalisation factor $\alpha$ is given by
\begin{equation}
  \alpha^{-1}=\mathbf{F}^{\mathrm T}\mathbf{C}^{-1}\mathbf{F} \;.
\end{equation}
The estimate has the variance
\begin{equation}
  \sigma^2=\int\frac{d^2k}{(2\pi)^2}\Psi^{\mathrm T}\mathbf{C}\Psi \;.
\end{equation}

According to Eq.~(\ref{eq:mtc_filter}), the matched filter depends on the
noise, taking advantage of its correlation between all bands. In the special
case $C_{\nu_1\nu_2}=C_{\nu_1}\delta(\nu_1-\nu_2)$, i.e.~in the case of
uncorrelated noise, the matrix $\mathbf{C}$ and thus also its inverse are
diagonal, and the optimal filter for each band had the same shape as the
single-band filter defined in Section \ref{sect:sbf}. Just the normalisation
differes.

\begin{figure}[!ht]
  \centerline{\includegraphics[angle=-90,width=0.8\hsize]{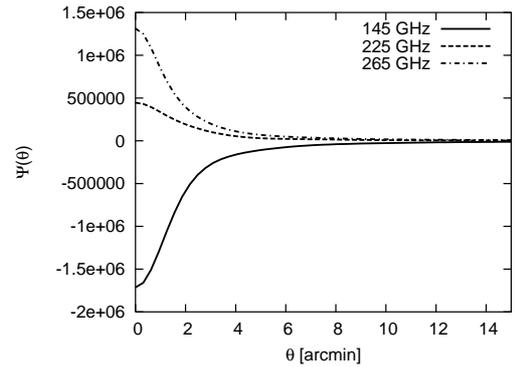}}
  \caption{The profile for the matched tSZ filter for a halo template with
    $M=10^{14}~M_\odot/h$ at $z=0.4$. The solid, dashed and dash-dotted lines
    refer to the 145~GHz, 225~GHz and 265~GHz bands, respectively.}
  \label{fig:7}
\end{figure}

Figure~\ref{fig:7} shows the matched-filter profile for the three bands of ACT
and a template with $M=10^{14}~M_\odot/h$ at $z=0.4$. It is evidently broader
than for a single band (see Fig.~\ref{fig:5}). Note also that the multi-band
SZ filter does not oscillate because the multi-band filter has complete
information on how to separate the signal from the noise pattern and therefore
does not need to compensate the background by oscillations. The matched filter
described in \cite{BMSCetal2006.1} shows oscillations (see Fig.~6)
because a different noise power spectrum was used to define the
filter, in particular Galactic foregrounds were added.

\section{Results: Statistics of the detections}
\label{sect:detections}

We now apply the single- and multi-band filters to our set of eleven SZ and
X-ray simulated maps and statistically analyse the samples obtained. We define
a \textit{detection} as a group of neighbouring pixels in the signal-to-noise
(hereafter S/N) maps, whose values exceed a certain threshold. The pixel with
the largest S/N value defines its position. A detection is true if it can be
associated to a halo present in the original $\Lambda$CDM simulation, while it
is spurious if caused by noise.

In order to reliably associate each detection with a halo from the parent
simulation, we apply the method described and used in \cite{FPAetal2007.1}, to
which we refer for details. Briefly, we produce sequences of maps by removing
sequentially one plane from the stack. If a halo is in fact located on an
individual plane, its detection must disappear from the map when this plane is
removed. If a detectable peak keeps appearing in all maps of a sequence, it
must originate from the noise.

\subsection{SZ single-band detections}

\begin{figure}[!ht]
  \centerline{
    \includegraphics[width=0.49\hsize]{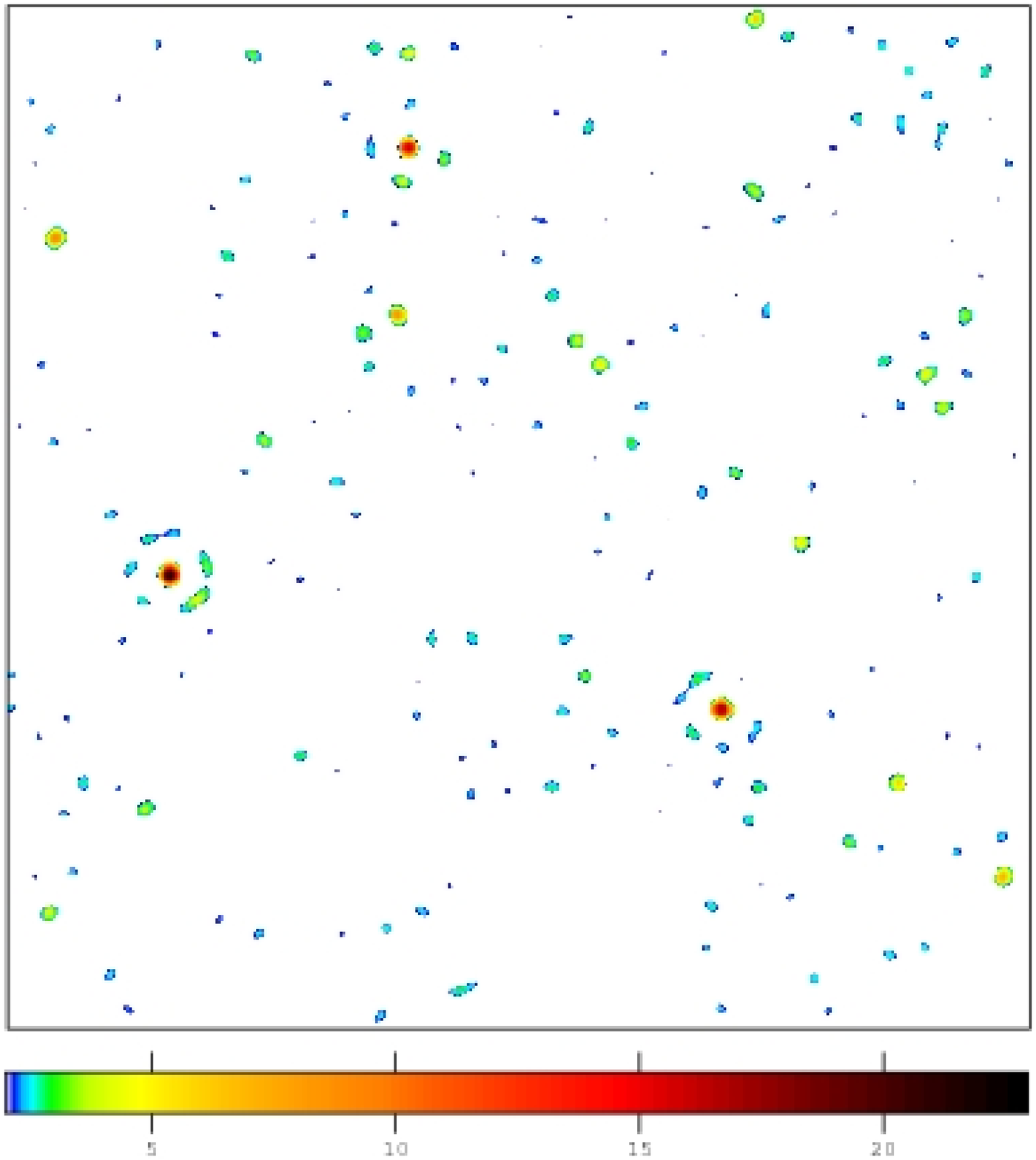}\hfill
    \includegraphics[width=0.49\hsize]{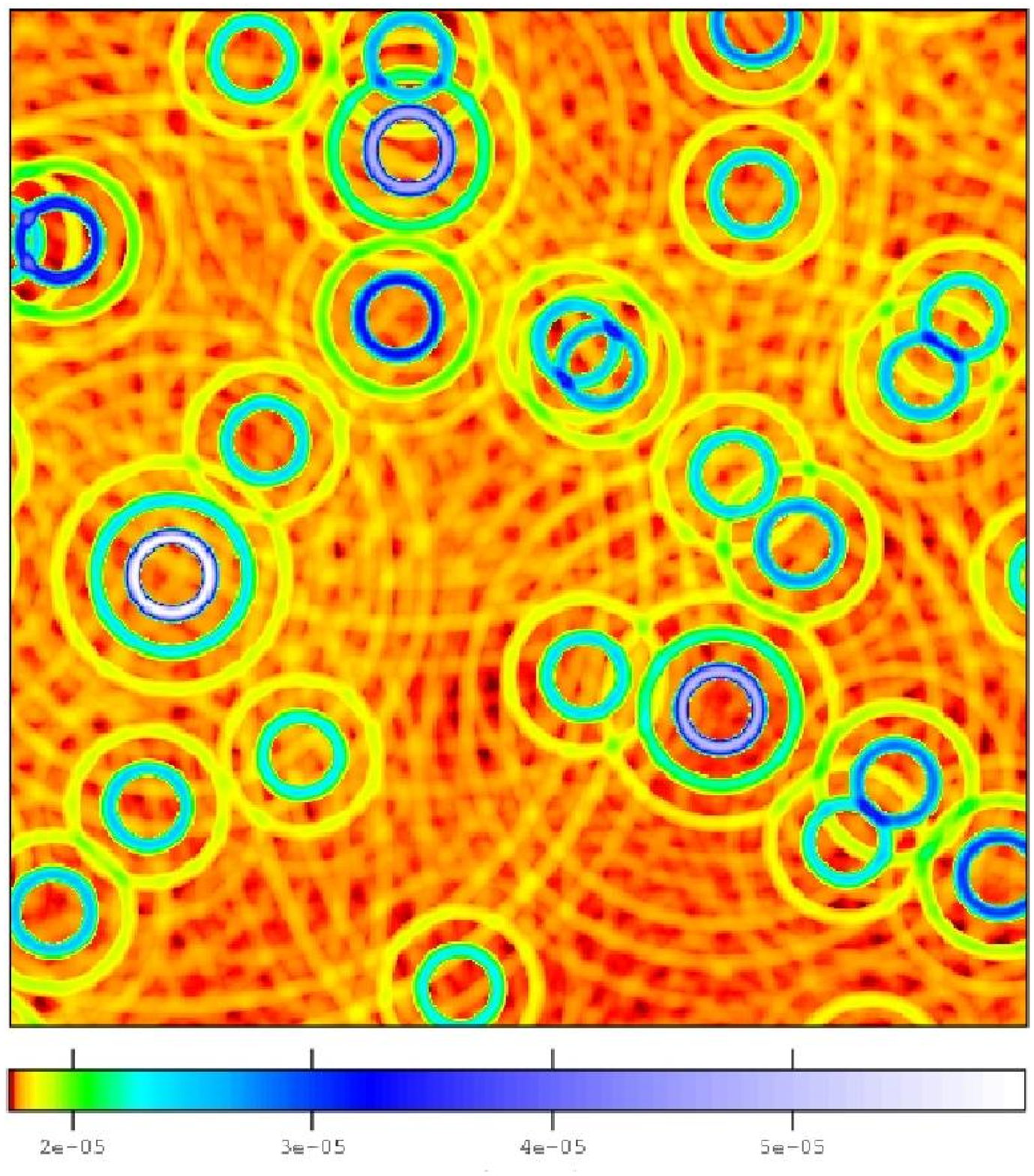}}
  \caption{Maps for the S/N ratio illustrating the procedure to
    suppress artificial structures in the filtered SZ maps using the
    correlated noise. The left panel shows the detections with
    $S/N\ge2$ as obtained with the filter, while the right panel shows
    the noise map including the damped oscillations of the single-band
    filter.}
  \label{fig:8}
\end{figure}

Due to the oscillations of the single-band filter (see the upper panels of
Fig.~\ref{fig:5}), detections consist of positive S/N peaks surrounded by
negative and positive ring-like structures. Typically, these rings are
fragmented by noise and appear as secondary S/N peaks surrounding the most
prominent detections. These secondary peaks can easily be confused with true
detections. Some examples can be seen in the left panel of
Fig.~\ref{fig:8}. Consequently, we need to include these correlated noise
structures in our noise model.

\begin{figure}[!ht]
  \centerline{\includegraphics[width=0.6\hsize]{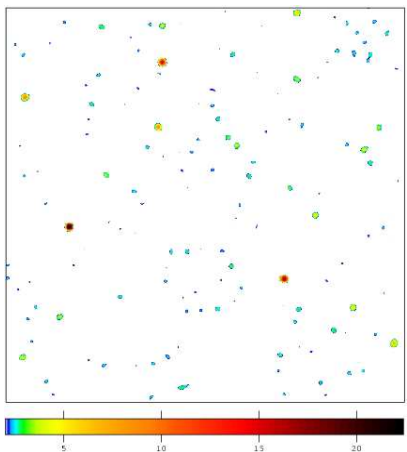}}
  \caption{S/N map, as shown in Fig.~\ref{fig:8}, after removal of the
    spurious ring artifacts. The side length of the map is 2.5 degrees.}
  \label{fig:9}
\end{figure}

This is possible because we can easily estimate their expected shapes and
amplitudes, which derive from the convolution of the signal with the
oscillatory pattern of the filter (see Fig.~\ref{fig:5}). The noise map is
thus obtained adopting an iterative approach. First, we compute the S/N maps
by applying the filter and estimating the noise. Second, we identify
the detection having the highest S/N value, compute the correlated
noise pattern as explained above and include it into the noise
map. Then, we use this new noise estimate to compute an updated S/N
map, where we identify new detections searching for local S/N
maxima. We can iterate this procedure until no more detection is found
with S/N exceeding a given threshold. The noise map resulting from
this procedure is shown in the right panel of Fig.~\ref{fig:8}. The
resulting S/N map obtained using the correlated noise pattern is shown
in Fig.~\ref{fig:9}, where, thank to the use of the correlated noise,
we notice that the fragmented rings are efficiently suppressed.

\begin{figure}[!ht]
  \centering{
    \includegraphics[angle=-90,width=0.49\hsize]{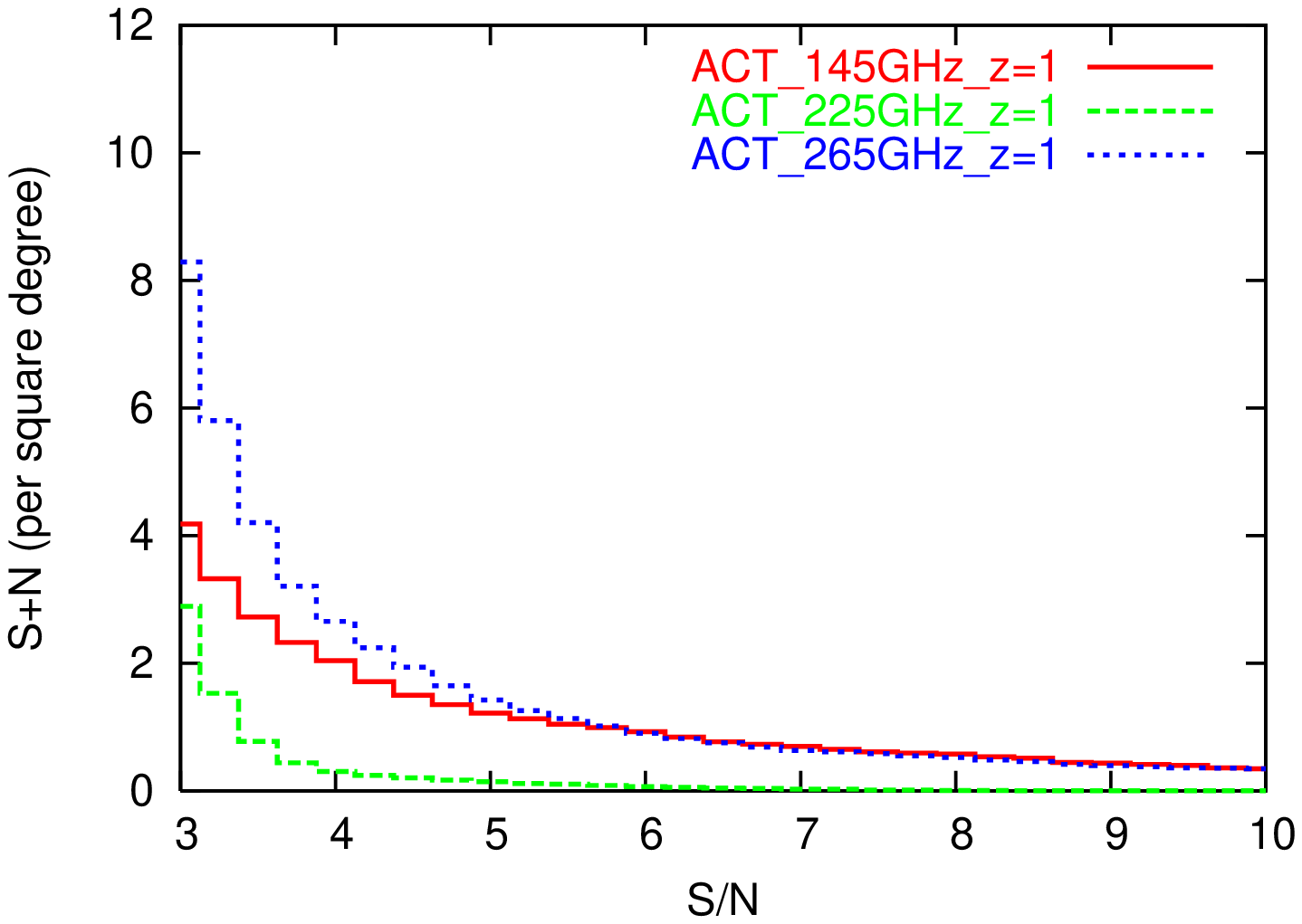}\hfill
    \includegraphics[angle=-90,width=0.49\hsize]{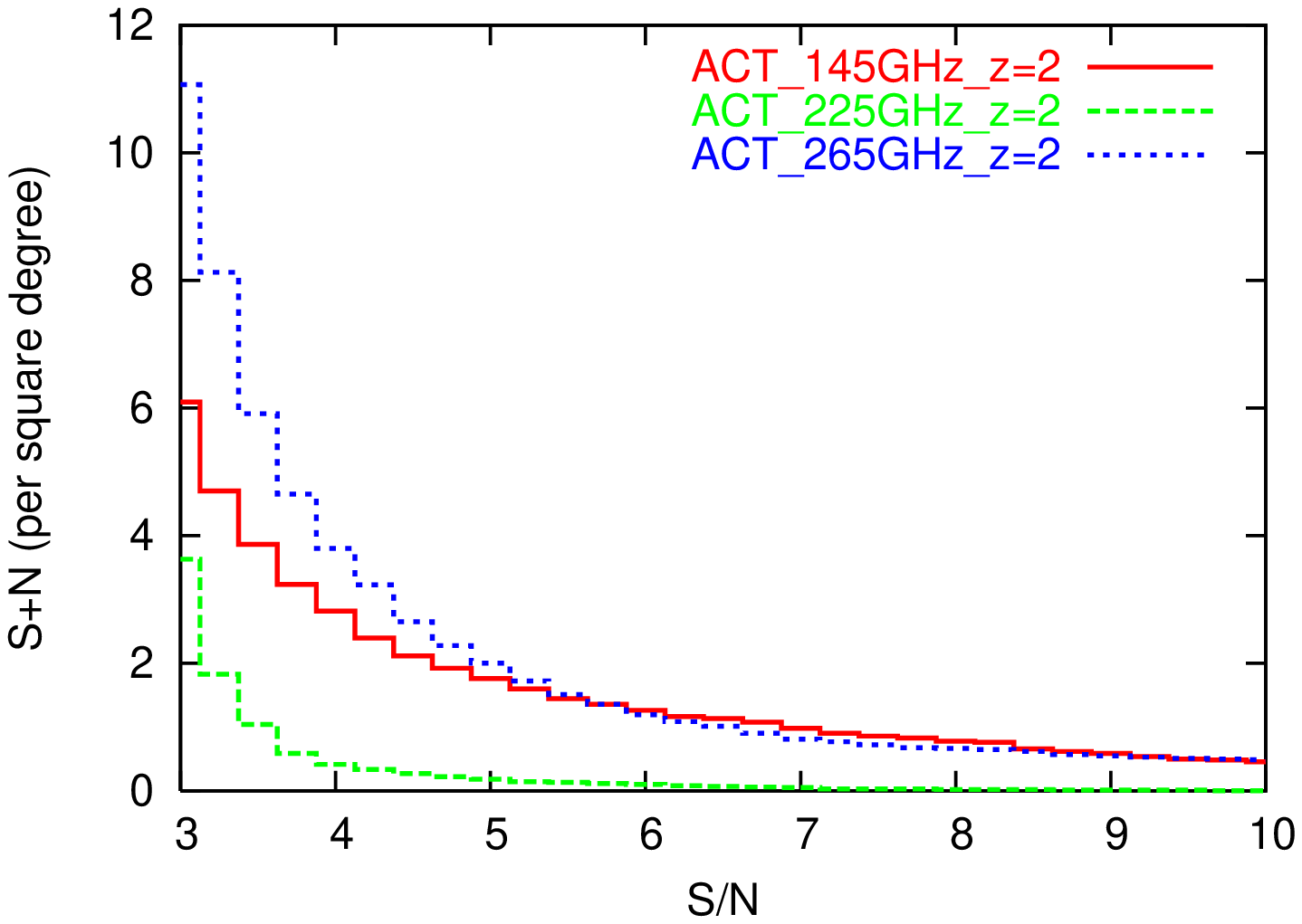}
    \includegraphics[angle=-90,width=0.49\hsize]{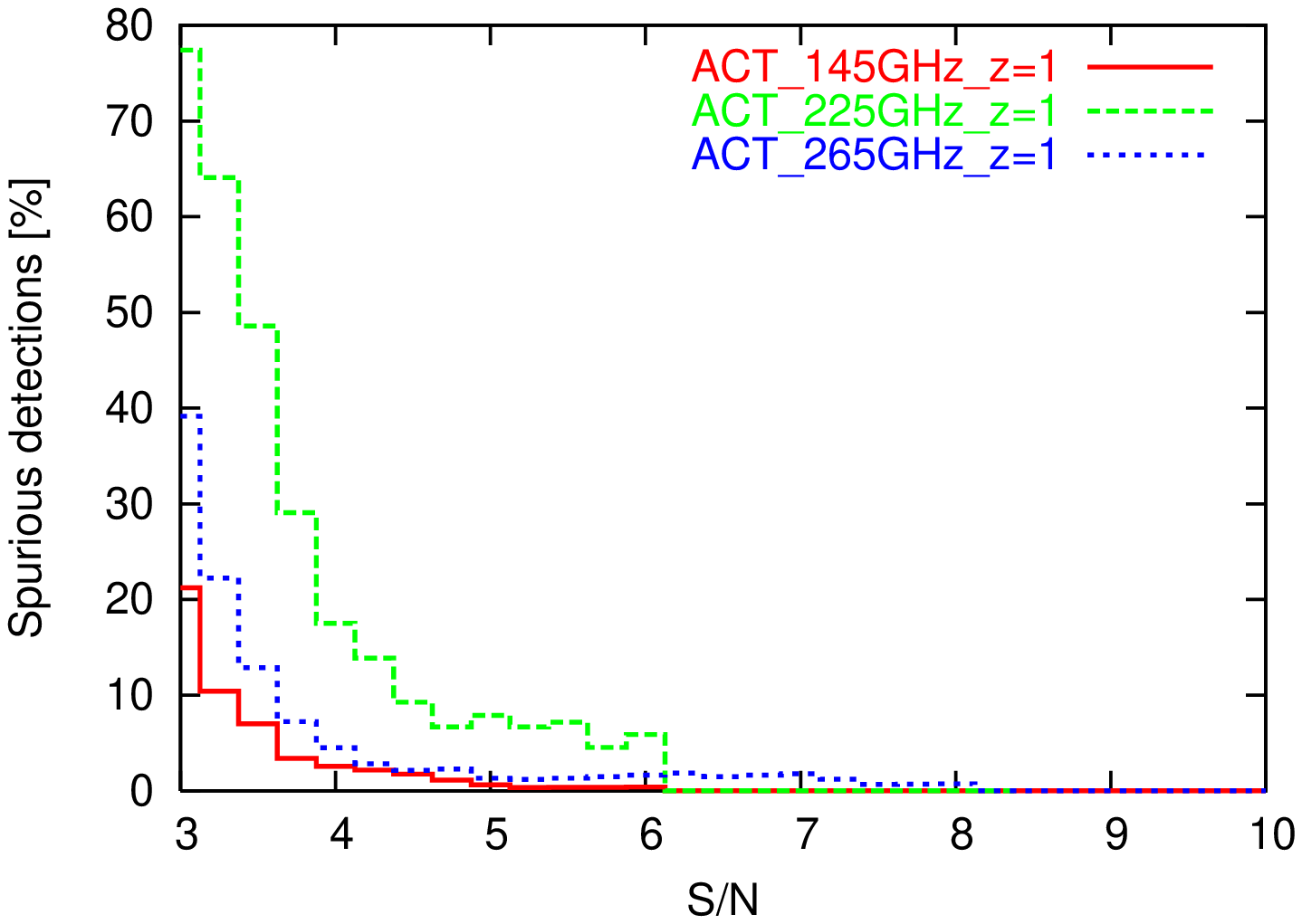}\hfill
    \includegraphics[angle=-90,width=0.49\hsize]{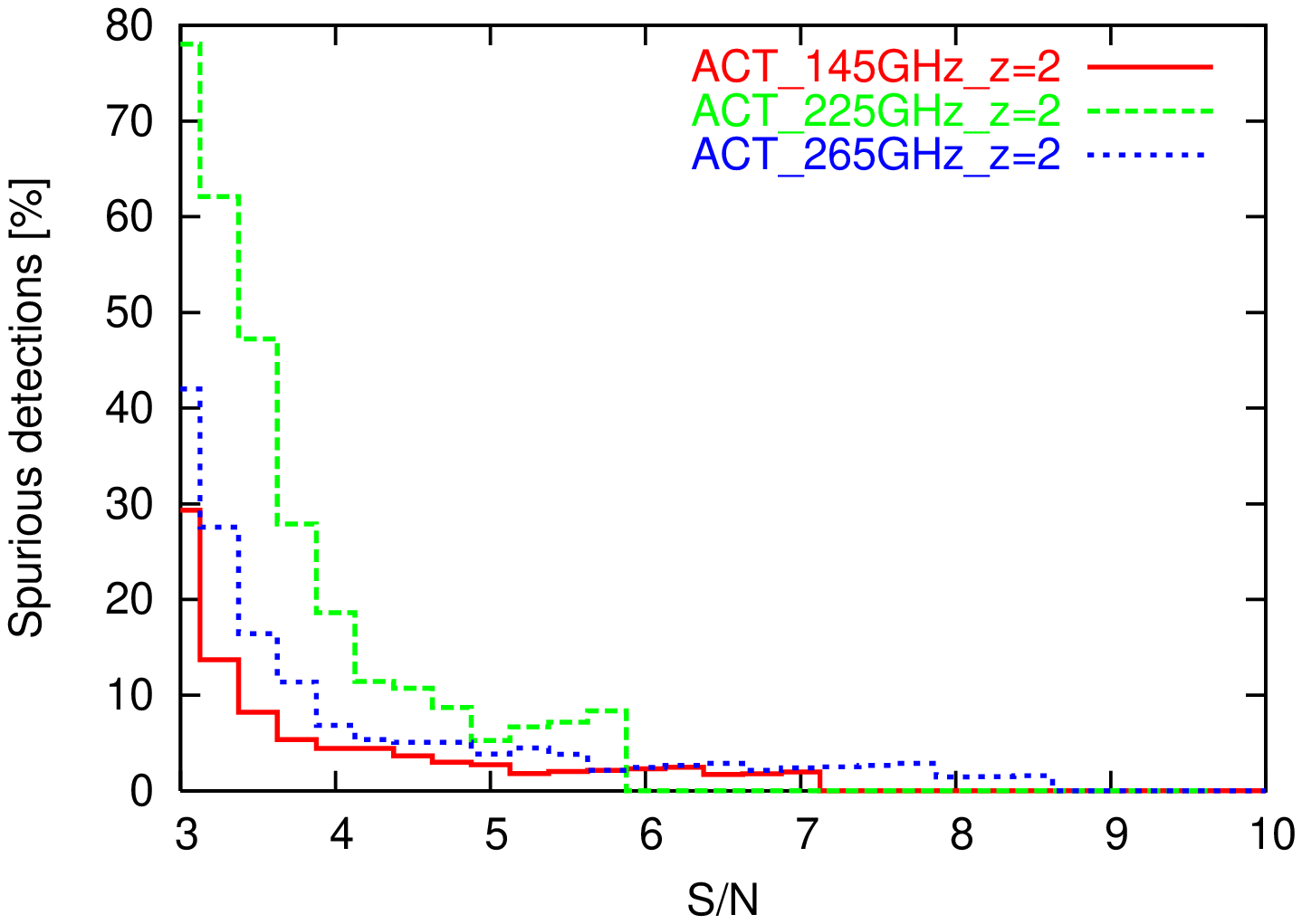}}
  \caption{The total number of detections per square degree (upper panels) and
    the fraction of spurious detections (bottom panels) as functions of the
    $S/N$ ratio. The results are averaged over eleven realisations and shown
    for light cones extending to redshifts $z=1$ (left panels) and $z=2$
    (right panels). Different line styles refer to the different bands
    considered, as indicated in the plots.}
  \label{fig:10}
\end{figure}

The upper panels of Fig.~\ref{fig:10} show the total number of detections per
square degree in our ACT simulations. Results are shown for all three channels
and for light cones limited to $z=1$ and $z=2$ (left and right panels,
respectively). The corresponding fractions of spurious detections are
displayed in the bottom panels. Results are averaged over eleven maps covering
$24.5\,\rm{deg^2}$ and $9.5\,\rm{deg^2}$ and renormalized to
$1\,\rm{deg^2}$ for limiting redshifts $z=1$ and $z=2$,
respectively. As expected, we find very few detections at
$\nu=225$~GHz where the tSZ effect is very weak.

Since the SZ effects depend only mildly on redshift, we find more objects in a
light-cone extending up to redshift $z=2$ than for $z=1$.  In particular, at
$\nu=145$~GHz and $\nu=265$~GHz, we find at least $30\%$ more detections in
the deeper light cone. The main differences for the results in these two bands
are at low signal-to-noise ratios, where more structures are detected in the
highest-frequency band, owed to the higher sensitivity of the instrument. For
$S/N\ge 5$, the number of detections is similar in the two bands.

Because of the low tSZ signal, the fraction of spurious detections at
$\nu=225$~GHz is very high ($\approx 80\%$ for $S/N\approx 3$) and virtually
independent of the limiting redshift of the light-cones. For the maps at
$\nu=145$~GHz and $\nu=265$~GHz at $z=1$ ($z=2$), we find that $20\%$ and
$40\%$ ($30\%$ and $40\%$) of the detections at $S/N\approx 3$ are spurious,
respectively. For $S/N>4$, the fraction of spurious detections drops to a few
per cent. Including matter up to redshift $z=2$ into the light cones slightly
increases the fraction of spurious detections because unresolved structures
overlap. The number of detections with high $S/N$ does not increase with the
depth of the light cones because the massive structures responsible for them
are mainly located at low redshifts, but the number of low-$S/N$ detections is
increased by up to a factor of five for the least massive detectable
halos.

\begin{figure}[!t]
  \centering{
    \includegraphics[angle=-90,width=0.49\hsize]{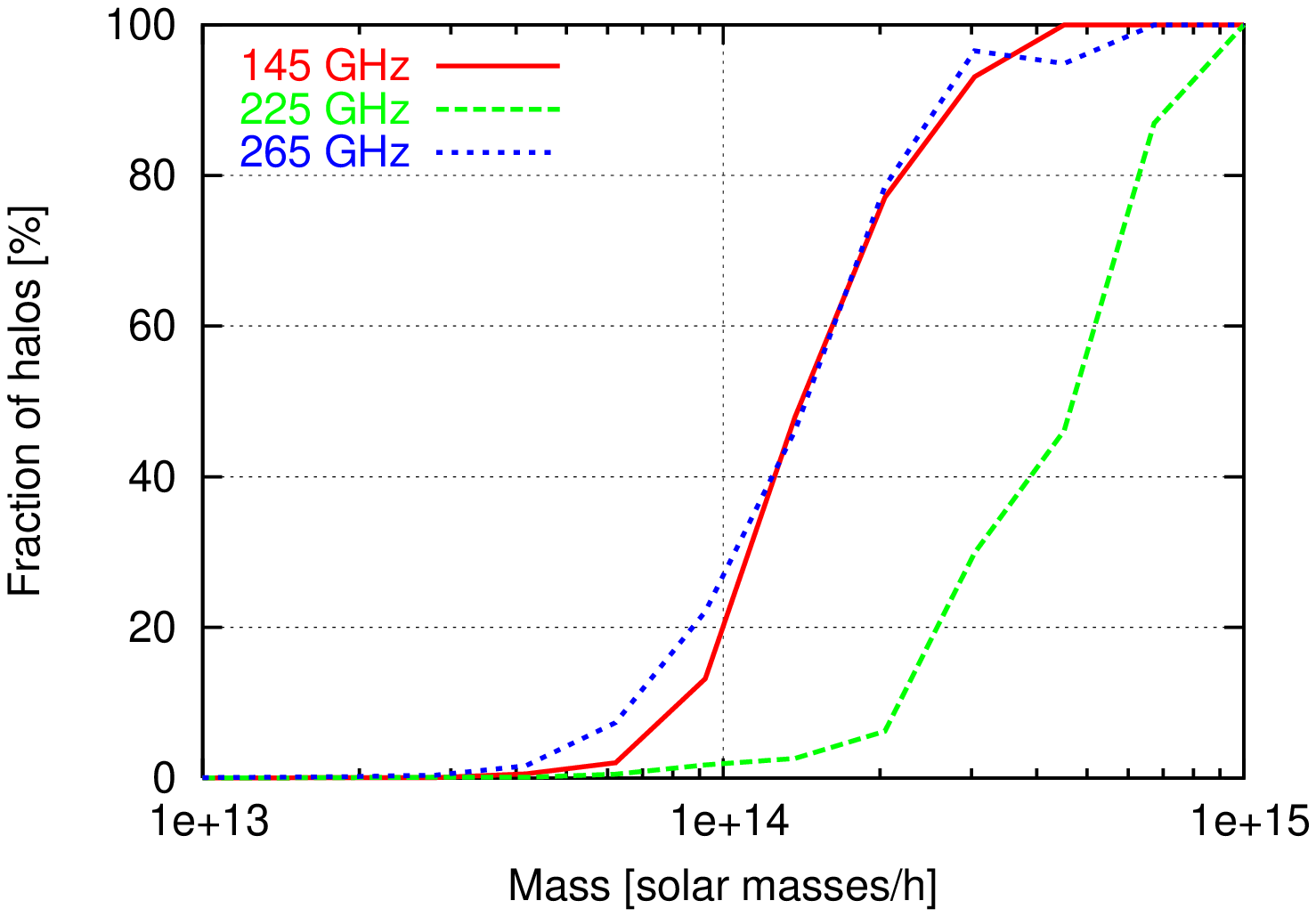}\hfill
    \includegraphics[angle=-90,width=0.49\hsize]{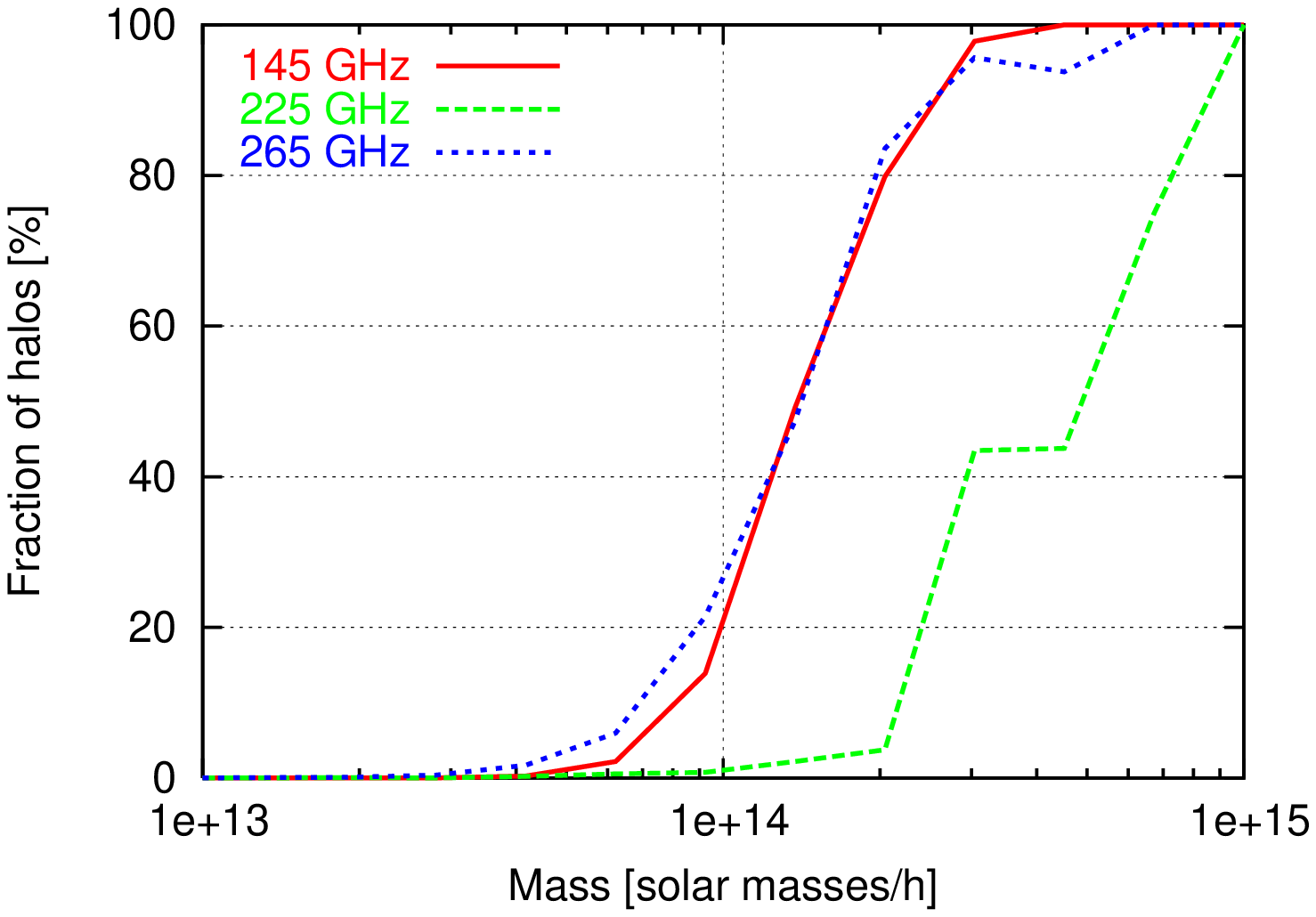}
    \includegraphics[angle=-90,width=0.49\hsize]{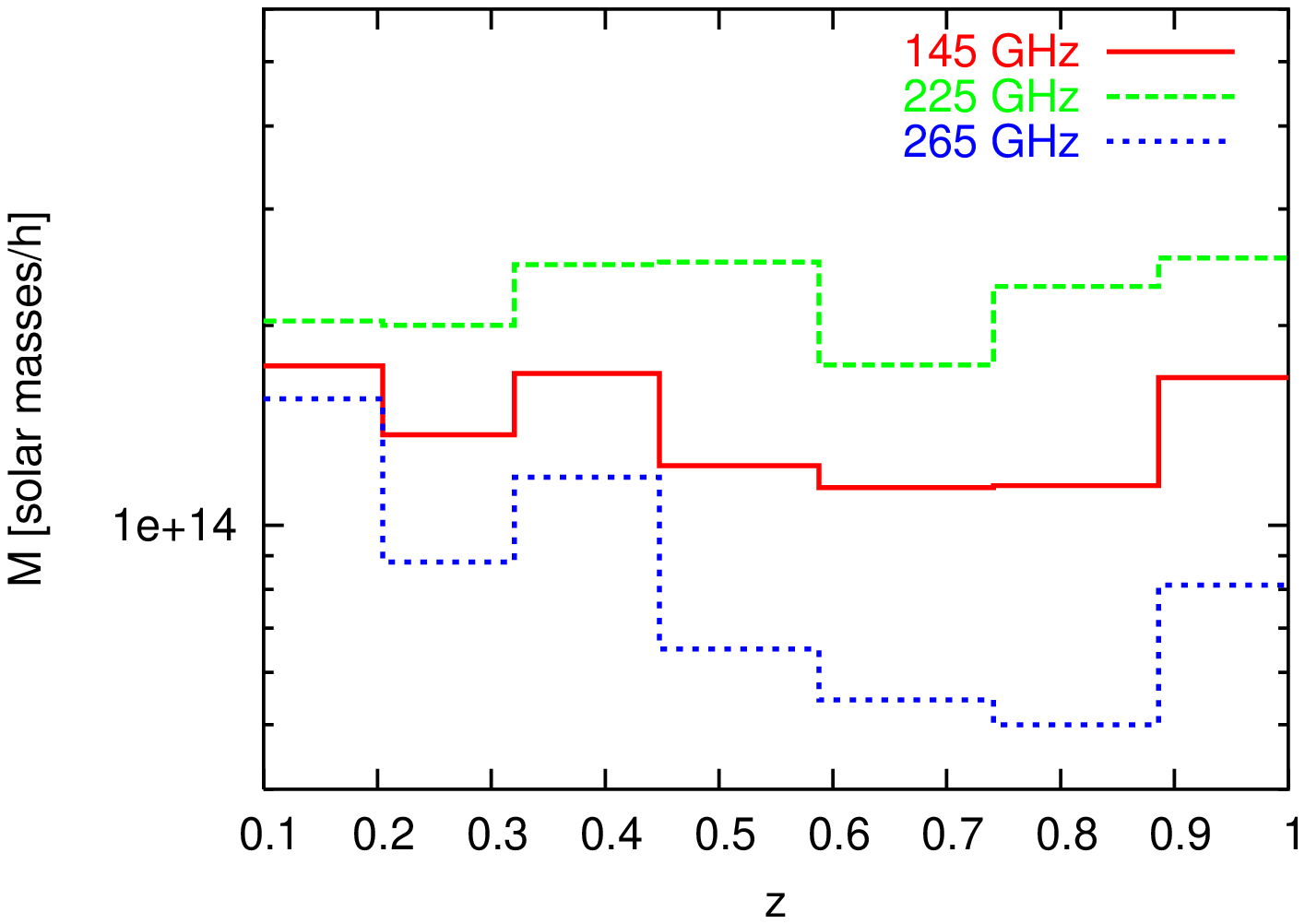}\hfill
    \includegraphics[angle=-90,width=0.49\hsize]{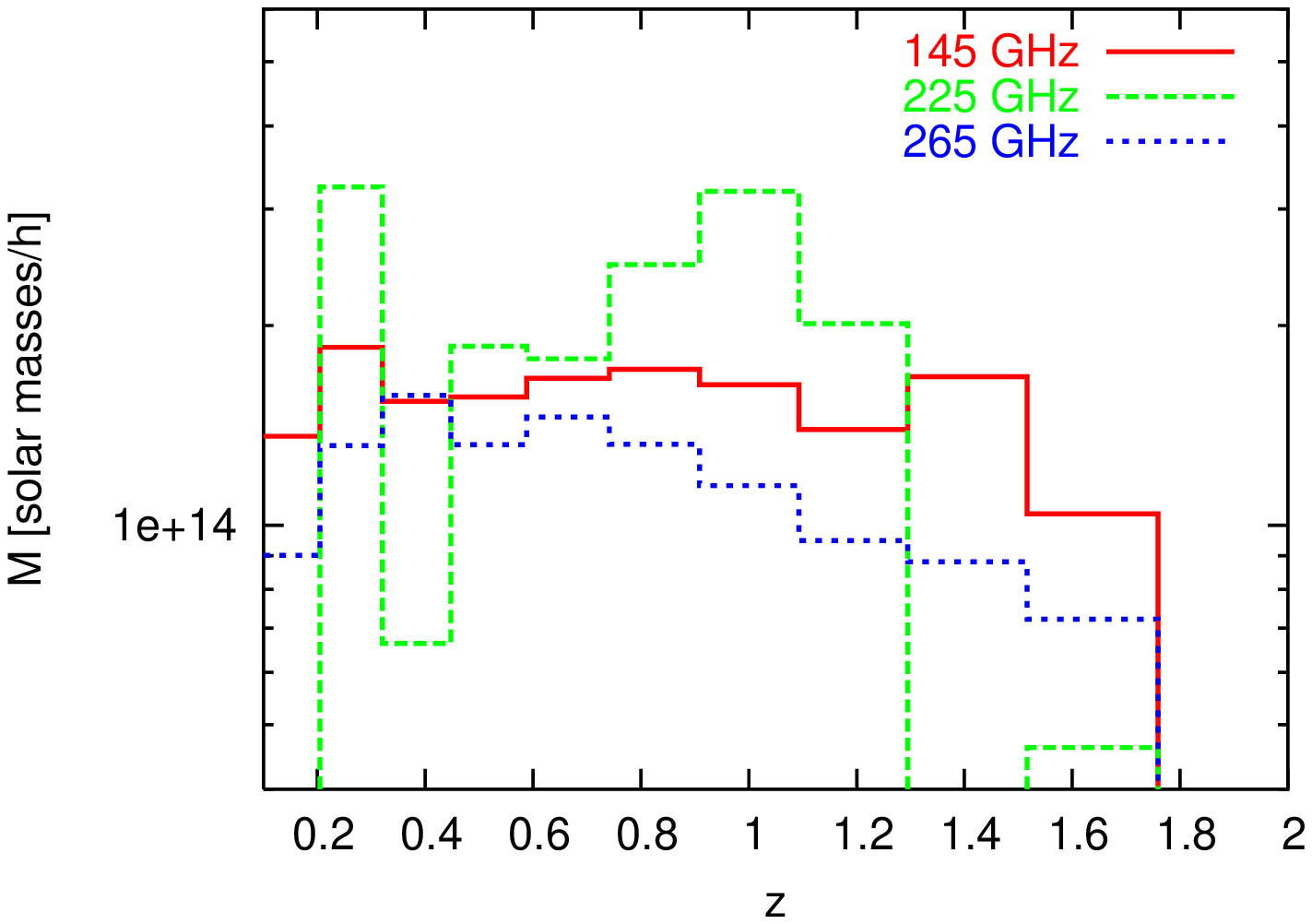}}
  \caption{Upper panels: the fraction of detected halos as a function of the
    halo mass. Lower panels: the sensitivity of the method expressed in terms
    of the minimum detectable mass as a function of redshift. Results are
    shown for light-cones extending to $z=1$ and $z=2$ in the left and right
    panels, respectively. Different line styles refer to the three different
    ACT bands, as labeled. The curves represent average over eleven
    realisations.}
  \label{fig:11}
\end{figure}

The upper panels of Fig.~\ref{fig:11} show the fraction of halos detected in
different bands as a function of mass. The left and right panels show the
results for light cones up to $z=1$ and $z=2$, respectively. The values shown
are obtained after averaging over eleven realisations. The lower panels of
Fig.~\ref{fig:11} present the sensitivity of the detections, expressed in
terms of the minimum detectable mass as a function of redshift. To reduce the
noise, this was computed combining two subsequent planes in the light cone and
averaging the mass of the ten least massive halos detected in the
corresponding redshift interval.

Since the tSZ effect depends only mildly on the source redshift, the
completeness for detections up to $z=1$ and $z=2$ is essentially the same. The
two channels at 145~GHz and 265~GHz perform equally well, while the
completeness is much lower at 225~GHz, as expected. The minimum detected mass
is $M\simeq 7\times 10^{13}M_\odot/h$ and the completeness reaches $100\%$ for
masses $M\simeq 3\times 10^{14}M_\odot/h$. The curve is quite steep, showing
that the completeness drops rapidly as the mass decreases. For halo masses
$\ge M\simeq 1.5\times 10^{14}M_\odot/h$, the completeness has already dropped
to $\sim50\%$. The minimum detectable halo mass is essentially constant for
all $z$, reflecting the near-independence of the tSZ effect on
distance. Again, the method is much less sensitive at 225~GHz than at the two
other frequencies. In addition, the filter tends to find halos with lower mass
at 265~GHz than at 145~GHz (blue and red curves, respectively), explaining the
higher completeness in the former band. This is due to the fact that the band
at 265~GHz has a better $FWHM$ (see Table \ref{tab:ACT}) and the noise is
therefore lower (see Eq.~\ref{eq:Inoise}).

Similar results were found by \cite{MLOetal2006.1} using the same approach to
detect clusters in simulated Planck data. For low detection limits
(corresponding to low $S/N$ ratios), \cite{MLOetal2006.1} found a very high
fraction of spurious detections, decreasing quite rapidly towards higher flux
limits. However, results can only be compared at a qualitative level because
the instrument and the band frequencies differ greatly. In particular, only
two channels used by \cite{MLOetal2006.1} have a frequency range close to
those of ACT. Moreover, we study $S/N$ ratios rather than fluxes and account
for the correlated ring patterns due to the oscillatory filter behaviour in
the noise model.

\subsection{Statistics of SZ multi-band detections}

\begin{figure}[!ht]
  \centering{
    \includegraphics[width=0.45\hsize]{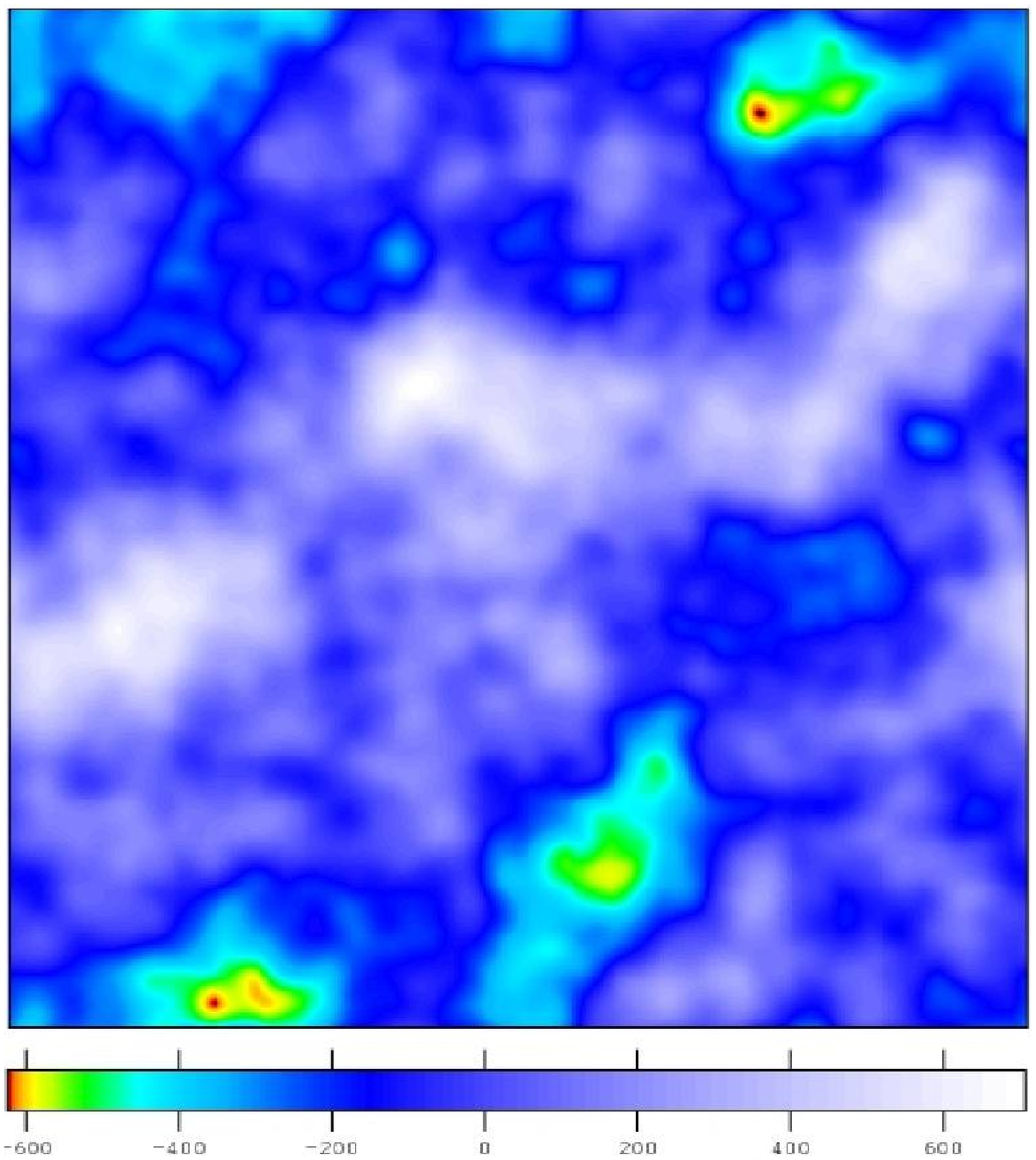}\hfill
    \includegraphics[width=0.45\hsize]{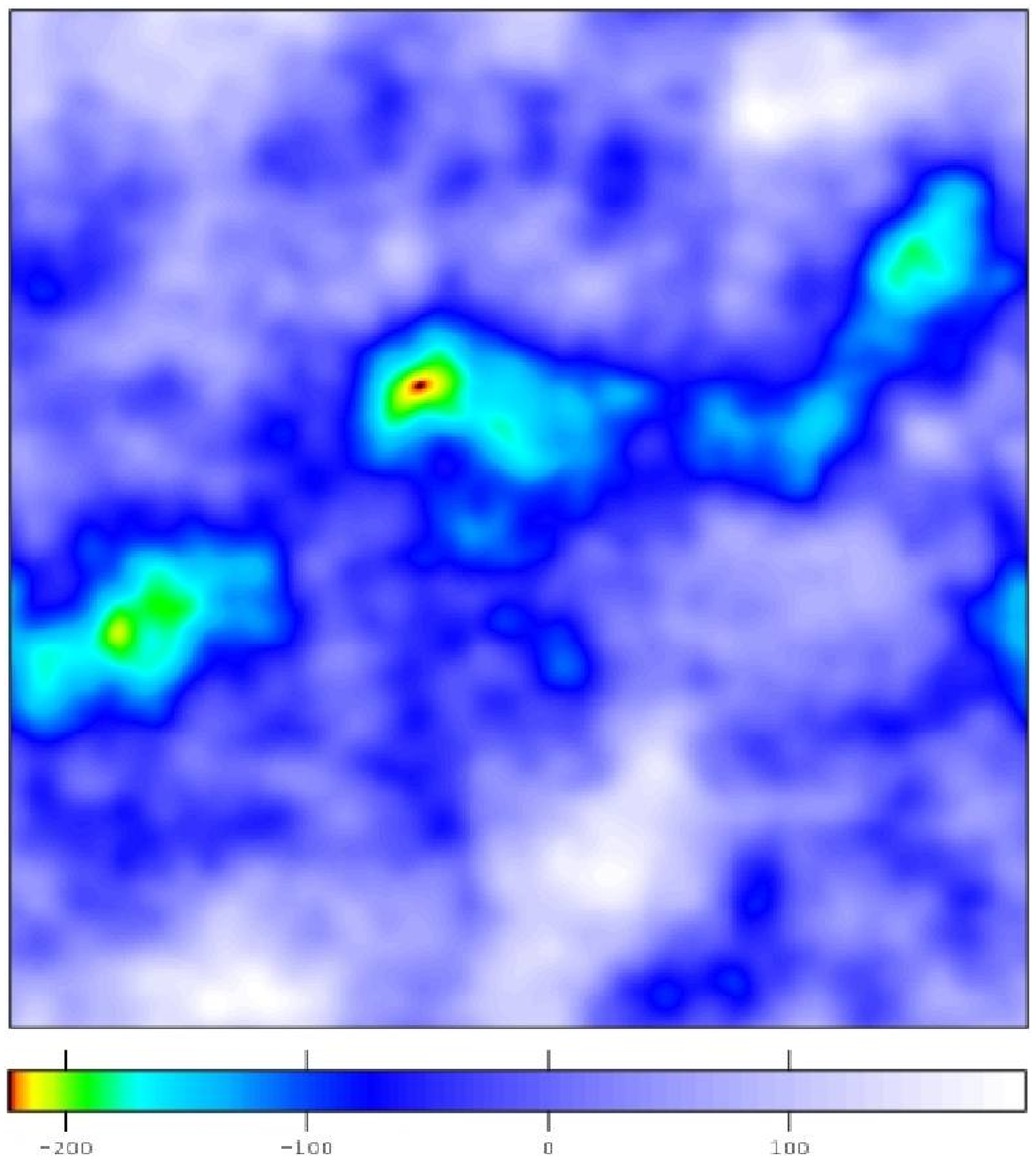}
    \includegraphics[width=0.45\hsize]{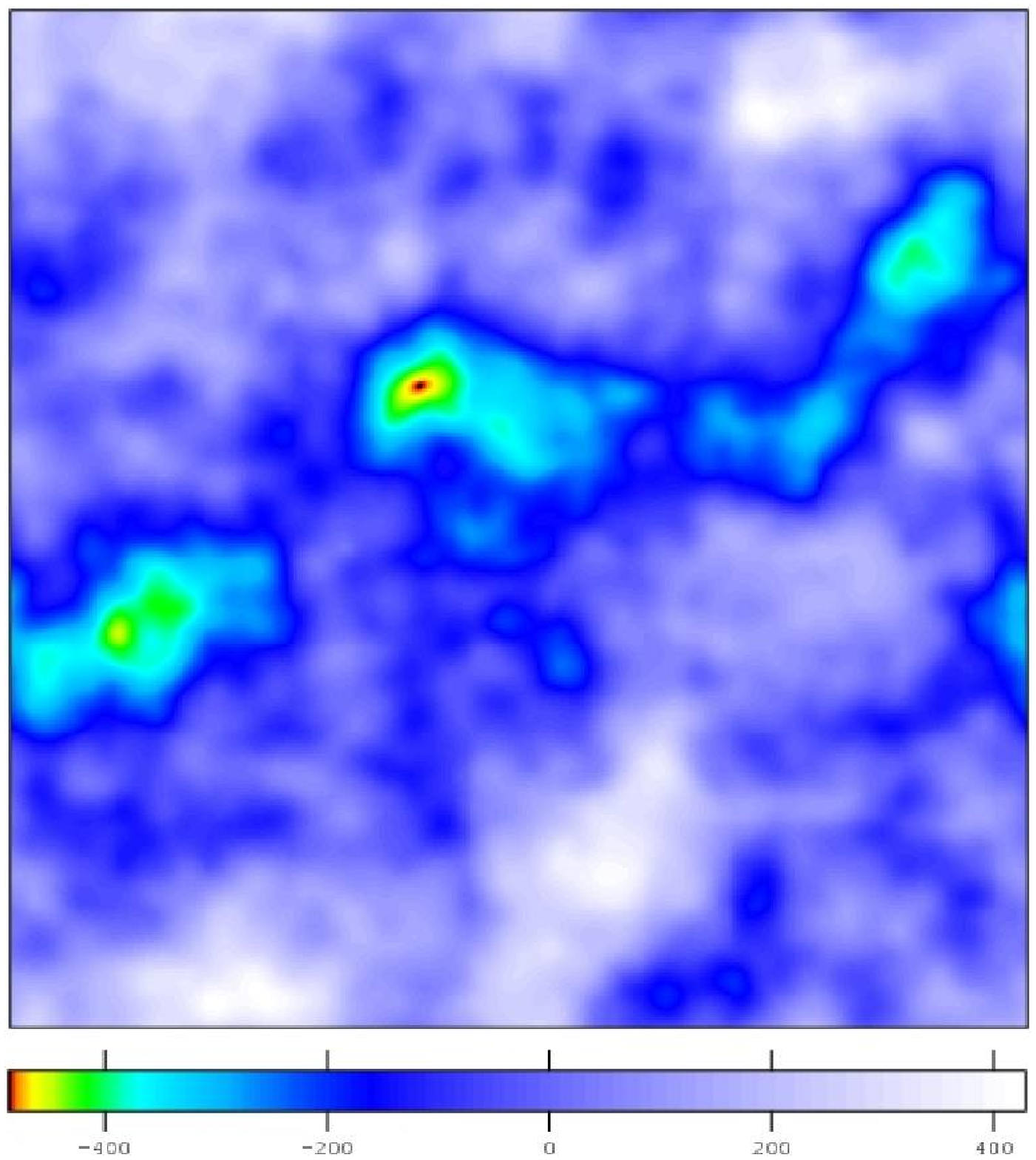}\hfill
    \includegraphics[width=0.45\hsize]{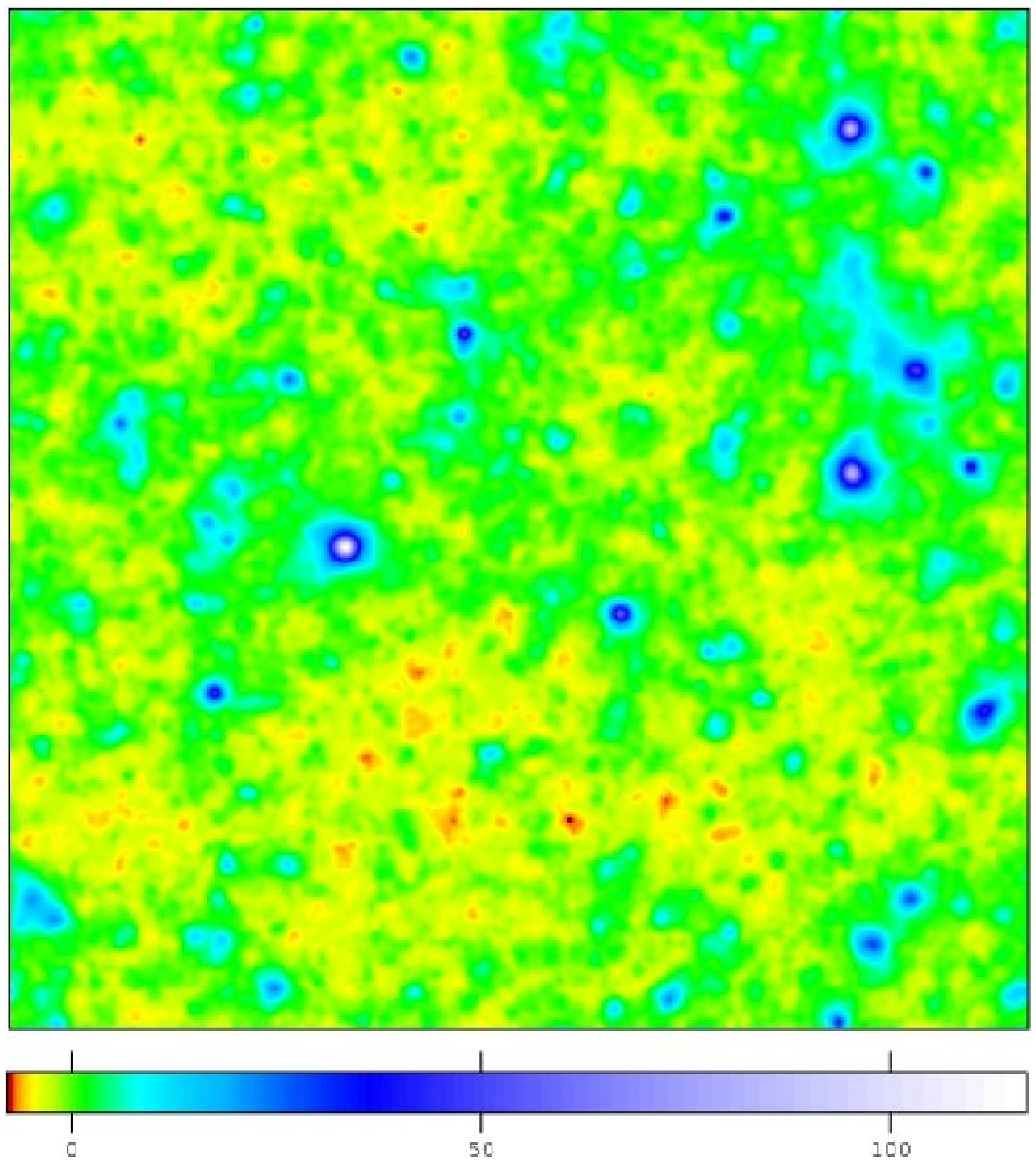}}
  \caption{Example of the application of the multi-band matched filter to a
    S/N map. The starting input is the noisy map in the right panel of
    Fig.~\ref{fig:3}. Different panels refer to the maps filtered at different
    frequencies: 145~GHz (upper left panel), 225~GHz (upper right panel),
    265~GHz (lower left panel). The final S/N map obtained by summing the
    previous three maps is shown in the lower right panel.}
  \label{fig:12}
\end{figure}

Figure~\ref{fig:12} shows an example of the S/N maps obtained by
applying each component of the multi-band filter to the three channels
at 145~GHz (upper left panel), 225~GHz (upper right panel) and 265 GHz
(bottom left panel). The final result is obtained by combining them
using Eq.~(\ref{eq:Aest}) to give the S/N map of the multi-band
detection in the bottom right panel.

\begin{figure}[!ht]
  \centering{
    \includegraphics[angle=-90,width=0.49\hsize]{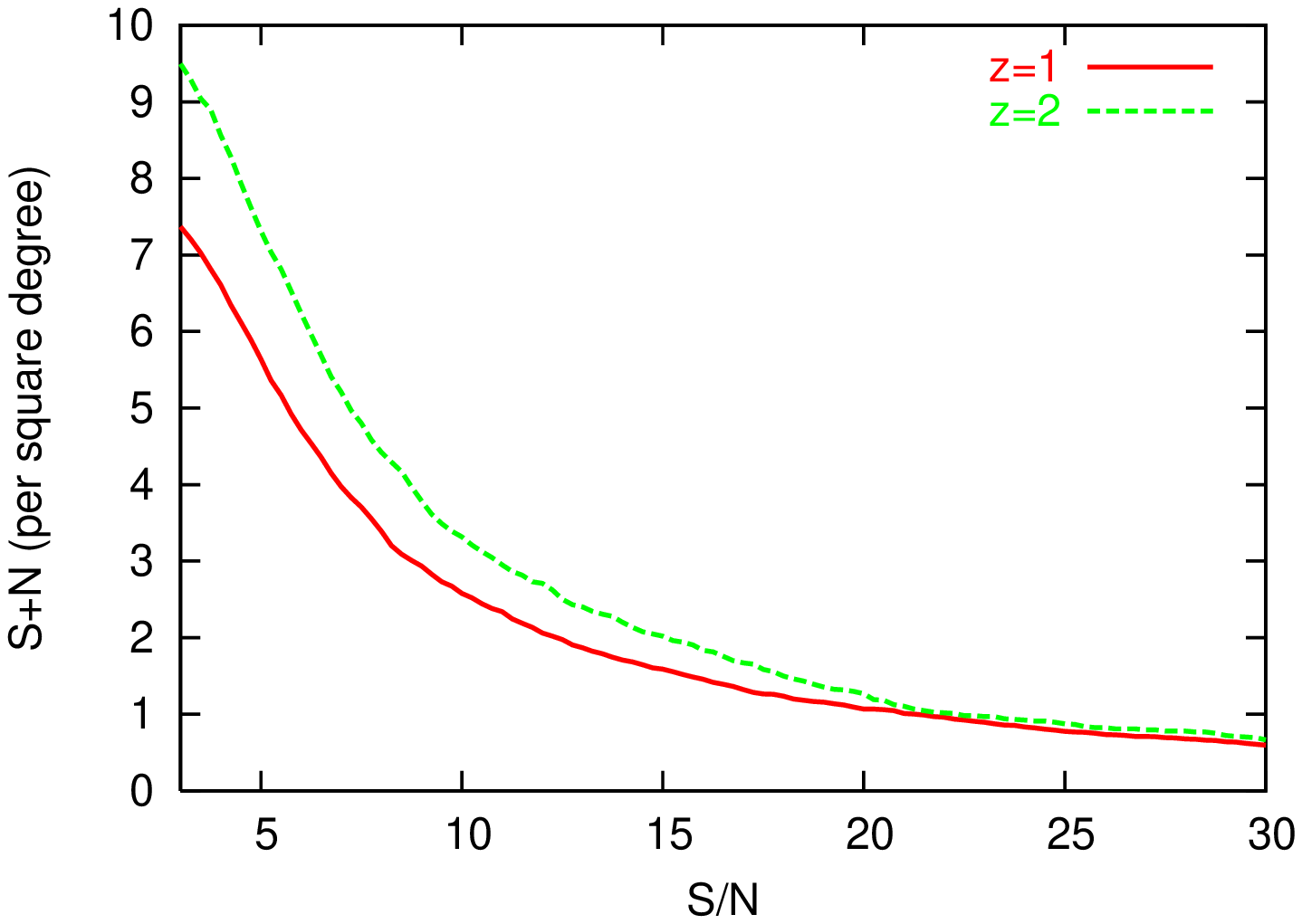}\hfill
    \includegraphics[angle=-90,width=0.49\hsize]{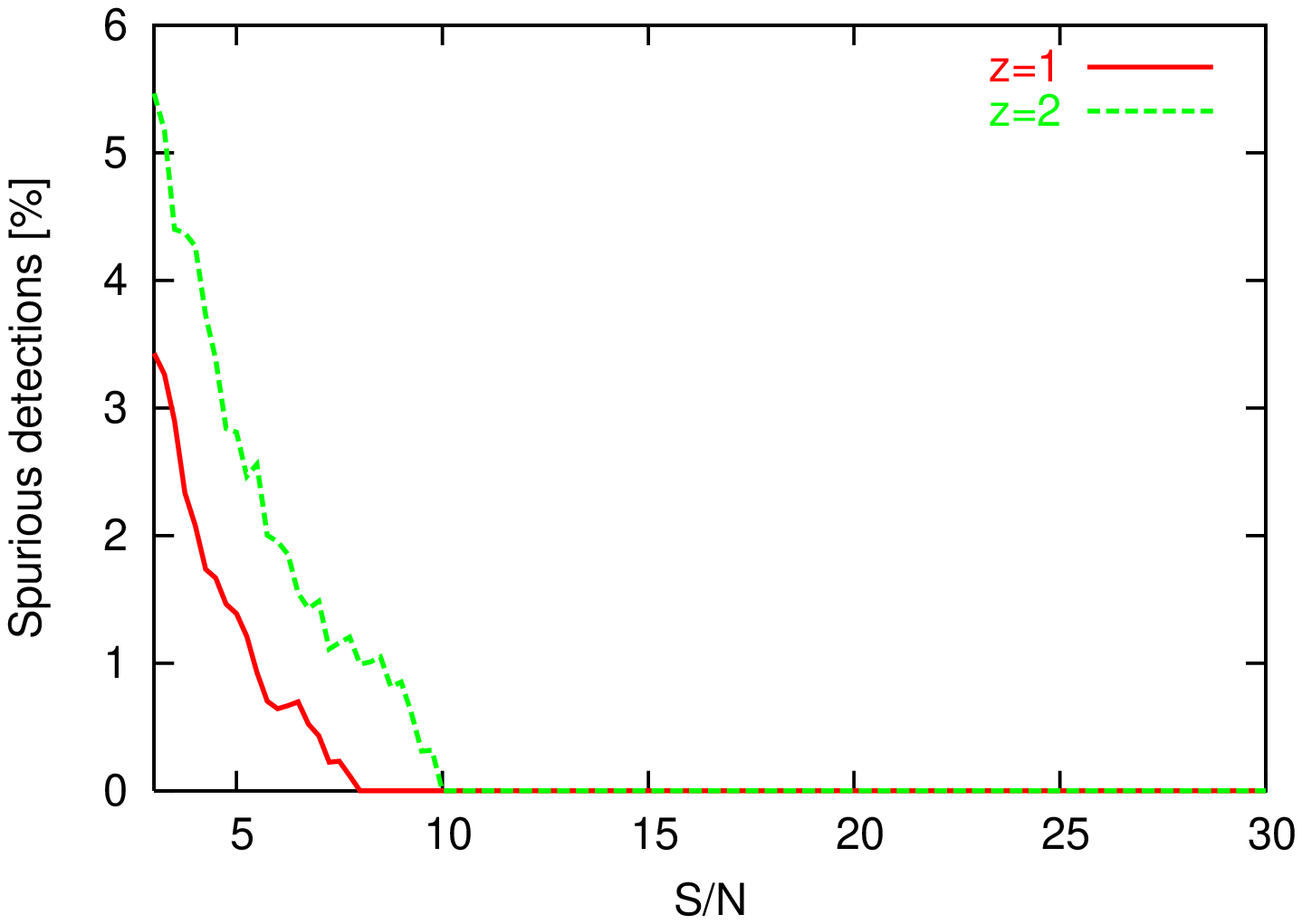}
    \includegraphics[angle=-90,width=0.49\hsize]{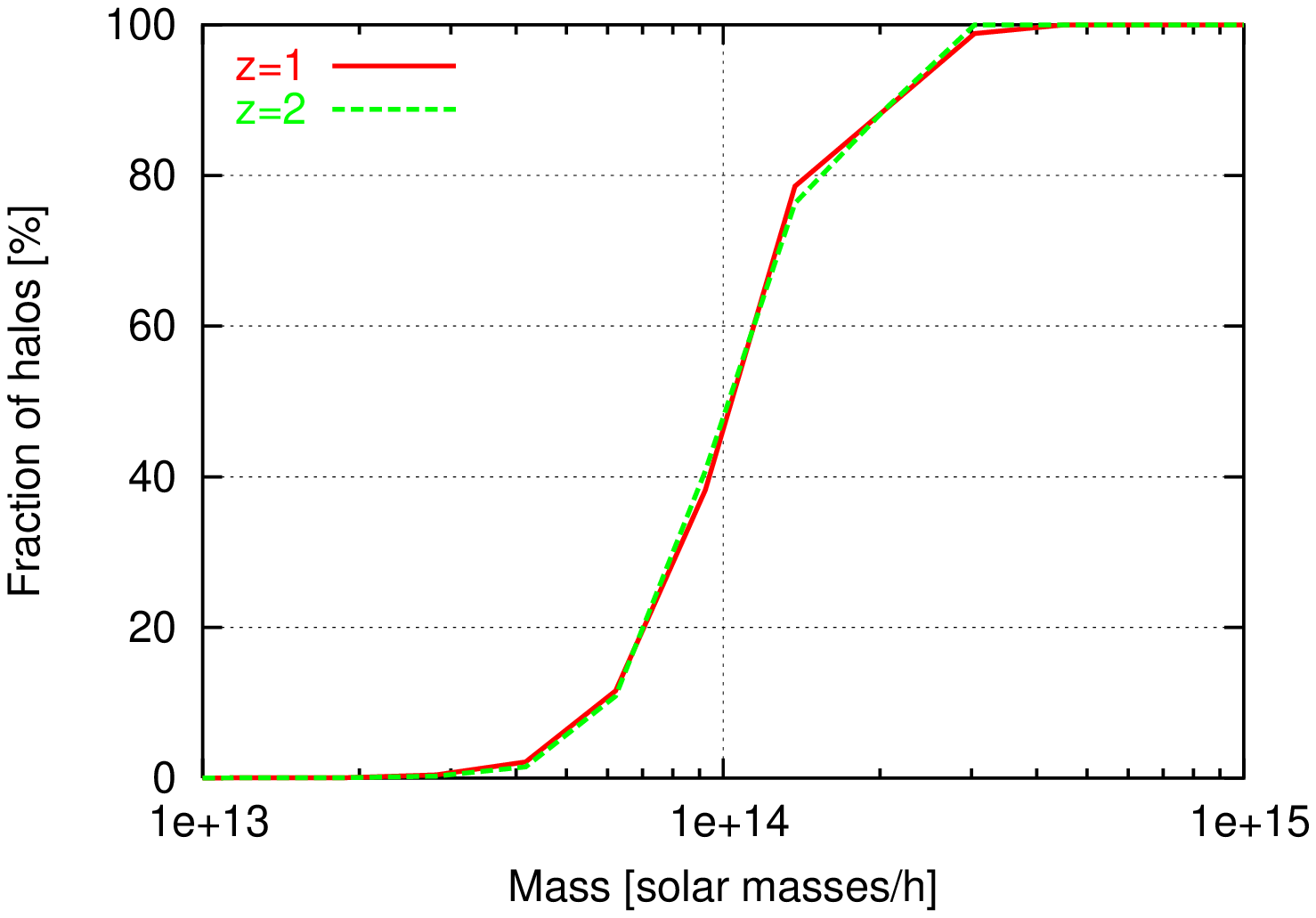}\hfill
    \includegraphics[angle=-90,width=0.49\hsize]{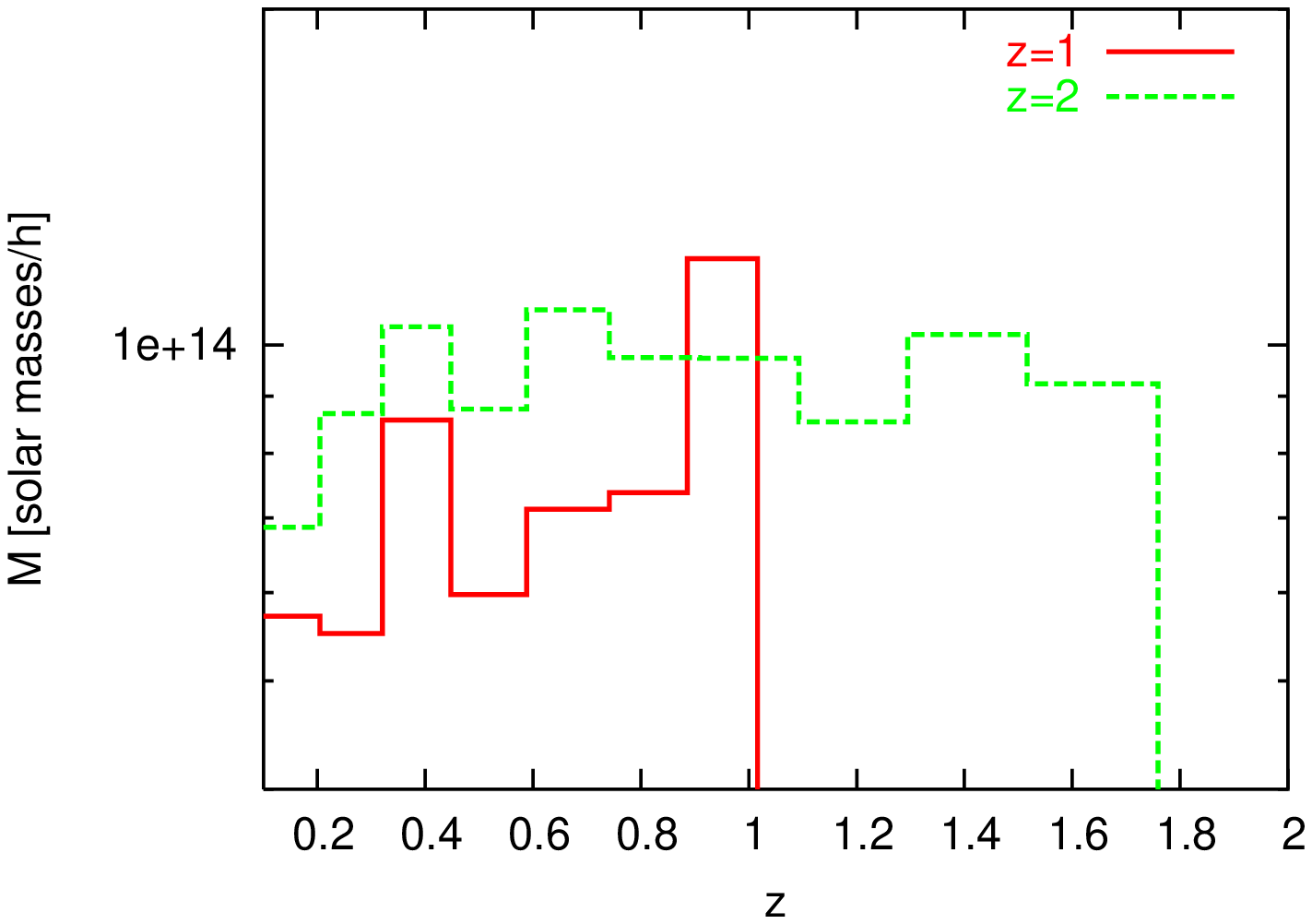}}
  \caption{Upper left panel: the total number of detections per square degree
    as a function of the $S/N$ ratio, obtained with the multi-band
    filter. Upper right panel: the fraction of spurious detections as a
    function of the $S/N$ ratio. Bottom left panel: the fraction of halos
    detected as a function of mass. Bottom-right panel: the minimum detectable
    mass as a function of redshift. Solid and dotted curves show results for
    light cones extending to $z=1$ and $z=2$, respectively.}
  \label{fig:13}
\end{figure}

Compared to the results obtained with the single-band filter, the S/N ratios
are now enhanced by at least a factor of three. As before, the S/N peaks in
the maps were identified with SExtractor \citep{EBESAR1996.1} which allows
easy de-blending of nearby detections. The number counts per square degree as
a function of the S/N threshold are shown in the upper left panel of
Fig.~\ref{fig:13}. The solid and dashed lines refer to light-cones extending
up to $z=1$ and $z=2$, respectively.

The lower number of total detections respect to the single-band filter
at 265~GHz, has not to be interpreted as a signal of a poorer
performance of the multi-band filter, but it has to be ascribed partly
to the fact that the multi-band filter is wider and increases the
confusion which is not entirely removed by SExtractor, but most
importantly because of the lower sample contamination. In fact, the
fraction of spurious detections is much lower for the multi-band than
for the single-band filter. We find that the number of real,
multi-band detections is higher by almost one order of magnitude for
$S/N\approx 3$ and by a factor of five for $S/N\approx 5$ than the
single-band detections. These results do not depend on the depth of
the light cones.

The top-right panel of Fig.~\ref{fig:13} shows the fraction of spurious
detections in the final catalogue. Again, the great advantage of multi-band
observations is obvious. While the fraction of spurious detections in the
265~GHz channel alone is $\sim 40\%$ at $S/N=3$ and $\sim 5\%$ at $S/N\sim 5$,
multi-band filtering reduces the fraction of spurious detection to the level
of a few percent even at very low $S/N$ ratios. Thus, the vast majority of the
detections in the top-right panel is due to real structures. The completeness
of the catalogue is shown in the bottom left panel of Fig.~\ref{fig:13}. We
find that $20\%$, $50\%$ and $80\%$ of the halos with masses exceeding
$7\times10^{13}\,M_\odot/h$, $\sim 10^{14}\,M_\odot/h$ and $2\times
10^{14}\,M_\odot/h$ contained in the light cones are detected by the
multi-band filter, respectively. In contrast, the single-band filter detects
only $\sim 20\%$ of the halos with masses exceeding $10^{14}\,M_\odot/h$ in
the 145~GHz and in the 265~GHz channels. Finally, we present in the
bottom right panel of Fig.~\ref{fig:13} the minimum halo mass
detectable for the multi-band filter as a function of redshift. Again,
the curves show a very weak dependence on redshift, although the
values seem to weakly increase with redshift. The mass threshold
ranges from $\sim 6\times 10^{13}\, M_{\odot}/h$ at $z=0$ to $\sim
9\times 10^{13}\,M_\odot/h$ at $z=1.8$, which confirms that the
multi-band filter also improves the sensitivity.

We conclude this section by comparing our results to an independent study of
multi-band filtering. \cite{JBMEetal2006.1} adopted a multi-band matched
filter for analysing Monte-Carlo simulations. Mimicking SZ observations with
the South-Pole-Telescope (SPT), they performed a statistical study of the
detections comparable to ours. Since the SPT bands and FHWM are similar to
those of ACT, a fair comparison between our and their results is
possible. They found 6 detections per square degree at $S/N>5$, very close to
the 8 detections per square degree we find. The lowest masses they can detect
is three times higher than we find for ACT. This is expected because of the
higher noise level of SPT compared to ACT. The contamination by spurious
detections in the catalogues found by \cite{JBMEetal2006.1} is also similar to
ours.

\subsection{Statistics of X-ray detections}

We now turn to the halo detections in the simulated X-ray maps. We show
results only for light-cones extending up to $z=1$, because the number of
X-ray sources at higher redshifts is very low, and the corresponding plots are
very similar to those for $z=1$.

\begin{figure}[!ht]
  \centering{
    \includegraphics[angle=-90,width=0.49\hsize]{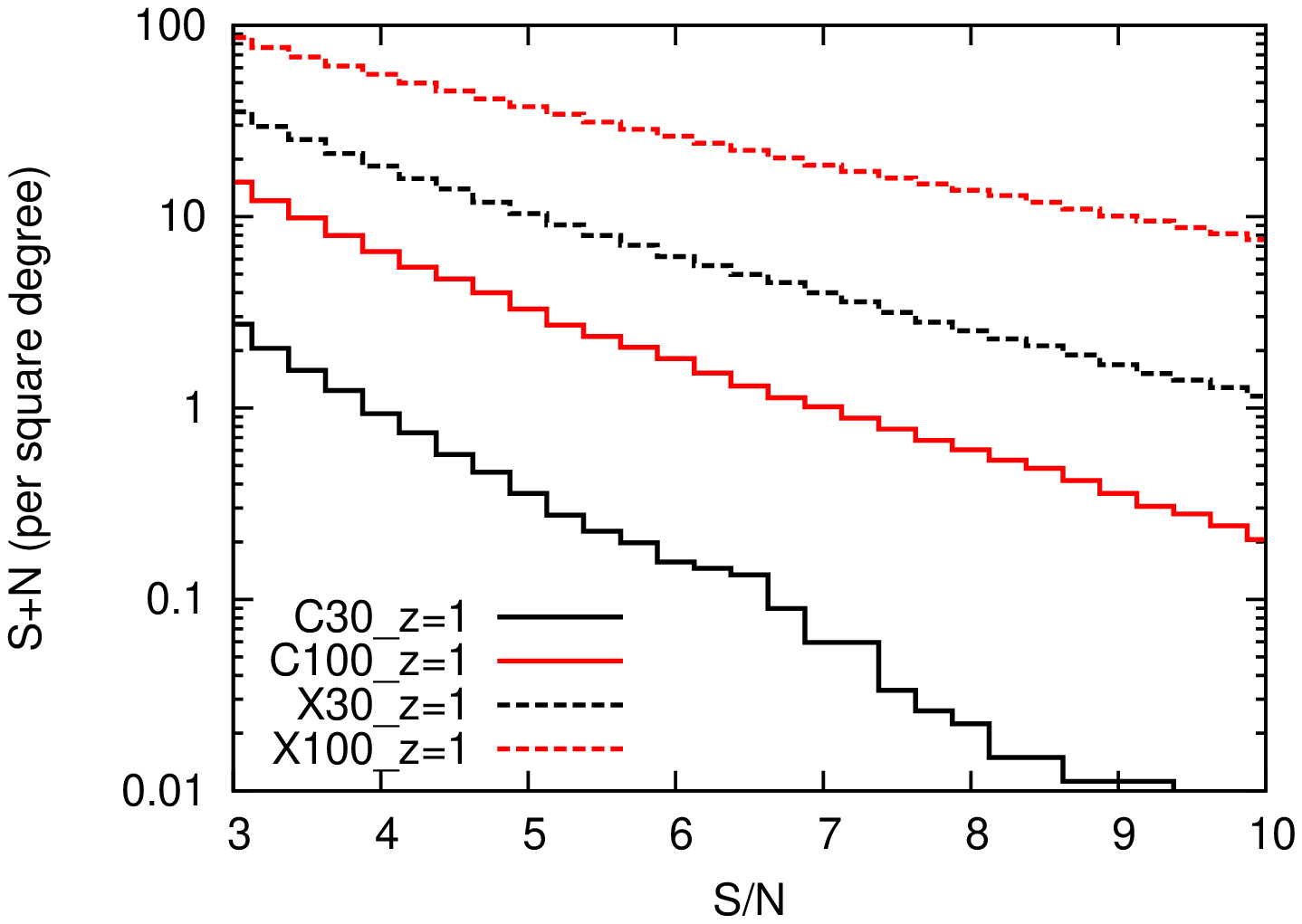}\hfill
    \includegraphics[angle=-90,width=0.49\hsize]{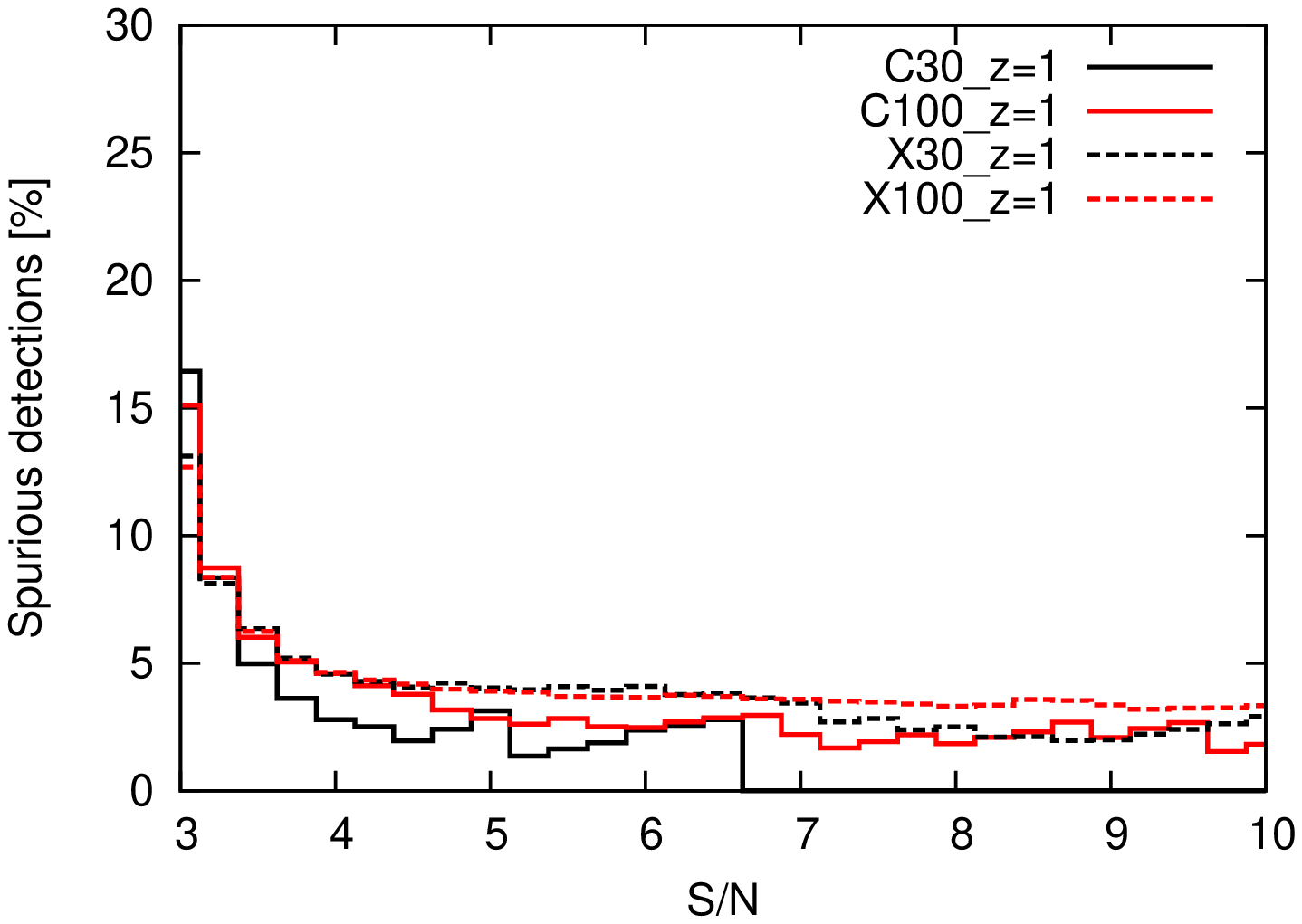}
    \includegraphics[angle=-90,width=0.49\hsize]{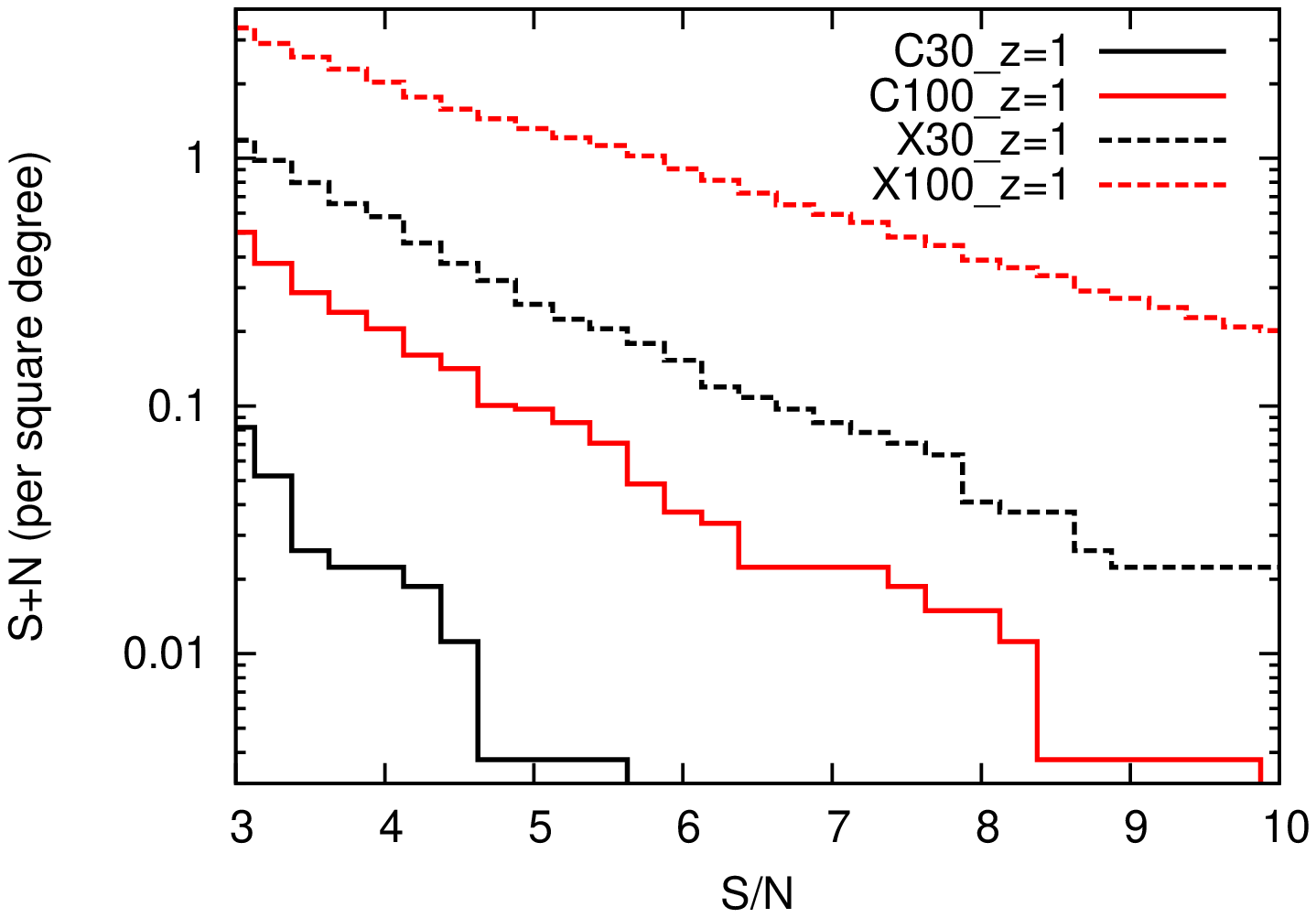}\hfill
    \includegraphics[angle=-90,width=0.49\hsize]{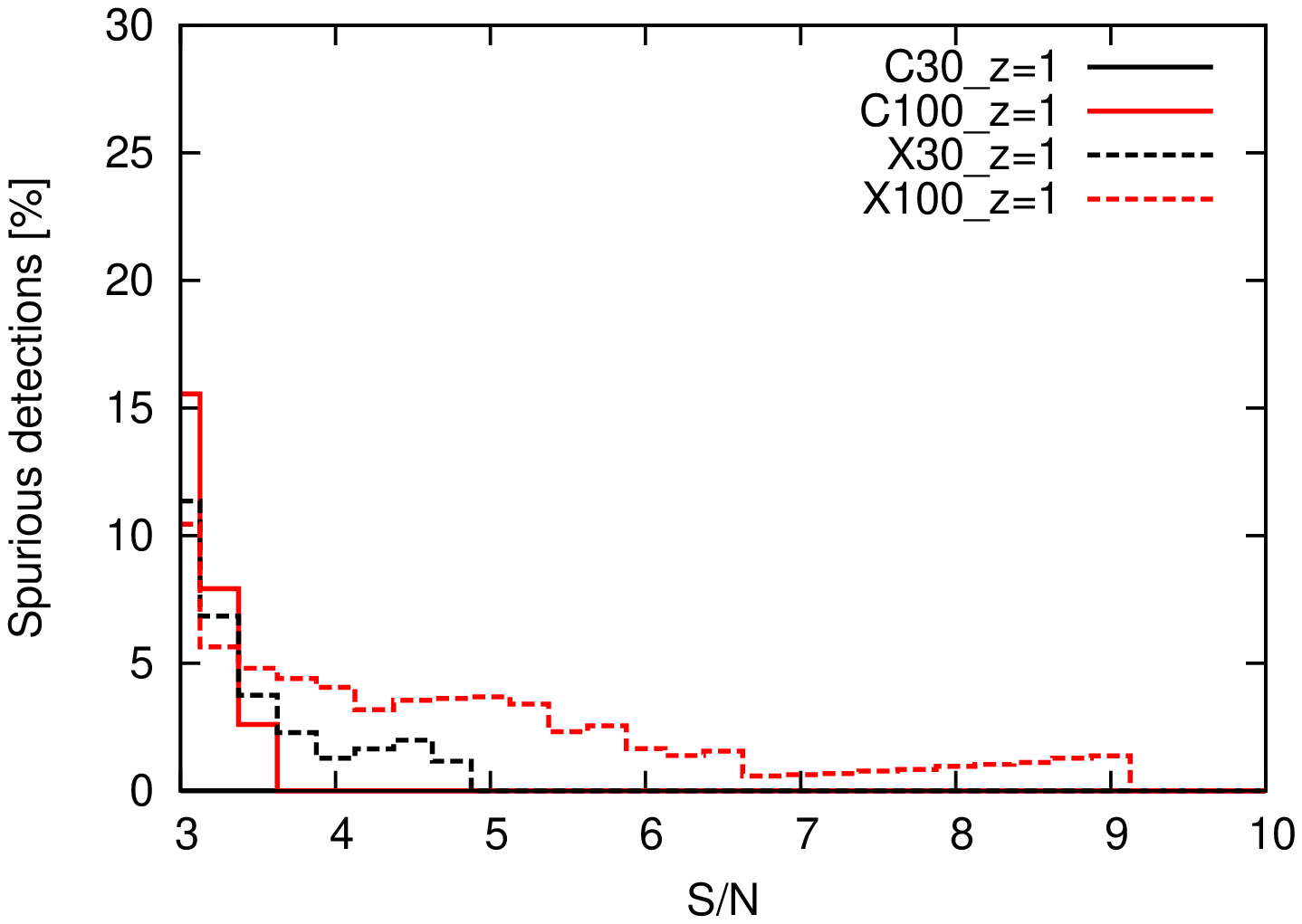}
    \includegraphics[angle=-90,width=0.49\hsize]{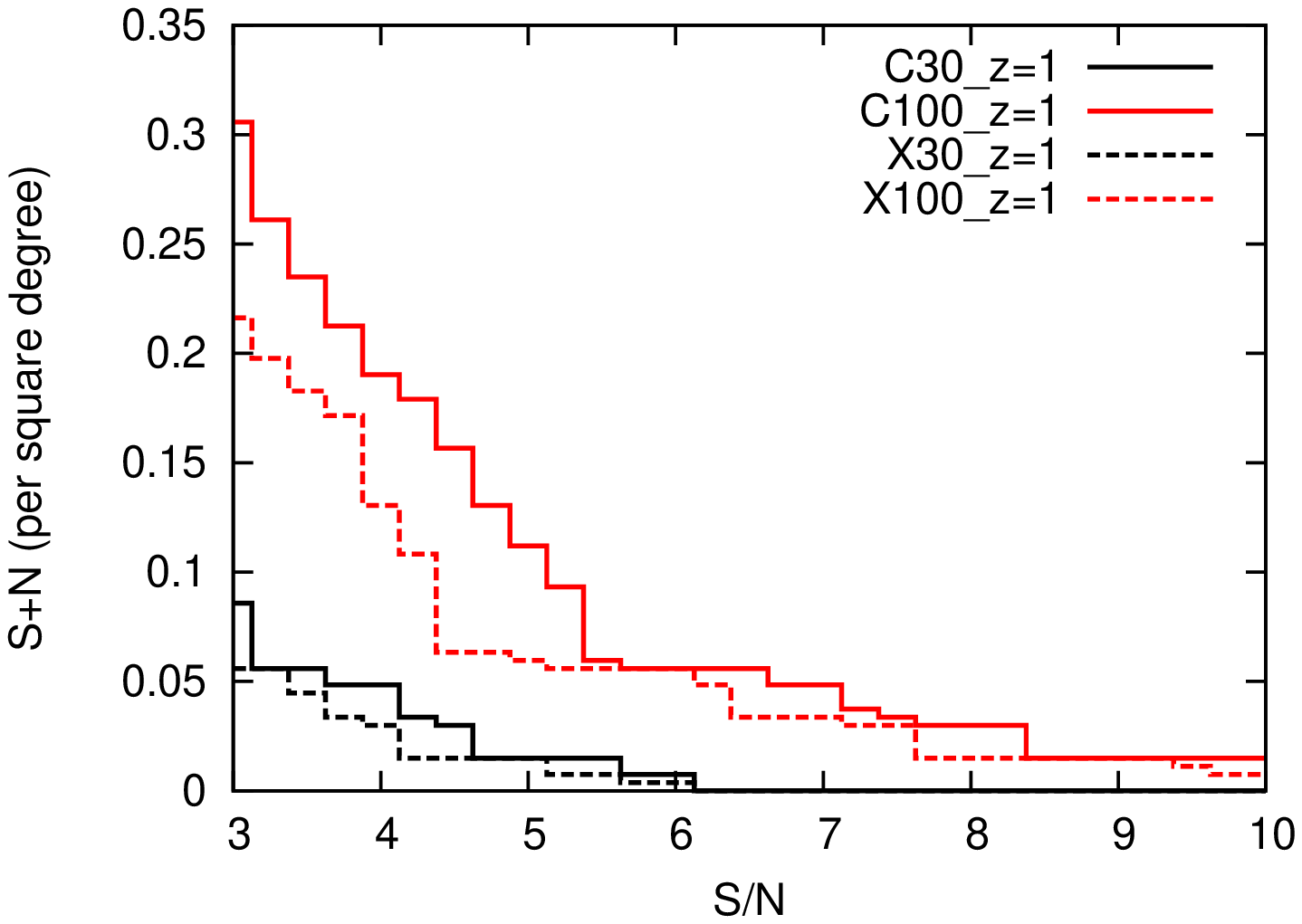}\hfill
    \includegraphics[angle=-90,width=0.49\hsize]{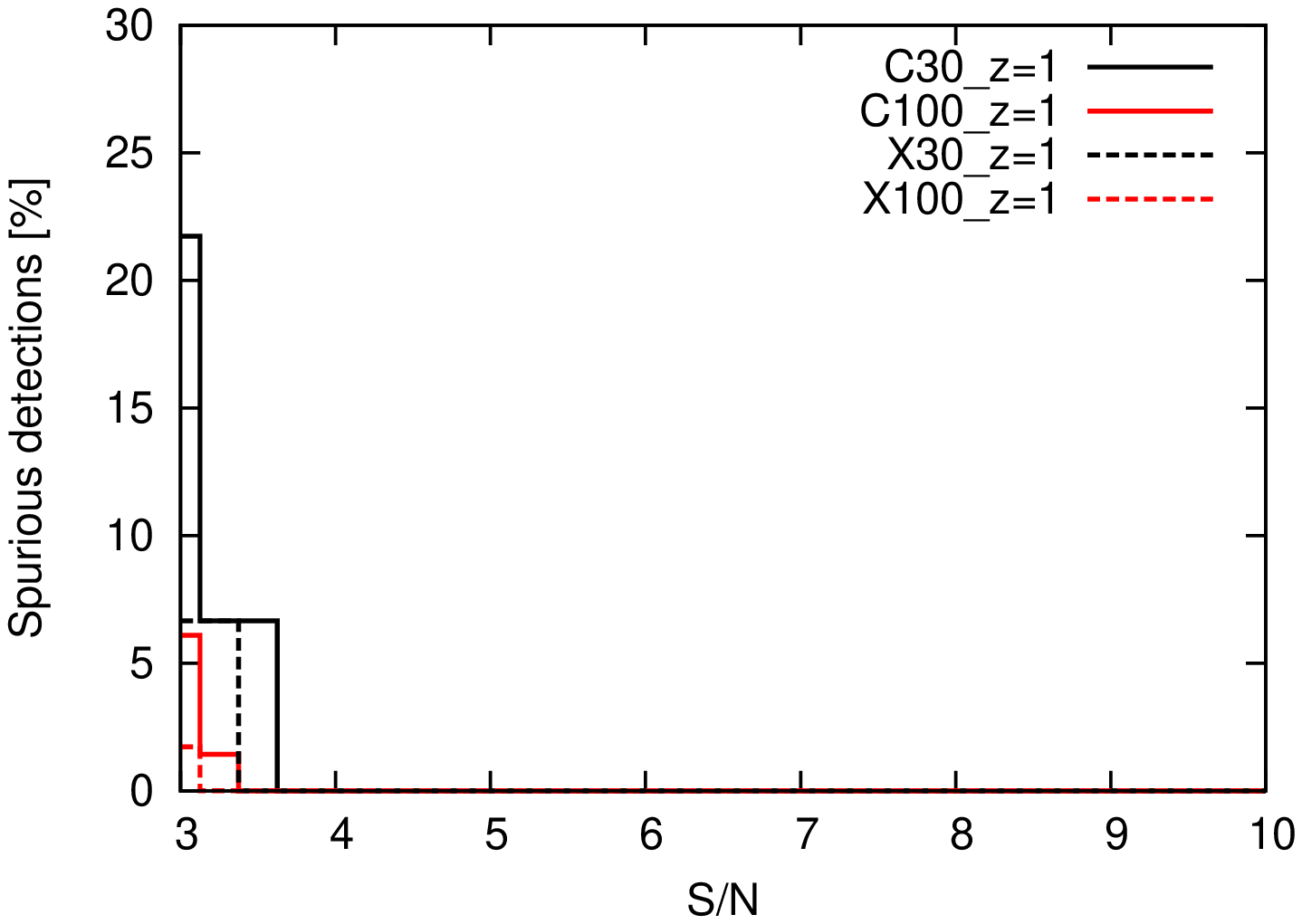}}
  \caption{Total number of X-ray detections per square degree (left panels)
    and fraction of spurious detections (right panels) as a function of the
    $S/N$ ratio. The results are shown for three different bands analysed:
    soft band (upper panel), hard band (middle panel), hardest band (bottom
    panel). The light cones extend to $z=1$. Different line styles correspond
    to Chandra and XMM-{\em Newton}, respectively), different colours to
    different exposure times (black and red curves are for 30 ks and 100 ks,
    respectively). The results are averaged over eleven realisations.}
  \label{fig:14}
\end{figure}

The left panels of Fig.~\ref{fig:14} show the total number of detections per
square degree. From top to bottom, results are shown for the soft, hard and
hardest bands. They represent the average over eleven different map
realisations and numbers are normalized to $1\,\rm{deg^2}$. The number of
detections found for XMM-{\em Newton} is at least five times higher
than with Chandra. Although XMM-{\em Newton} has a higher background
level than Chandra, its better performance is due to the different
effective area of the instrument (see Table \ref{tab:Xtel}). The
number of detections decreases quite rapidly with the S/N
threshold. We find few objects with high signal-to-noise ratios also
in the high-energy band.

The fraction of spurious detections is presented in the right panels of
Fig.~\ref{fig:14}. We notice that it is always below $20\%$ in all bands for
detections with $S/N\gtrsim3$, and that it drops to a constant value
$\approx5\%$ independent of exposure time and instrument. This is due to few
spurious peaks having quite high $S/N$ ratio. Since also the number of real
objects with high S/N ratio is approximately constant over a large range of
S/N, their ratio is almost constant. These few, but prominent spurious
detections are due to regions in the maps where several low-mass halos are
approximately aligned along the line-of-sight.

\begin{figure}[!ht]
  \centering{
    \includegraphics[angle=-90,width=0.49\hsize]{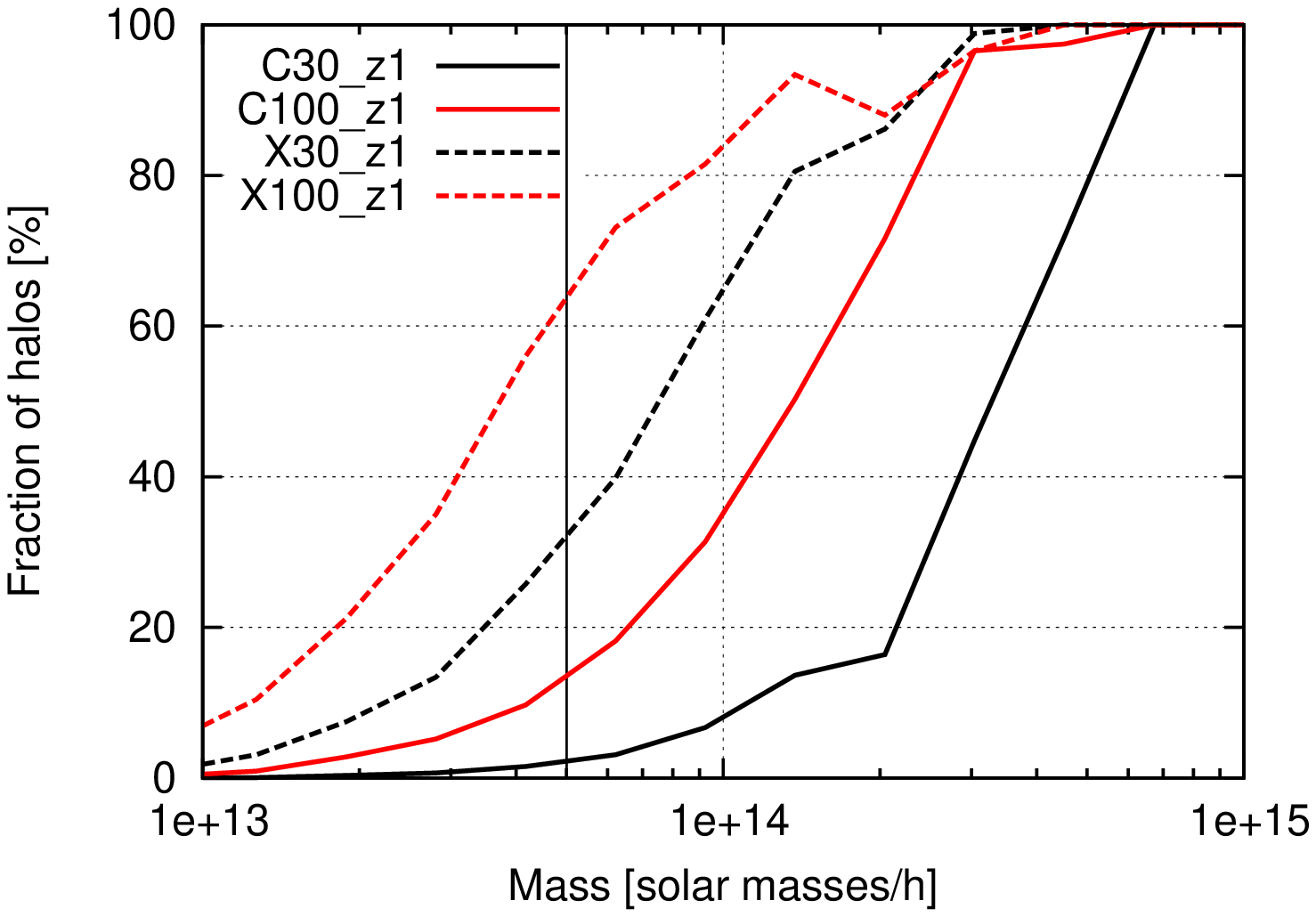}\hfill
    \includegraphics[angle=-90,width=0.49\hsize]{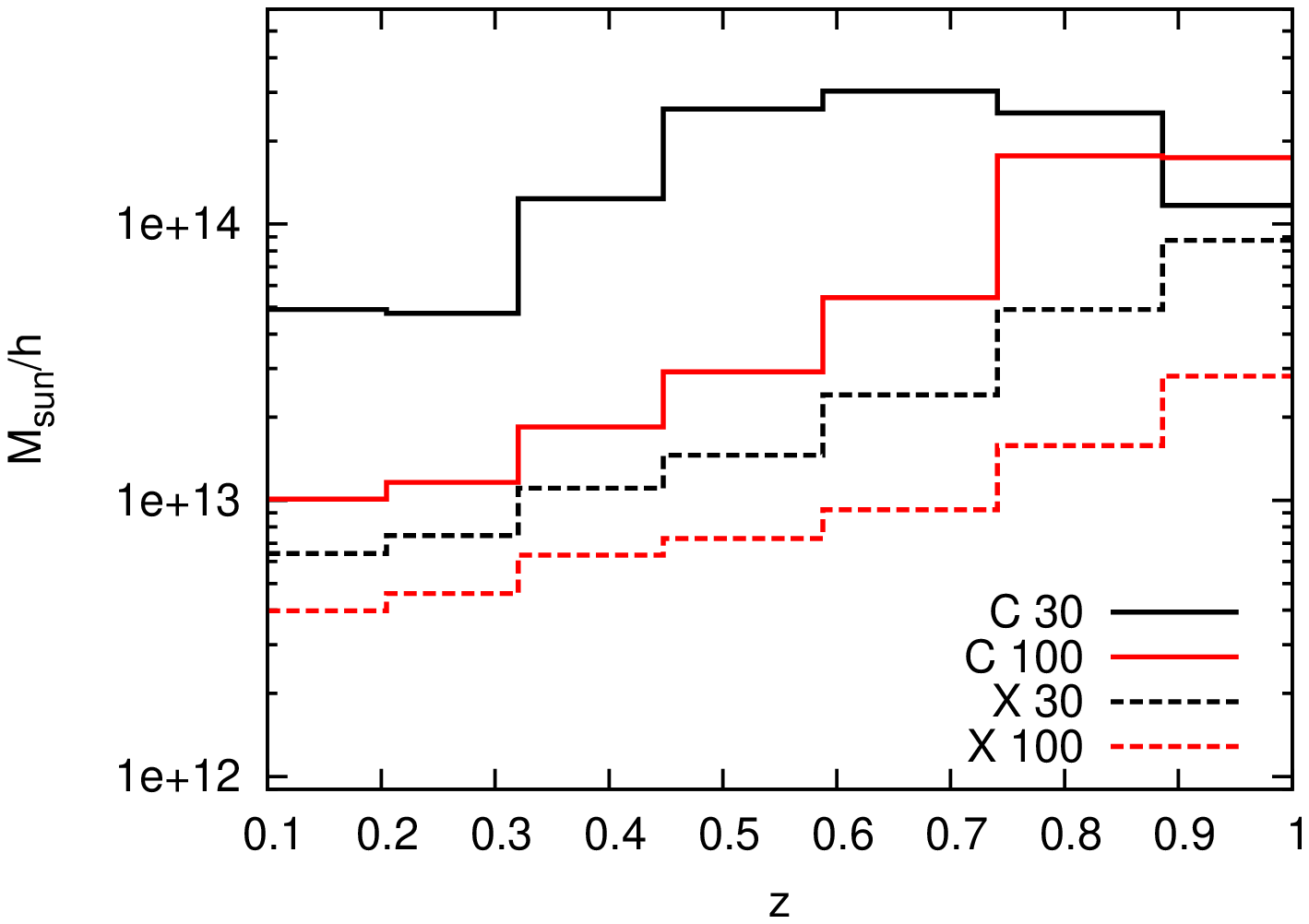}
    \includegraphics[angle=-90,width=0.49\hsize]{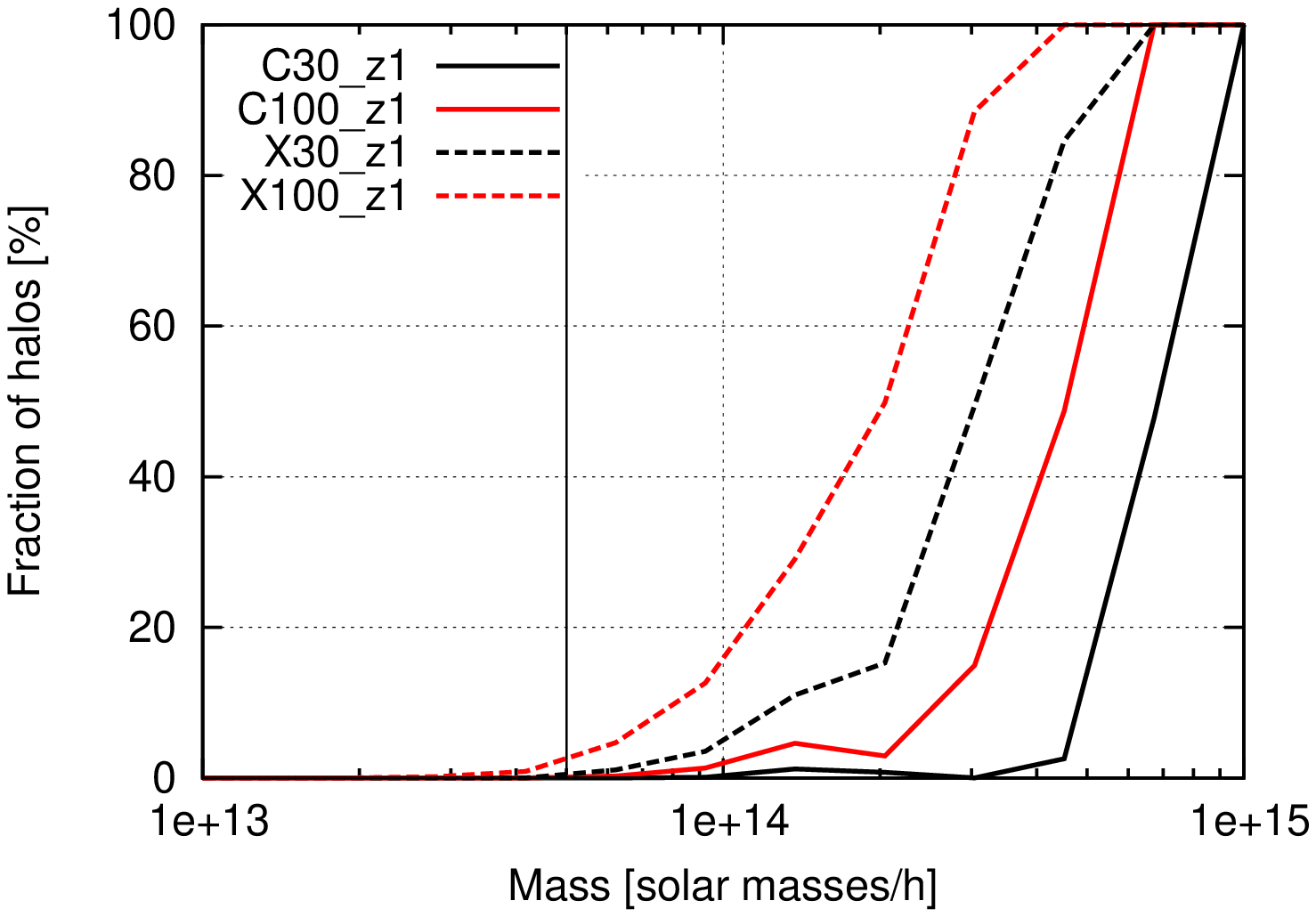}\hfill
    \includegraphics[angle=-90,width=0.49\hsize]{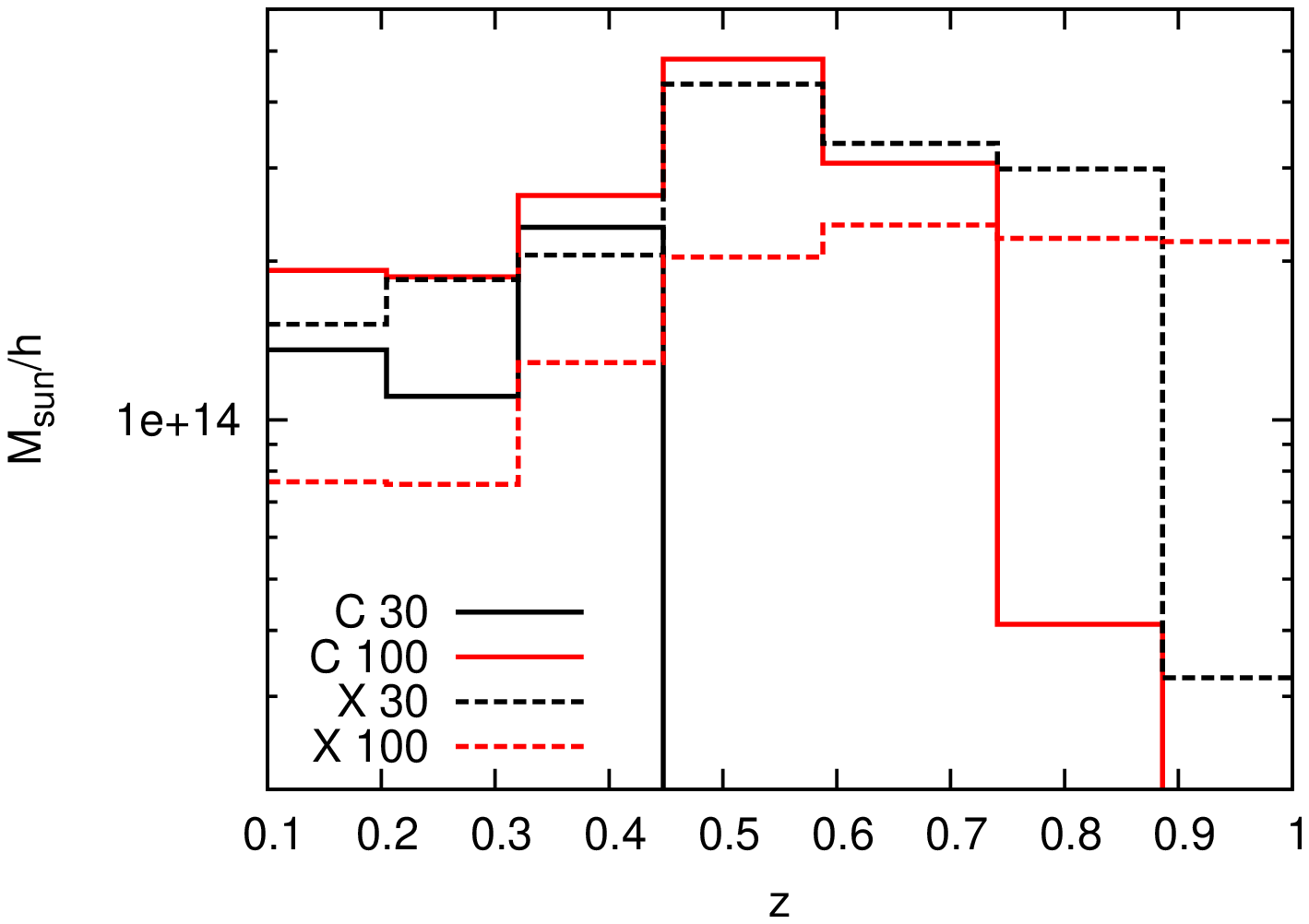}
    \includegraphics[angle=-90,width=0.49\hsize]{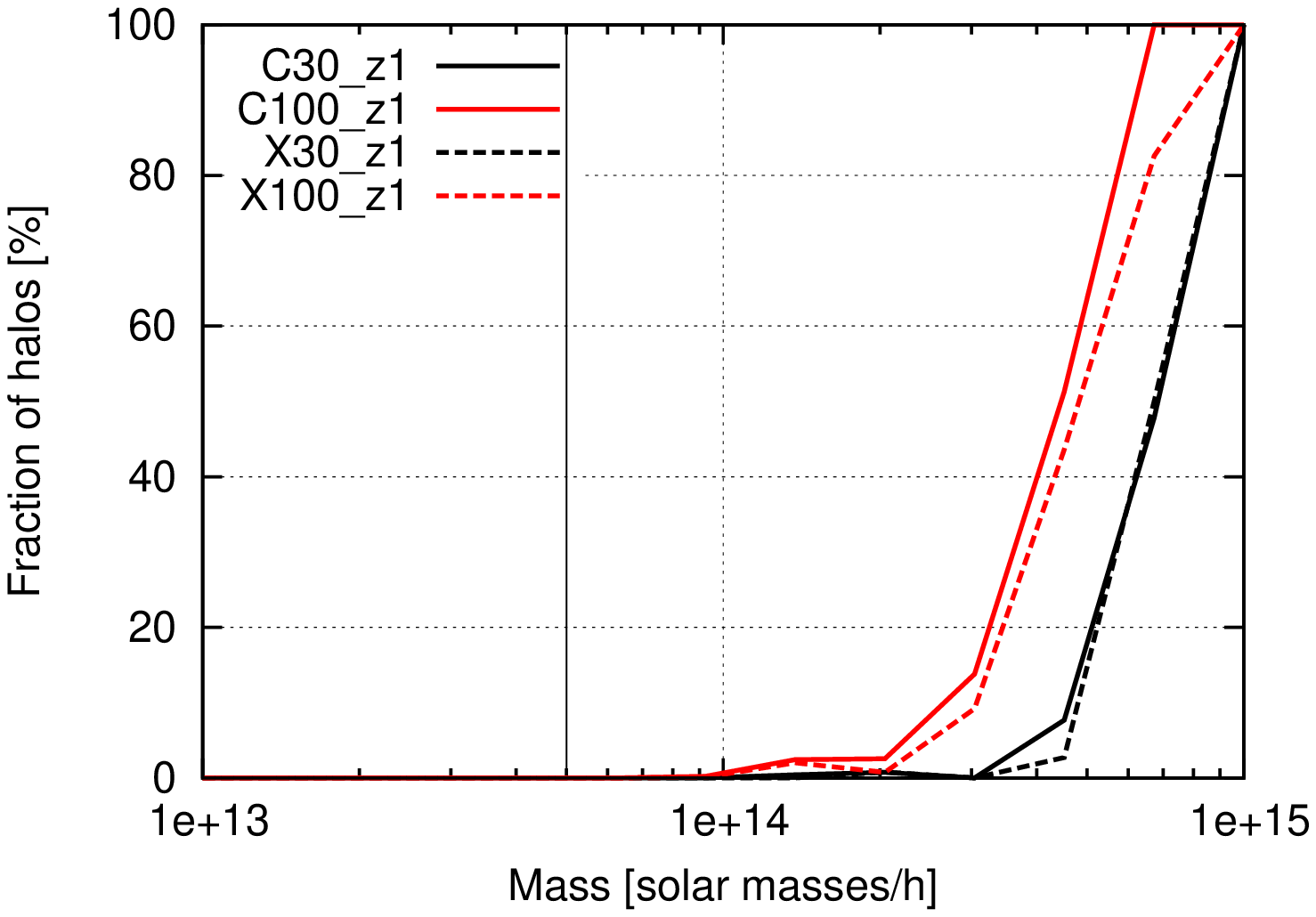}\hfill
    \includegraphics[angle=-90,width=0.49\hsize]{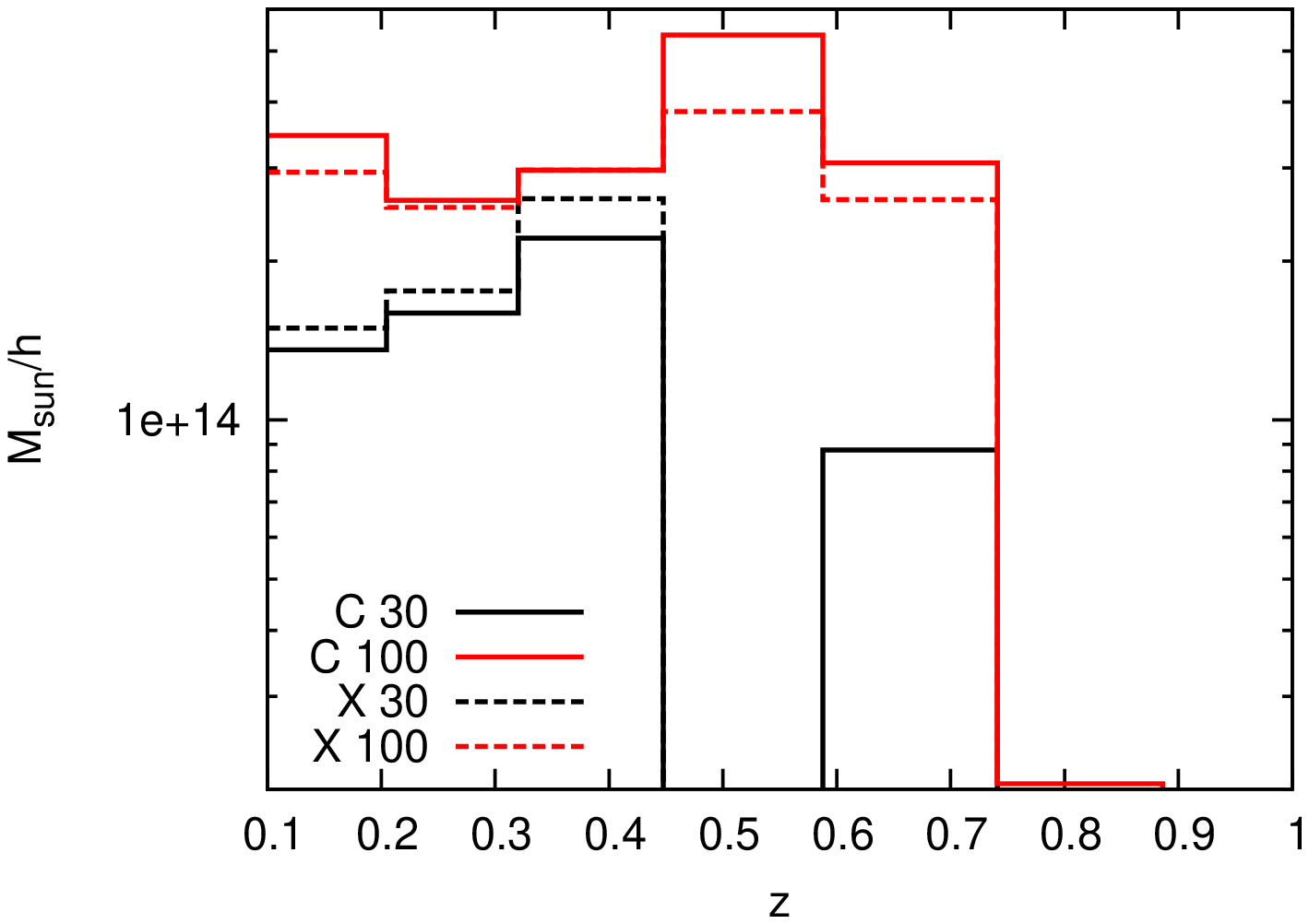}}
  \caption{Fraction of halos detected as a function of halo mass (left panels)
    and sensitivity of the method (expressed in terms of minimum detectable
    mass, right panels) in three different X-ray bands: soft band (upper
    panel), hard (middle panel), and hardest (bottom panel). Light cones
    extend to $z=1$. Line-styles and colours refer to different instruments
    and exposure times, as in Fig.~\ref{fig:14}. The results are averaged over
    eleven realisations.}
  \label{fig:15}
\end{figure}

Figure~\ref{fig:15} shows the completeness of the detections. In the soft and
hard bands, the detection catalogues for XMM-{\em Newton} are always more
complete than for Chandra. In the soft band, the completeness decreases much
more slowly as a function of mass than in the harder bands. The completeness
reaches $100\%$ for masses $M\gtrsim 2\times 10^{14}~M_\odot/h$, and it is
$\sim 50\%$ at $M\sim 3\times 10^{13}~M_\odot/h$ for XMM-{\em Newton} with an
exposure time of 100 ks, while it is of the order of few percent for
Chandra. In the hard band, Chandra detects only halos with mass $M>2-3\times
10^{14}~M_\odot/h$, and the detection threshold drops by a factor four with
XMM-{\em Newton}. Results for the soft band are similar, but the softer
photons of lower-mass halos lower the detection limit to $M\approx
10^{13}~M_\odot/h$. Conversely, only halos with masses exceeding a few times
$10^{14}~M_\odot/h$ can be detected in the hardest band. There, the
completeness for Chandra is almost as high or even larger than for XMM-{\em
  Newton} for an integration time of 100~ks, because only very massive halos
are detected, and Chandra has a better resolution than XMM-{\em Newton} (see
Table \ref{tab:Xtel}). As the X-ray emission depends mostly on the
particles of the halo core, it might be that X-ray properties for
halos of mass $M\approx 10^{13}\,M_\odot/h$ are not very well
converged, therefore reliable results can only be safely inferred to
halos of mass $M\approx 5\times 10^{13}\,M_\odot/h$. The limit is
shown in the left panels of Fig.~\ref{fig:15} by the black vertical
solid line.

The right panels of Fig.~\ref{fig:15} show the sensitivity for the X-ray
detections. In the soft band, Chandra performs worse than XMM-{\em Newton} and
thus requires larger masses for detection. In the hard and hardest bands, the
sensitivity at low redshifts is similar for both satellites, even if Chandra
performs slightly better because of its better resolution. Chandra finds halos
only up to $z\sim 0.4$ in the hard band (for an integration time of 30 ks),
while XMM-{\em Newton} reaches $z\sim 0.9$.

As expected, only the most massive clusters can be identified at high
redshifts. In particular we note that the detection threshold increases
approximately by a factor of 5 between $z\approx 0.1$ and $z\approx 1$. The
mass of detected halos also increases in the hard and hardest bands because
only the most massive halos emit such energetic photons. Finally, we recall
that multi-band filtering is pointless for X-ray analyses because the noise in
the different bands is uncorrelated, hence a multi-band filter would reduce to
a single-band filter.

Compared to \cite{AFIetal2007.1}, who used a wavelet scale-wise reconstruction
of the image \citep[for details see][]{AVIetal1998.1}, we find a higher
contamination in our sample, due to the fact that unlike \cite{AFIetal2007.1}
we do not use any optical information, but the amount of X-ray detections with
$S/N>3$ is similar.

\section{Correlation between X-ray and multi-band SZ detections}
\label{sect:correlation}

We now describe the properties of common detections in the X-ray and
SZ maps. We only consider X-ray detections in the soft band because
their number was highest, and correlate them with multi-band SZ
detections. A major problem is caused by the different angular scales
of the detections in the different maps. Objects detected in X-rays
are much smaller and more numerous. Thus, many of them may overlap
with the same SZ detection according to the exposure time. For
XMM-{\em Newton} with an integration time of 100 ks, we found that a
single SZ detection on average overlaps with at least ten X-ray
detections. To solve this problem we only considered the brightest
X-ray detection contained in the portion of the map associated to the
SZ detection.

\begin{figure}[!ht]
  \centering{
    \includegraphics[angle=-90,width=0.49\hsize]{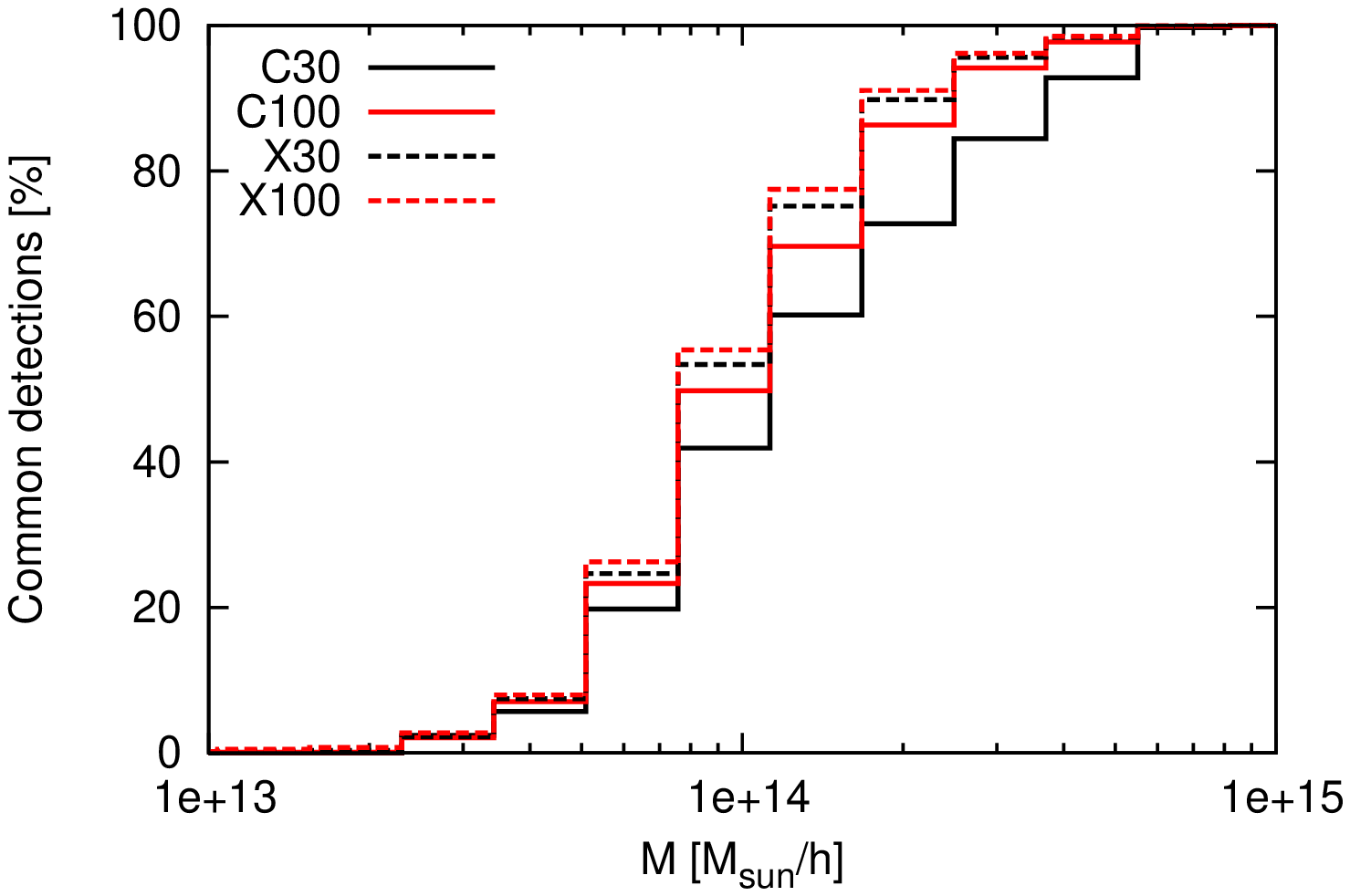}\hfill
    \includegraphics[angle=-90,width=0.49\hsize]{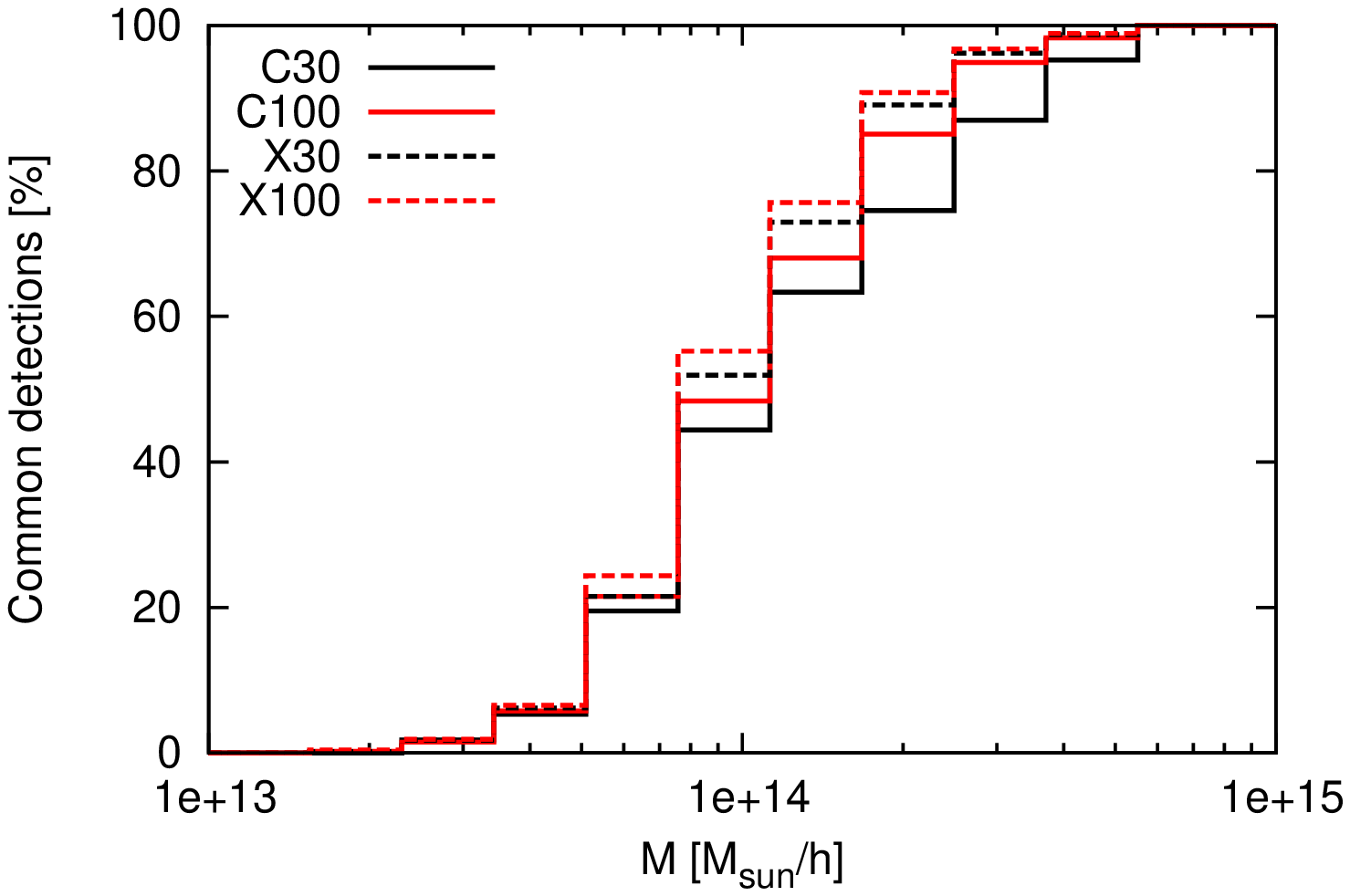}
    \includegraphics[angle=-90,width=0.49\hsize]{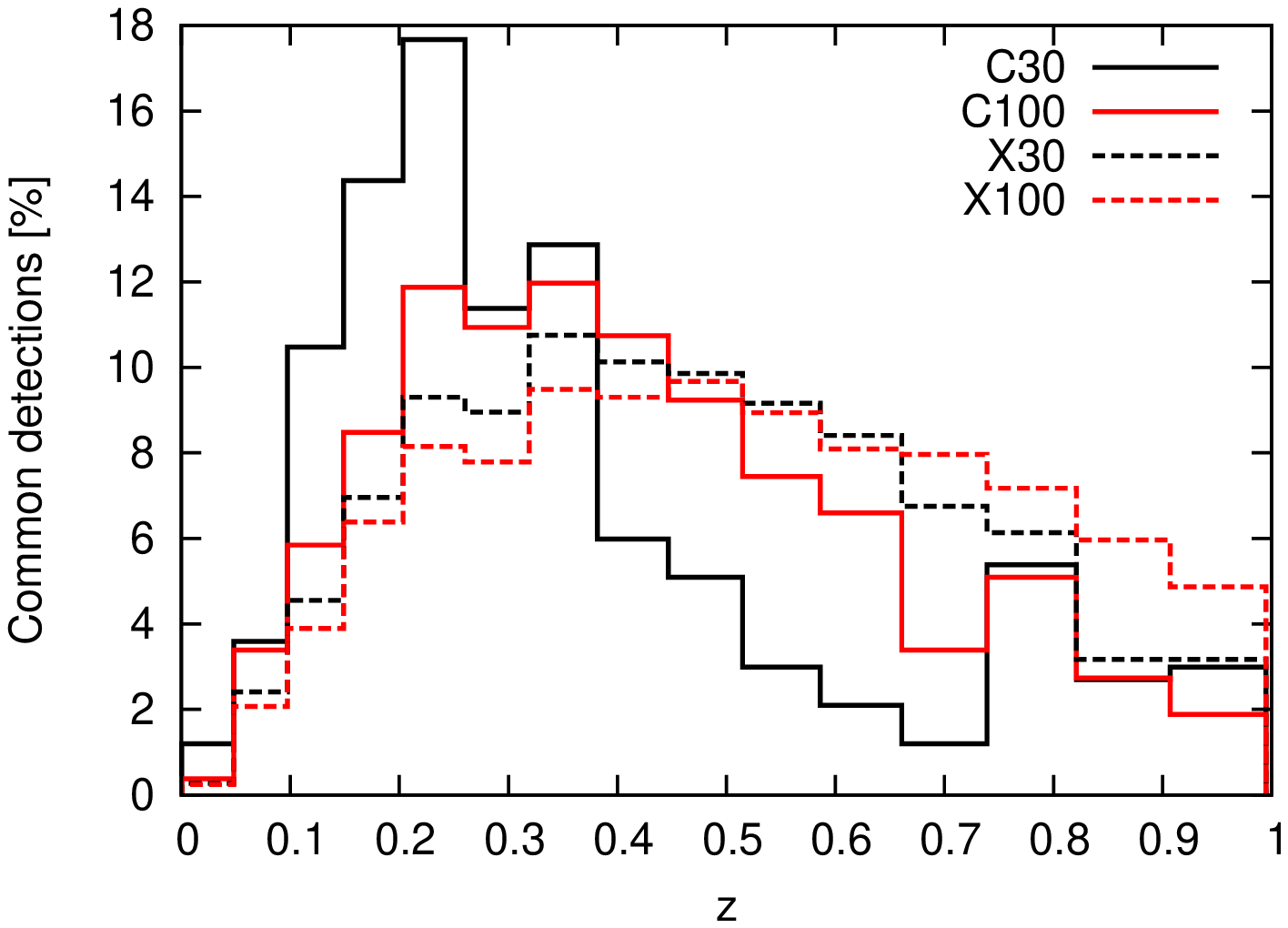}\hfill
    \includegraphics[angle=-90,width=0.49\hsize]{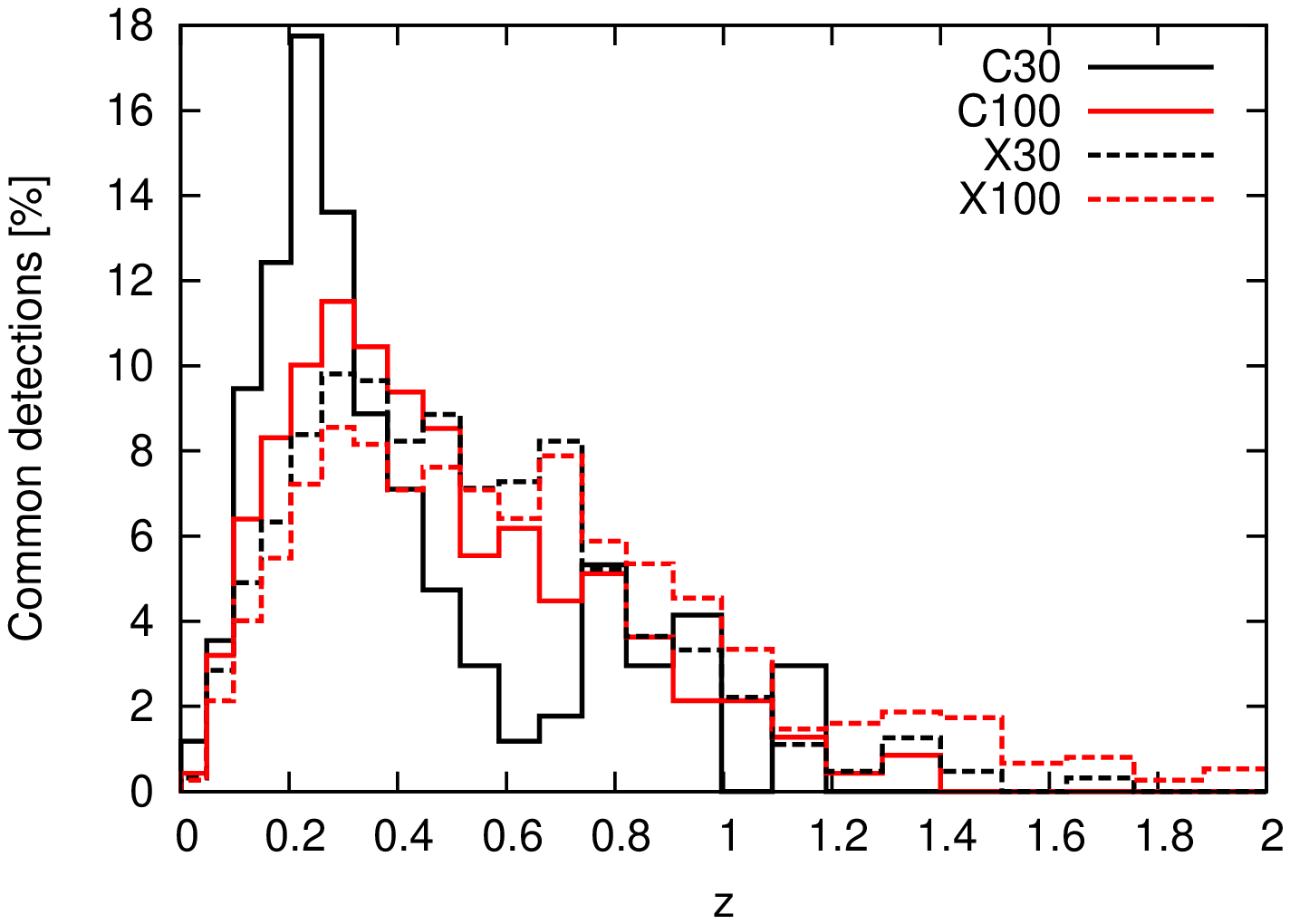}}
  \caption{Top panels: cumulative mass distribution for halos detected in both
    X-rays and SZ maps. Results are shown for light cones extending to $z=1$
    (left panels) and $z=2$ (right panels). Bottom panels: redshift
    distributions for the same halos. The solid line refers to Chandra, the
    dotted line to XMM-{\em Newton}. The black and red curves refer to the two
    different integration times (30 and 100 ks, respectively).}
  \label{fig:16}
\end{figure}

Figure~\ref{fig:16} shows the percentage of true detections in the X-ray
observations which have a SZ counterpart. We show the cumulative percentage as
a function of halo mass (top panels), and a differential distribution as a
function of redshift of the detected halos (bottom panels). The left and right
panels refer to light-cones with limiting redshifts $z=1$ and $z=2$,
respectively. Different line styles correspond to different instruments in the
X-ray observations, while different colors refer to different integration
times. The curves are normalised to the total number of clusters present in
both X-ray and SZ catalogues. The mass distribution of the common detections
reflects the mass sensitivity of the SZ multi-band filter (see
Fig.~\ref{fig:13}), while the redshift selection reflects the dependence of
the X-ray sensitivity on redshift, which drops towards high redshift.

In the lower panels of Fig.~\ref{fig:16}, we notice that the maximum fraction
of common detections occurs approximately at $z=0.3$, due to the X-ray
sensitivity. Several X-ray detections will be masked by a single SZ detection
and thus not seen. This implies that combining SZ with X-ray maps will result
in a halo number strongly biased by the angular extent of the SZ detection. A
given SZ halo will mask X-ray clusters projected into its close neighbourhood,
independently of their redshift. This can be problematic for studies of
substructures because it is impossible to identify substructures in the SZ
map, and it will become difficult to study clusters at intermediate
redshifts. This will be addressed in more detail in future work.

\begin{figure}[!ht]
    \centering{
      \includegraphics[width=0.49\hsize]{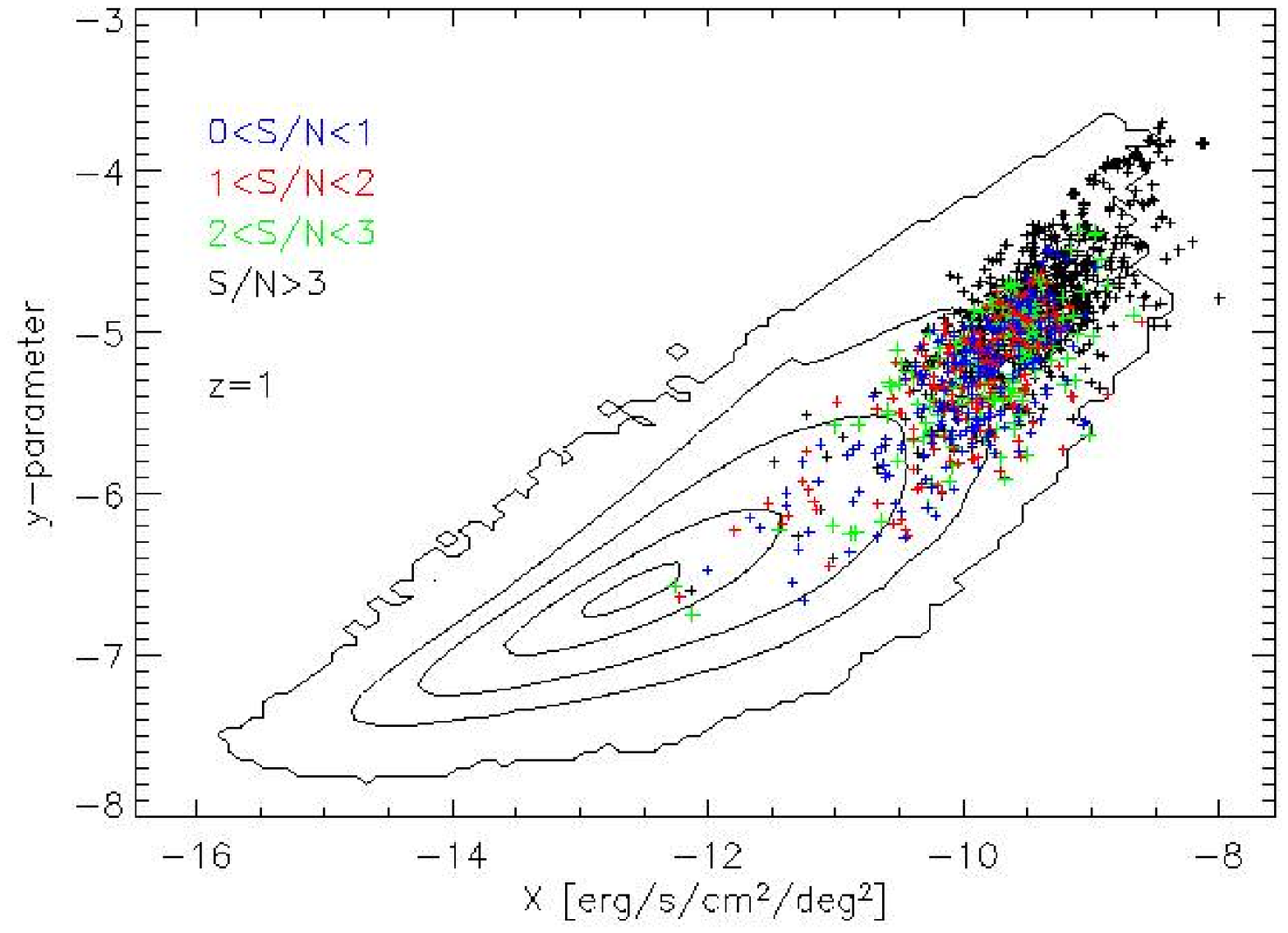}\hfill
      \includegraphics[width=0.49\hsize]{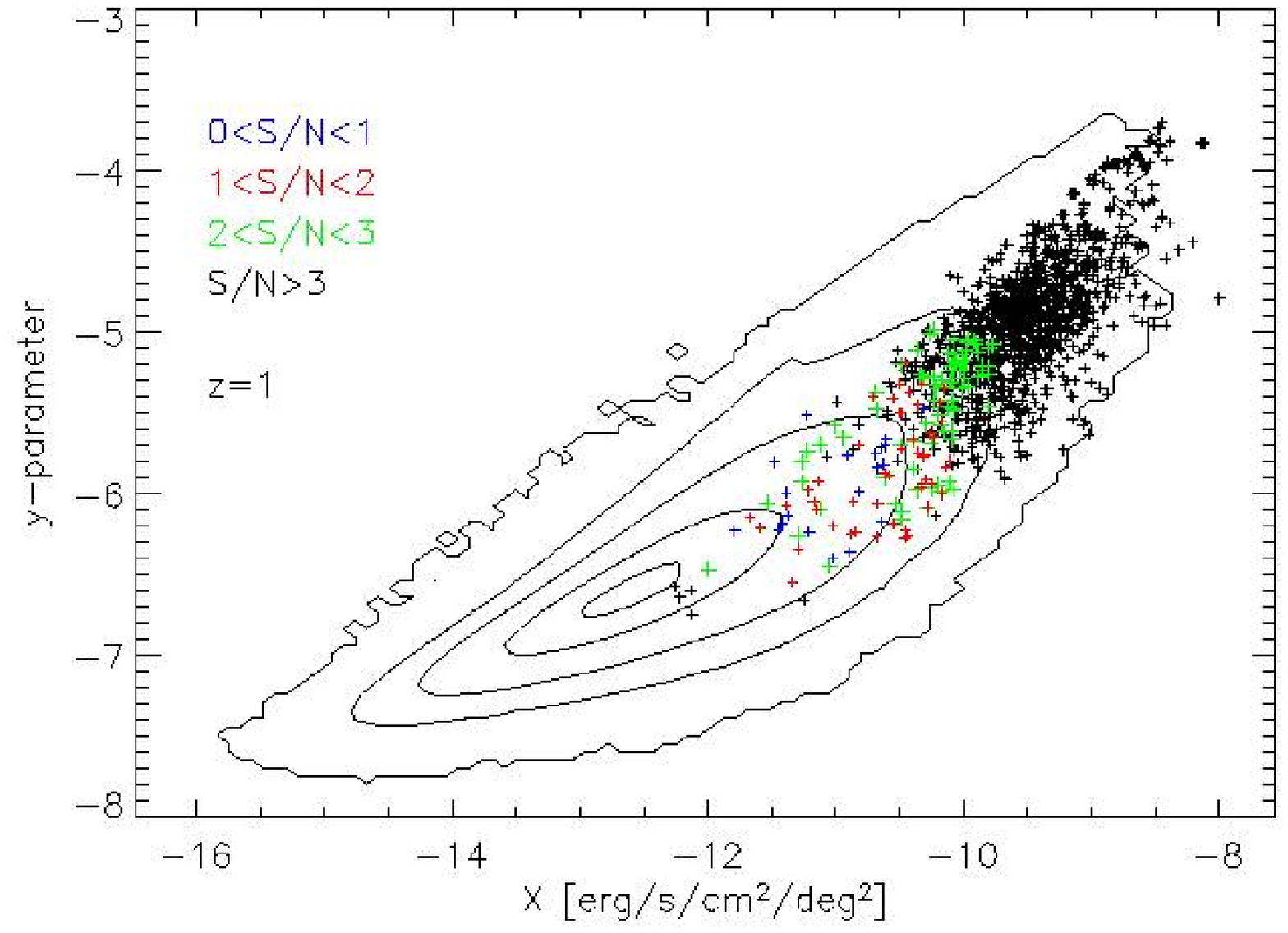}\\
      \includegraphics[width=0.49\hsize]{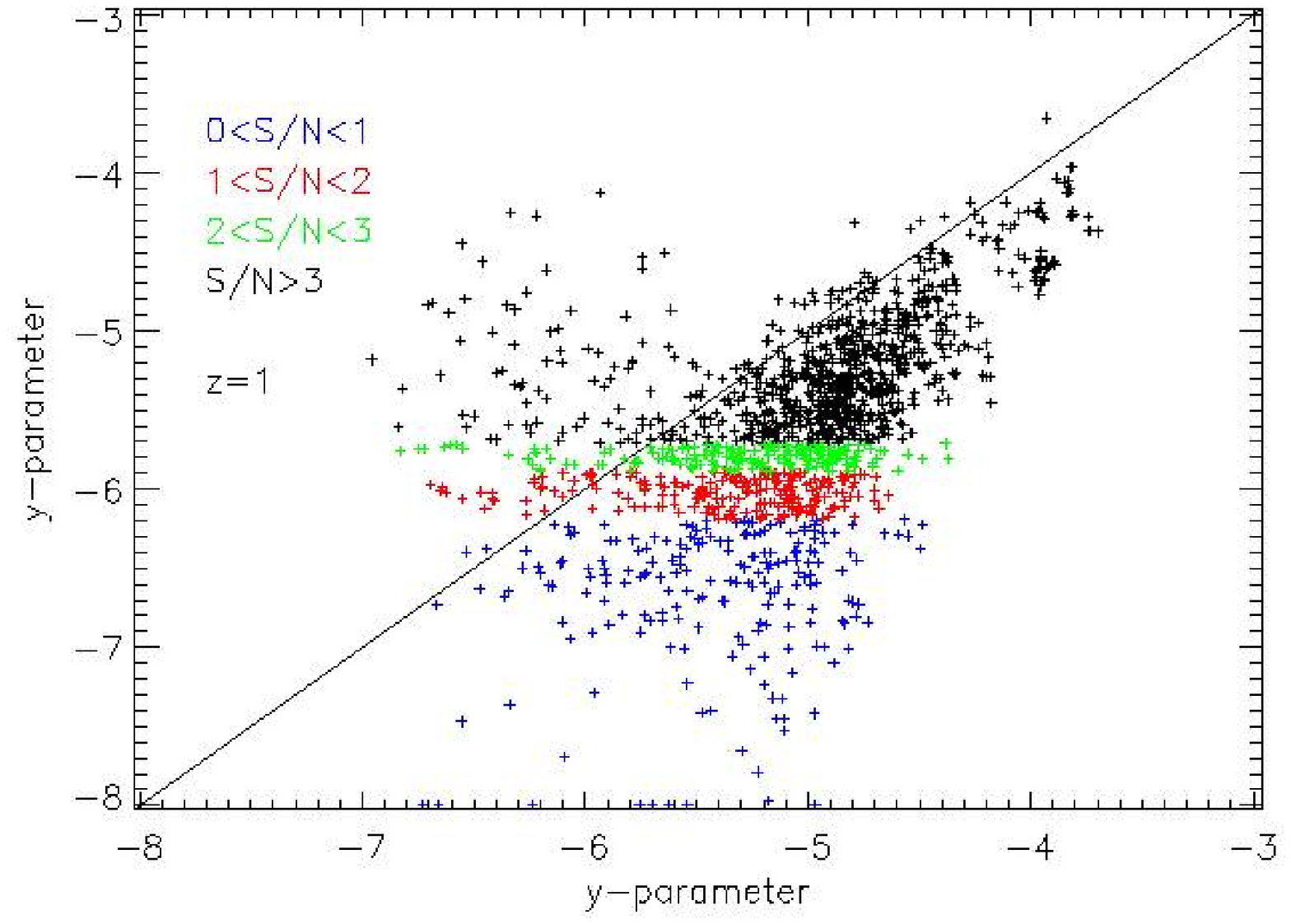}\hfill
      \includegraphics[width=0.49\hsize]{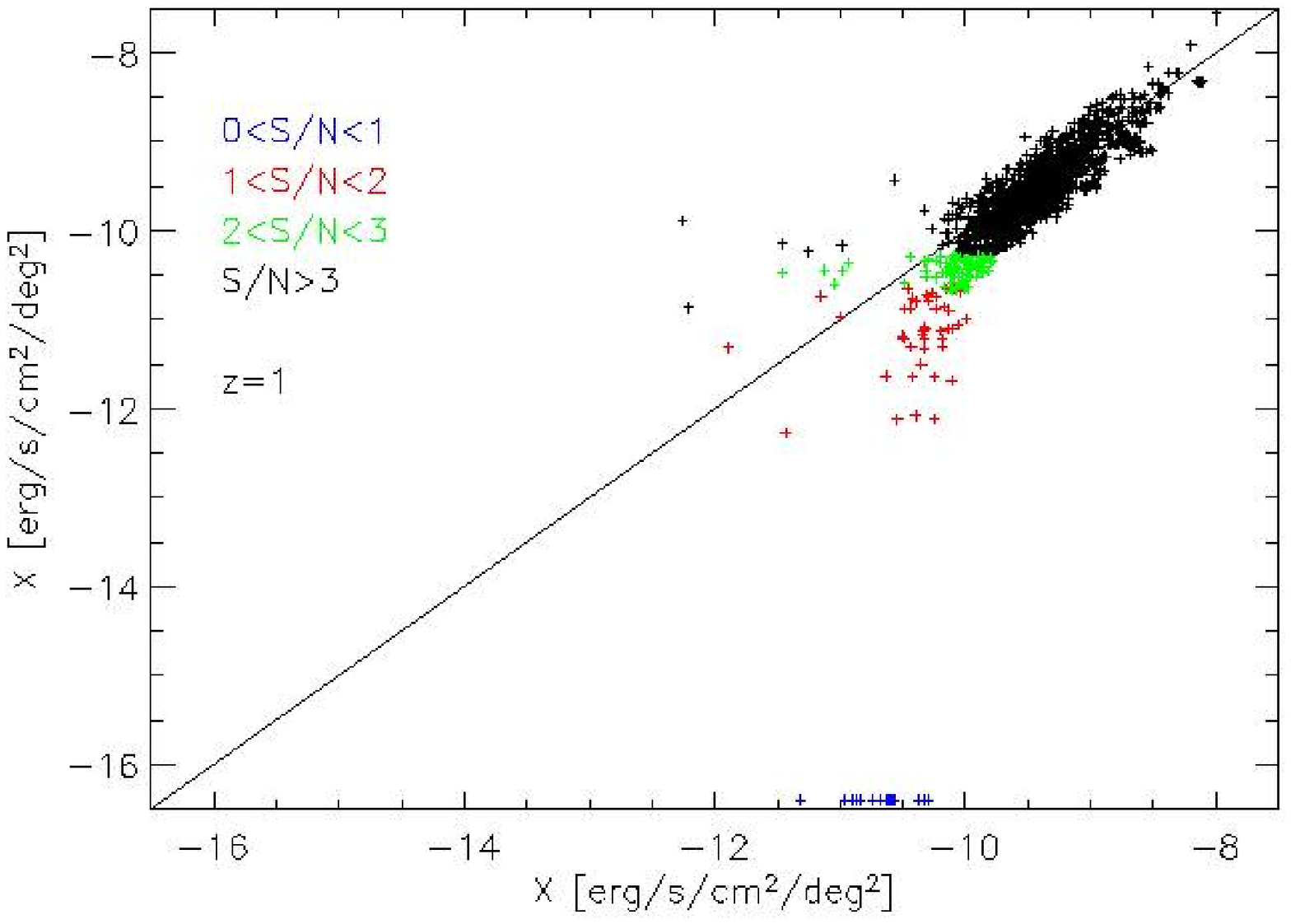}}
  \caption{Upper panels: correlation between the tSZ and X-ray signals
    ($y$- and $x$-axes, respectively). The results are compared to the
    distribution of common detections, and are shown as a function of
    their S/N ratios (blue for $0<S/N<1$, red for $1<S/N<2$, green for
    $2<S/N<3$, and black for $S/N>3$). The left and right panels refer
    to the S/N values obtained using the multi-band SZ  and the
    (soft-band) X-ray filter, respectively. Lower panels: correlation
    between the values in the filtered maps ($y$-axis) and the
    original ones ($x$-axis) for the multi-band SZ filter (left panel)
    and the (soft-band) X-ray filter (right panel). Again detections
    are binned using their S/N ratios. The diagonal line would
    correspond to a perfect agreement between original and filtered
    maps. All plots refer to results for light-cones extending up to
    $z=1$.}
  \label{fig:17}
\end{figure}

The upper panels of Fig.~\ref{fig:17} show the correlation
between the X-ray and tSZ signals of the common real detections
grouped according to their S/N ratio significance, overlaid with the
theoretical SZ and X-rays correlation as obtained in the original
simulation. The analysis is limited to the soft band for XMM-{\em
  Newton} and an exposure time of 100 ks.

As expected, detections with $S/N>3$ occupy the upper right part of the
contours ($y>10^{-6}$ and $X>10^{-11}$~erg/s/cm$^2$/deg$^2$) because
they correspond to the most massive (and therefore brighter) halos. We
also note that we do not find any detection with $y<10^{-7}$ and
$X<10^{-13}$~erg/s/cm$^2$/deg$^2$ because these values of the flux
correspond to objects whose mass is of the order of the minimum mass
that the filters can detect.

The lower panels show, the correlation between the filtered maps
(y-axis) and the original maps (x-axis). On the left, we show the SZ
multi-frequency filter, on the right the X-ray maps. We notice a
substantial scatter for low Compton-$y$ parameters in the SZ compared
to the X-ray maps, corresponding to detections with very low S/N
ratio. We also note that, while the filtered X-ray maps show unbiased
values for $S/N>3$, this is not strictly true for the multi-band SZ
filter: the estimated amplitudes are biased towards lower
values. This may be due to the fact that the $\beta$-model does not
exactly reproduce the signal distribution of the SZ-halos. Finally, we
found that the signal-to-noise ratios of the common SZ and X-rays
detections are uncorrelated.

\section{Conclusions}
\label{sect:conclusions}

We have tested the performance of linear filtering in optimising and combining
galaxy-cluster detections in X-ray and SZ observations. Based on a large
$N$-body hydrodynamical cosmological simulation, we have studied the halo
selection function, the contamination of halo catalogues and the mass
sensitivity of the filtering technique. Our work extends earlier studies by
taking gas physics properly into account and treating SZ and X-ray objects
simultaneously with similar techniques.

We constructed light cones up to redshifts 1 and 2 and simulated typical
observations of the X-ray emission (modelled after the XMM-{\em Newton} and
Chandra observatories) and the SZ signal (adapted to ACT) produced by the
matter contained in the light cones. Our set of 11 different realisations
covers total areas of $\sim 264~\rm{deg^2}$ and $\sim 106~\rm{deg^2}$
for limiting redshifts $z=1$ and $z=2$, respectively.

We constructed and used single- and multi-band matched filters for halo
detections in the synthetic SZ maps, and a single-band filter for the X-ray
detections. We identified halos as S/N peaks in the filtered images and
compared them with the parent distribution of dark matter halos contained in
the original cosmological simulation.

We find that multi-band filtering considerably reduces the number of spurious
SZ detections and increases the fraction of halos reliably identified in the
light cones. For instance, $50\%$ of the halos with masses $M\simeq
10^{14}M_\odot/h$ are found when using a multi-band filter, compared to $30\%$
with the single-band filter. The mass sensitivity of the multi-band filter is
approximately four times higher than for the single-band filter. As expected,
the sensitivity is virtually independent of redshift.

The fraction of spurious detections in the X-ray maps is typically low, and
almost constant (about $5\%$) for a large range of S/N ratios. On the whole,
the X-ray catalogues are more complete than the SZ catalogues, even when
multi-band SZ filters are used. However, the redshift distribution of the
halos detected through their X-ray emission is strongly peaked at low
redshifts, and only massive halos can be detected at high redshift. This
emphasises the importance of using complementary methods.

Weak-lensing halos detected with the optimal filter proposed by
\cite{MMAetal2005.1} reach a similar mass sensitivity as SZ and (soft) X-ray
detections at redshifts where the lensing efficiency is highest, but clusters
can be detected up to high redshifts. The contamination of weak-lensing halo
catalogues is comparable to the X-ray catalogues, while the SZ contamination
is much lower. Weak-lensing halo catalogues are similarly complete as SZ
catalogues. For more quantitative details, see \cite{FPAetal2007.1}.

Catalogues of common SZ and X-ray detections show that their mass function
closely follows that of the SZ detections (due to the mass sensitivity of the
filter), while their redshift distribution is very similar to that of the
X-ray detections because X-ray clusters can be observed more reliably at low
and intermediate redshifts.

Finally, we have shown that the detections represent the
simulated halos in a virtually unbiased manner (see lower panels in
Fig.~\ref{fig:17}).

\acknowledgements{We are grateful to Mauro Roncarelli for helpful suggestions
  on the simulation of SZ and X-ray maps. We thank Stefano Borgani and
  Giuseppe Murante for providing the outputs of the hydrodynamic simulation
  used in this work. Computations have been performed using the IBM-SP5 at
  Cineca (Consorzio Interuniversitario del Nord-Est per il Calcolo
  Automatico), Bologna, with CPU time assigned under an INAF-CINECA grant. We
  acknowledge financial contributions from contracts ASI-INAF I/023/05/0,
  ASI-INAF I/088/06/0 and INFN PD51. This work was supported by the Deutsche
  Forschungsgemeinschaft (DFG) under the grants BA 1369/5-1 and 1369/5-2 and
  through the Transregio Sonderforschungsbereich TR 33, as well as by the DAAD
  and CRUI through their Vigoni program.}

\bibliography{SZ_X_Halo_Correlation}
\bibliographystyle{aa}

\end{document}